\documentclass[galaxies,review,accept,moreauthors,pdftex,12pt,a4paper]{mdpi}
\usepackage{amsmath}
\usepackage{amsfonts}
\usepackage{amssymb}
\lastpage{x}
\doinum{10.3390/------}
\pubvolume{xx}
\pubyear{2014}
\history{Received: xx / Accepted: xx / Published: xx}

\Title{Self-gravitating systems in Extended Gravity}

\Author{Arturo Stabile $^{1,}$* and Salvatore  Capozziello $^{2,}$\,$^{3}$\,$^{4,}$*}

\address{%
$^{1}$ Dipartimento di Ingegneria, Universit\`{a} del Sannio, Palazzo Dell'Aquila Bosco Lucarelli, Corso Garibaldi, 107 - 82100, Benevento, Italy,\\
$^{2}$ Dipartimento di Fisica, Universit\`{a} di Napoli "Federico II", Complesso Universitario di Monte Sant'Angelo, Edificio G, Via Cinthia, I-80126, Napoli, Italy,\\
$^{3}$ INFN Sezione di Napoli, Complesso Universitario  di Monte Sant' Angelo, Ed. G, Via Cinthia, I-80126, Napoli, Italy,\\
$^{4}$ Gran Sasso Science Institute (INFN), Via F. Crispi 7, I-67100, L' Aquila, Italy.}

\corres{arturo.stabile@gmail.com and capozziello@na.infn.it}

\abstract{
Starting from the weak field limit, we discuss astrophysical applications of Extended Theories of Gravity where higher order curvature invariants and scalar fields are considered  by generalizing the Hilbert-Einstein action linear in the Ricci curvature scalar $R$.  Results are compared to General Relativity in the hypothesis that Dark Matter contributions to the dynamics can be neglected thanks to modified gravity. In particular, we consider   stellar hydrostatic equilibrium, galactic rotation curves,  and gravitational lensing. Finally,  we discuss the weak field limit in  the Jordan and Einstein frames pointing out how effective quantities, as gravitational potentials, transform from one frame to the other  and the interpretation of results can completely change accordingly. }


\begin{document}


\section{Introduction}


The today observed  Universe appears spatially flat and undergoing an accelerated expansion. Several observational data  probe this pictures \cite{riess, ast, clo, spe,carrol,sahini}  but two unrevealed ingredients are needed in order to achieve  this dynamical scenario, namely Dark Matter (DM) at galactic and extragalactic scales  and  Dark Energy (DE) at cosmological scales. In particular,  the dynamical  evolution of self-gravitating structures can be figured out  by the standard Newtonian gravity, but DM is needed  to obtain agreement with observations \cite{NFW}.

On the other hand, the effort to give a physical explanation to the
today observed cosmic acceleration has attracted a lot of  interest in the so-called Extended Theories of Gravity \cite{report,repsergei,olmo_palatini}, considered  a viable mechanism to explain the cosmic acceleration by extending the
geometric sector  without  introducing 
DM and DE. Other issues, at  astrophysical level, can be framed in the same approach \cite{annalen} and then,
apart the cosmological dynamics, a systematic analysis of such
theories urges at short scale and in the weak field  limit. Extended Theories of Gravity are then a new paradigm for gravitational physics  aimed  to address shortcomings coming  out at ultra-violet and infra-red scales. 

While it is very natural to extend General
Relativity (GR) to theories with additional geometric degrees of
freedom, recent attempts focused on the old idea of modifying the gravitational Lagrangian in
 taking into account generic functions of curvature invariants and leading to higher-order field
equations. Despite of  the positive results of GR,
the study of possible modifications of Einstein's theory
 has a long history which reaches back to the
early 1920s \cite{weyl_1,weyl_2,pauli,bach,edd,lan,buc,bic}. The proposed early amendments of Einstein's
theory were aimed toward the unification of gravity with
the other interactions of physics. Besides,
the presence of  Big Bang singularity, the flatness and
horizon problems \cite{guth} led to the statement that
Cosmological Standard Model, based on GR and Standard Model of
Particles, is inadequate to describe the universe in
extreme regimes. From Quantum Field Theory point of view, GR is a
\emph{classical}  theory  which does not work  at fundamental level in order to achieve a full and comprehensive quantum description of
spacetime and  gravity.

Despite of  modifications, the spirit of GR should be preserved since the only
effective request is that the Hilbert-Einstein action is not given a priori. This is the reason why one can deals  with {\it Extended Theories of Gravity} and not with 
{\it Alternative Gravity}. 
One of the possibilities is that gravitational interaction  could act in
different ways at different scales. In any case, the robust results of GR, at local and Solar System scales,  have to be preserved for consistency (see \cite{book} for a detailed discussion).  This is the case of $f(R)$-gravity which reduces, in principle,  to GR as soon as $f(R)\,\rightarrow\,R$.

On the other hand, the strong gravity regime \cite{psaltis}  is
another way to check the viability of these theories. In general, the formation and the evolution of stars can be considered as suitable test-beds for
Alternative Theories of Gravity.  Considering the case of $f(R)$-gravity,  divergences
stemming from the functional form of $f(R)$ may prevent the existence of
relativistic stars in these theories \cite{briscese}. There are
also numerical solutions corresponding to static star configurations with strong gravitational fields \cite{babi_1,babi_2} where the choice of the
equation of state  is crucial for the existence of solutions.

Some observed stellar systems  are incompatible with the
standard models of stellar structure. We refer to anomalous neutron stars, the so called 
"magnetars"  \cite{mag}  with masses larger than their expected  Volkoff mass.
It  seems that, on particular length scales, the gravitational force is larger
or smaller than the corresponding GR value \cite{neutron1,neutron2}.

A lot of research work is  pointing out that physics of compact objects could be the {\it experimentum crucis} to retain or rule out Alternative Theories of Gravity since precise observational data could give extremely good signatures for the models
(see for example \cite{jaime,santos1, yazadjiev, staykov, orellana, cooney, arapoglu, alavirad, ganguly, farinelli}).

Other motivations to modify gravity  come from the issue of a full
recovering of the Mach principle which leads to assume a varying
gravitational coupling. The principle states that the local
inertial frame is determined by the averaging process of  motion of
distant astronomical objects \cite{bondi}. This fact implies that
the gravitational coupling can be scale-dependent and related to
some scalar field. As a consequence, the concept of {\it inertia}
and the Equivalence Principle have to be revised. For example, the
Brans-Dicke theory \cite{bra-dic} is a preliminary attempt to define
an alternative theory of gravity: it takes into
account a variable Newton gravitational coupling, whose dynamics
is governed by a scalar field non-minimally coupled to the
geometry. In such a way, Mach's principle is better implemented
\cite{bra-dic, cap-der-rub-scu, sciama}. 

As already mentioned, corrections to the gravitational Lagrangian
were already considered by several authors \cite{weyl_2, edd, lan}
soon after the GR was proposed. From a conceptual viewpoint, there are no reason \emph{a priori}  to
restrict the gravitational Lagrangian to
a linear function of the Ricci scalar, minimally coupled with
matter \cite{mag-fer-fra}. Since all curvature invariants are at
least second order differential, the corrective terms in the field
equations will be always at least fourth order. That is why one calls them
higher order terms (with respect to  GR).

Due to the increasing complexity of the field equations, the main amount of works is related to achieve  some formally
equivalent theories, where  the reduction of  the order of 
field equations can be obtained  by considering  metric and 
connections as independent fields \cite{ama-elg-mot-mul,
mag-fer-fra, all-bor-fra, sot1, sot-lib}. In addition, many
authors exploited the formal relation  to scalar-tensor
theories to make some statements about the weak field regime
 \cite{dam-esp}. Moreover other authors discussed a systematic analysis of such theories   in the low energy limit \cite{olmo1, olmo2, olmo3, dam-esp, clifton, odintsov, PRD, PRD1, PRD2, Stabile_Capozziello, CCCT}. In particular, Fourth Order Gravity has been studied in the Newtonian limit (weak field and small velocity) and in the Minkowskian limit (e.g. gravitational waves) \cite{minko}. In the former case, one finds  modifications of gravitational potential, while in the latter,  massive gravitational   wave modes come out \cite{quadrupolo}. 
 
However,  the weak field limit of such  theories have to be tested against realistic self-gravitating structures. Galactic rotation curves, stellar hydrodynamics and gravitational lensing appear natural candidates as test-bed experiments \cite{CCT, BHL, BHL1, stabile_scelza}.
 
In this paper,  starting from Newtonian and post-Newtonian approximations of Extended Gravity,  we match the outcomes with some typical astrophysical structures. The Newtonian limit of the general Fourth Order Gravity is considered by  generalizing  $f(R)$ models  with  generic functions containing  other  curvature invariants, as   \emph{Ricci square} ($R_{\alpha\beta}R^{\alpha\beta}$) and \emph{Riemann square} ($R_{\alpha\beta\gamma\delta}R^{\alpha\beta\gamma\delta}$). The spherically symmetric solutions of metric tensor yet present Yukawa-like behaviors, but, in general, one has two characteristic lengths.  
The Gauss-Bonnet invariant ${\cal G}=R^2-4R_{\mu\nu}R^{\mu\nu}+R_{\mu\nu\lambda\sigma}R^{\mu\nu\lambda\sigma}\,,$ and the Weyl invariant $C_{\alpha\beta\gamma\delta}C^{\alpha\beta\gamma\delta}$ can be reduced to the above cases.

The plan of the review is the following. In Sec. \ref{general_remarks} we summarize the basic ingredients of the Newtonian and the post-Newtonian limits \cite{PRD1}.
Sec. \ref{new_limit} is devoted to  the Newtonian limit of  Fourth Order Gravity  \cite{PRD2}, while, in Sec. \ref{new_limit_pointlike_sol},  we discuss the solutions generated by  pointlike sources with and without gauge conditions. A brief analysis of the free parameters of  the theory is performed. 

In Sec. \ref{hydro_new} we briefly review the classical hydrostatic problem for  stellar structures and in Sec. \ref{hydro_mod} we derive the modified Poisson equation in the framework of the Newtonian limit of  $f(R)$-gravity. Then the modified Lan\'{e}-Emden equation is obtained in Sec. \ref{hydro_solutions}.  Analytic and numerical solutions are discussed  \cite{hydrostatic}.

Galactic rotation curves are considered  in Sec. \ref{gal_rotation}. The main properties of  galactic metric are discussed  and, in particular,   the violation of the Gauss theorem due to the corrections to the Newtonian potential. In Sec. \ref{mass_model}, we discuss models for the mass distribution of  galactic components. A qualitative and quantitative comparison between the experimental data and the theoretical predictions of rotation curve \cite{stabile_scelza} is obtained for our Galaxy and for NGC 3198 in Sec. \ref{curve_FOG} by fixing numerically the free parameters of the theory. The motion of massive particles  is  discussed in Sec. \ref{CCT_theory}:  analogies and differences between general  Fourth Order Gravity and  $R^n$-gravity are pointed out.  Fourth Order Gravity is also considered with and without DM. The outcomes are compared with  GR. The rotation curves results higher or  lower than the same curves in  GR, if  the squared Ricci scalar contribution is dominant or not with respect to the contribution of the squared Ricci tensor. However by modifying gravity, also the spatial description of DM could undergoes modifications and the free parameters of model  assume different values.

Gravitational lensing  is analyzed in the case of  pointlike sources (Sec. \ref{GL_pointlike}) and extended sources (Sec. \ref{GL_extended}). An important outcome is found in the case of thin lenses: the bending of photons in  $f(R)$-gravity is the same as  in GR \cite{stabile_stabile, jetzer}. If we add  the squared Ricci tensor  contribution,  we obtain a shift of the image positions  (Sec. \ref{GL_lens_equation}).

Finally we take into account the weak field limit of scalar-tensor gravity \cite{stabile_stabile_cap} in the Jordan frame (sec. \ref{ST_JF}) and  compare it with the analogous in the Einstein frame (sec. \ref{ST_EF}). Specifically, we consider how physical quantities, like gravitational potentials, derived in the Newtonian approximation for the same scalar-tensor theory, behave in the Jordan and in the Einstein frame. The approach allows to discriminate features that are invariant under conformal transformations and gives contributions  in the debate of selecting the true physical frame.The particular case of $f(R)$-gravity is considered in Sec. \ref{ST_f(R)}.

Discussions and conclusions are drawn in Sec. \ref{concl}.


\section{ Newtonian and  post-Newtonian approximations}\label{general_remarks}


It is worth discussing  some general issues on the
Newtonian and post-Newtonian limits. Basically there are some
general features one has to take into account when approaching
these limits, whatever the underlying theory of gravitation is.

If one consider a system of gravitationally interacting particles
of  mass $\bar{M}$, the kinetic energy
$\frac{1}{2}\bar{M}\bar{v}^2$ will be, roughly, of the same order
of magnitude as the typical potential energy
$U=G\bar{M}^2/\bar{r}$, with $\bar{M}$, $\bar{r}$, and $\bar{v}$
the typical average values of  masses, separations, and velocities
of these particles. As a consequence:

\begin{equation}
\bar{v^2}\sim \frac{G\bar{M}}{\bar{r}}\,,
\end{equation}
(for instance, a test particle in a circular orbit of radius $r$
about a central mass $M$ will have velocity $v$ given in Newtonian
mechanics by the exact formula $v^2=GM/r$.)

The post-Newtonian approximation can be described as a method for
obtaining the motion of the system to  higher than  first
order (approximation which coincides with the Newtonian mechanics)
with respect to the quantities $G\bar{M}/\bar{r}$ and $\bar{v}^2$
assumed small with respect to the squared light speed $c^2$. This
approximation is sometimes referred to as an expansion in inverse
powers of the light speed.

The typical values of the Newtonian gravitational potential $U$
are nowhere larger than $10^{-5}$ in the Solar System (in
geometrized units, $U/c^2$ is dimensionless). On the other hand,
planetary velocities satisfy the condition $\bar{v}^2\lesssim U$,
while\footnote{We consider here on the velocity $v$ in
units of the light speed $c$.} the matter pressure $p$ experienced
inside the Sun and the planets is generally smaller than the
matter gravitational energy density $\rho U$, in other words
\footnote{Typical values of $p/\rho$ are $\sim 10^{-5}$
in the Sun and  $\sim 10^{-10}$ in the Earth \cite{will}.}
$p/\rho\lesssim U$. Furthermore one must consider that even other
forms of energy in the Solar System (compressional energy,
radiation, thermal energy, etc.) have small intensities and the
specific energy density $\Pi$ (the ratio of the energy density to
the rest-mass density) is related to $U$ by $\Pi\lesssim U$ ($\Pi$
is $\sim 10^{-5}$ in the Sun and $\sim 10^{-9}$ in the Earth
\cite{will}). As matter of fact, one can consider that these
quantities, as function of the velocity, give second order
contributions

\begin{equation}
U\sim v^2\sim p/\rho\sim \Pi\sim\mathcal{O(2)}
\end{equation}
Therefore, the velocity $v$ gives $\mathcal{O}(1)$ terms in the velocity
expansions, $U^2$ is of order $\mathcal{O}(4)$, $U\,v$ of $\mathcal{O}(3)$, $U\,\Pi$ is of
$\mathcal{O}(4)$, and so on. Considering these approximations, one has

\begin{equation}
\frac{\partial}{\partial t}\sim\textbf{v}\cdot\nabla
\end{equation}
and

\begin{equation}
\frac{|\partial/\partial t|}{|\nabla|}\sim\mathcal{O}(1)
\end{equation}
Now, particles move along geodesics

\begin{equation}\label{geodesic}
\frac{d^2x^\mu}{ds^2}+\Gamma^\mu_{\sigma\tau}\frac{dx^\sigma}{ds}\frac{dx^\tau}{ds}=0
\end{equation}
where $ds\,=\,\sqrt{g_{\alpha\beta}dx^\alpha dx^\beta}$ is the relativistic distance. The equation (\ref{geodesic}) can be written in details as

\begin{equation}
\frac{d^2x^i}{dt^2}=-\Gamma^i_{tt}-2\Gamma^i_{tm}\frac{dx^m}{dt}-
\Gamma^i_{mn}\frac{dx^m}{dt}\frac{dx^n}{dt}+\biggl[\Gamma^t_{tt}+
2\Gamma^t_{tm}\frac{dx^m}{dt}+2\Gamma^t_{mn}\frac{dx^m}{dt}\frac{dx^n}{dt}\biggr]\frac{dx^i}{dt}
\end{equation}
where the set of coordinates adopted is
$x^\mu\,=\,(t,x^1,x^2,x^3)$. In the Newtonian approximation, that is vanishingly small
velocities and only  first-order terms in the difference between
$g_{\mu\nu}$ and the Minkowski metric $\eta_{\mu\nu}$, one obtains
that the particle motion equations reduce to the standard
result

\begin{equation}\label{geodesic_new}
\frac{d^2x^i}{dt^2}\simeq-\Gamma^i_{tt}\simeq-\frac{1}{2}\frac{\partial
g_{tt}}{\partial x^i}
\end{equation}
The quantity $1-g_{tt}$ is of order $G\bar{M}/\bar{r}$, so that
the Newtonian approximation gives
$d^2x^i/dt^2$ to the order
$G\bar{M}/\bar{r}^2$, that is, to the order $\bar{v}^2/r$. As a
consequence if we would like to search for the post-Newtonian
approximation, we need to compute
$d^2x^i/dt^2$ to the order
$\bar{v}^4/\bar{r}$. Due to the Equivalence Principle and the
differentiability of spacetime manifold, we expect that it should
be possible to find out a coordinate system in which the metric
tensor is nearly equal to the Minkowski one $\eta_{\mu\nu}$, the
correction being expandable in powers of
$G\bar{M}/\bar{r}\sim\bar{v}^2$. In other words one has to
consider the metric developed as follows\footnote{The Greek index runs from $0$ to $3$; the Latin index runs from $1$ to $3$.}

\begin{equation}\label{approx1}
\begin{array}{ll}g_{tt}(t,\textbf{x})\simeq1+g^{(2)}_{tt}(t,\textbf{x})+g^{(4)}_{tt}(t,\textbf{x})+\mathcal{O}(6)
\\\\g_{ti}(t,\textbf{x})\simeq g^{(3)}_{ti}(t,\textbf{x})+\mathcal{O}(5)\\\\
g_{ij}(t, \textbf{x})\simeq-\delta_{ij}+g^{(2)}_{ij}(t,
\textbf{x})+\mathcal{O}(4)
\end{array}
\end{equation}
where $\delta_{ij}$ is the Kronecker delta, and for the
controvariant form of $g_{\mu\nu}$, one has

\begin{equation}\label{approx2}
\begin{array}{ll}g^{tt}(t,\textbf{x})\simeq
1+g^{(2)tt}(t, \textbf{x})+g^{(4)tt}(t,
\textbf{x})+\mathcal{O}(6)
\\\\
g^{ti}(t,\textbf{x})\simeq g^{(3)ti}(t,\textbf{x})+\mathcal{O}(5)\\\\
g^{ij}(t,\textbf{x})\simeq-\delta^{ij}+g^{(2)ij}(t,{\textbf{x}})+\mathcal{O}(4)
\end{array}
\end{equation}
In evaluating $\Gamma^\mu_{\alpha\beta}$ we must take into account
that the scale of distance and time, in our systems, are
respectively set by $\bar{r}$ and $\bar{r}/\bar{v}$, thus the
space and time derivatives should be regarded as being of order

\begin{equation}
\frac{\partial}{\partial x^i}\sim\frac{1}{\bar{r}}\,, \ \ \ \ \ \
\ \frac{\partial}{\partial t}\sim\frac{\bar{v}}{\bar{r}}
\end{equation}
Using the above approximations (\ref{approx1}), (\ref{approx2}),
we have, from the definition of Christoffel symbols $\Gamma^\mu_{\alpha\beta}=\frac{1}{2}g^{\mu\sigma}(g_{\alpha\sigma,\beta}+g_{\beta\sigma,\alpha}
-g_{\alpha\beta,\sigma})$,

\begin{equation}
\begin{array}{ll} \begin{array}{ccc}
{\Gamma^{(3)}}^{t}_{tt}=\frac{1}{2}g^{(2),t}_{tt}\, & \,\,\, & {\Gamma^{(2)}}^{i}_{tt}=\frac{1}{2}g^{(2),i}_{tt} \\
& & \\
{\Gamma^{(2)}}^{i}_{jk}=\frac{1}{2}\biggl({g^{(2)}}^{,i}_{jk}-{g^{(2)}}^{i}_{j,k}-{g^{(2)}}^{i}_{k,j}\biggr)\,
& \,\,\, & {\Gamma^{(3)}}^{t}_{ij}=\frac{1}{2}\biggl
({g^{(3)}}^{t}_{i,j}+{g^{(3)}}^{t}_{j,i}-{g^{(3)}}^{,t}_{ij}\biggr) \\
& & \\
{\Gamma^{(3)}}^{i}_{tj}=\frac{1}{2}\biggl({g^{(3)}}^{,i}_{tj}-{g^{(3)}}^{i}_{t,j}-{g^{(2)}}^{i}_{j,t}\biggr)\,
& \,\,\, & {\Gamma^{(4)}}^{0}_{0i}=\frac{1}{2}\biggl
({g^{(4)}}^{t}_{t,i}+g^{(2)tt}g^{(2)}_{tt,i}\biggr) \\
& & \\
{\Gamma^{(4)}}^{i}_{tt}=\frac{1}{2}\biggl({g^{(4)}}^{,i}_{tt}+g^{(2)im}g^{(2)}_{tt,m}-2{g^{(3)}}^{i}_{t,t}\biggr)\, & \,\,\, &
{\Gamma^{(2)}}^{t}_{ti}= \frac{1}{2}{g^{(2)}}^{t}_{t,i}
\end{array}
\end{array}
\end{equation}
The Ricci tensor component are

\begin{equation}\label{PPN_ricci}
\begin{array}{ll}R^{(2)}_{tt}=\frac{1}{2}\triangle
g^{(2)}_{tt}\\\\R^{(4)}_{tt}=\frac{1}{2}\triangle
g^{(4)}_{tt}-\frac{1}{2}{g^{(2)}}^{mn}_{,m}g^{(2)}_{tt,n}-\frac{1}{2}{g^{(2)}}^{mn}g^{(2)}_{tt,mn}+\frac{1}{2}{g^{(2)}}^{m}_{m,tt}-\frac{1}{4}{g^{(2)}}^{t,m}_{t}g^{(2)}_
{tt,m}-\frac{1}{4}{g^{(2)}}^{m,n}_{m}g^{(2)}_{tt,n}-{g^{(3)}}^{m}_{t,mt}\\\\R^{(3)}_{ti}=\frac{1}{2}\triangle
g^{(3)}_{ti}-\frac{1}{2}{g^{(2)}}^{m}_{i,mt}-\frac{1}{2}{g^{(3)}}^{m}_{t,mi}+\frac{1}{2}{g^{(2)}}^{m}_{m,ti}\\\\R^{(2)}_{ij}=\frac{1}{2}\triangle
g^{(2)}_{ij}-\frac{1}{2}{g^{(2)}}^{m}_{i,mj}-\frac{1}{2}{g^{(2)}}^{m}_{j,mi}-\frac{1}{2}{g^{(2)}}^{t}_{t,ij}+\frac{1}{2}{g^{(2)}}^{m}_{m,ij}\end{array}
\end{equation}
with $\triangle$ and $\bigtriangledown$, respectively, the
Laplacian and the gradient in flat space. The inverse of the metric tensor is defined by means of the
equation

\begin{equation}
g^{\alpha\rho}g_{\rho\beta}=\delta^\alpha_\beta
\end{equation}
with $\delta^\alpha_\beta$ the Kronecker delta. The relations
among the higher than first order terms turn out to be

\begin{equation}
\begin{array}{ll}g^{(2)tt}(t,\textbf{x})=-g^{(2)}_{tt}(t,\textbf{x})\\\\g^{(4)tt}(t,\textbf{x})={g^{(2)}_{tt}(t,\textbf{x})}^2-g^{(4)}_{tt}
(t,\textbf{x})\\\\g^{(3)ti}(t,\textbf{x})=g^{(3)}_{ti}(t,\textbf{x})\\\\g^{(2)ij}(t,\textbf{x})=-g^{(2)}_{ij}(t,\textbf{x})
\end{array}
\end{equation}
Finally the Lagrangian of a particle in presence of a
gravitational field can be expressed as proportional to the
invariant distance $ds^{1/2}$, thus we have\,:

\begin{equation}
L=\biggl(g_{\rho\sigma}\frac{dx^\rho}{dt}\frac{dx^\sigma}{dt}\biggr)^{1/2}=\biggl(g_{tt}+2g_{tm}v^m+g_{mn}v^mv^n\biggr)^{1/2}=
\biggl(1+g^{(2)}_{tt}+g^{(4)}_{tt}+2g^{(3)}_{tm}v^m-\textbf{v}^2+g^{(2)}_{mn}v^mv^n\biggr)^{1/2}
\end{equation}
which, to the $\mathcal{O}(2)$ order, reduces to the classic Newtonian
Lagrangian of a test particle
$L_{\text{New}}=(1+g^{(2)}_{tt}-\textbf{v}^2)^{1/2}$,
where $\mathbf{v}=\frac{dx^m}{dt}\frac{dx_m}{dt}$. As matter
of fact, post-Newtonian physics has to involve higher than $\mathcal{O}(4)$
order terms in the Lagrangian.

An important remark concerns the odd-order perturbation terms $\mathcal{O}(1)$
or $\mathcal{O}(3)$. Since, these terms contain  odd powers of velocity
$\textbf{v}$ or of time derivatives, they are related to the
energy dissipation or absorption by the system. Nevertheless, the
mass-energy conservations  prevents the energy and mass losses
and, as a consequence, prevents, in the Newtonian limit, terms of
$\mathcal{O}(1)$ and $\mathcal{O}(3)$ orders in the  Lagrangian. If one takes into account
contributions  higher than $\mathcal{O}(4)$ order, different theories give
different predictions. GR, for example, due to the conservation of
post-Newtonian energy, forbids  terms of $\mathcal{O}(5)$ order; on the other
hand, terms of $\mathcal{O}(7)$ order can appear and are related to the energy
lost by means of the gravitational radiation.

On the matter side, \emph{i.e.}\ right-hand side of the field equations we start with the general definition of
the energy-momentum tensor of a perfect fluid with additional energy $\Pi$

\begin{eqnarray}
T_{\mu\nu}\,=\,(\rho+\Pi\rho+p)u_\mu u_\nu-p\,g_{\mu\nu}
\end{eqnarray}
The explicit form of the energy-momentum tensor can be derived as
follows

\begin{eqnarray}\label{PPN-tensor-matter}
\begin{array}{ll}
T_{tt}\,=\,\rho+\rho(v^2-2U+\Pi)+\rho\biggl[v^2\biggl(\frac{p}{\rho}+v^2+2V+\Pi\biggr)+\sigma-2\Pi
U\biggr]
\\\\
T_{ti}\,=\,-\rho v^i+\rho\biggl[-v^i\biggl(\frac{p}{\rho}+2V+v^2+\Pi\biggl)+h_{ti}\biggr]
\\\\
T_{ij}\,=\,\rho v^iv^j+p\delta_{ij}+\rho\biggl[v^iv^j\biggl(\Pi+\frac{p}{\rho}+4V+v^2+2U\biggr)
-2v^c\delta_{c(i}h_{0|j)}+2\frac{pV}{\rho}\delta_{ij}\biggr]
\end{array}
\end{eqnarray}
where $\rho$ is the density of energy.


\section{The Newtonian limit of $f(X,Y,Z)$-gravity}\label{new_limit}


Let us start with a general class of fourth order theories given
by the action

\begin{eqnarray}\label{HOGaction}
\mathcal{A}=\int d^{4}x\sqrt{-g}\biggl[f(X,Y,Z)+\mathcal{X}\mathcal{L}_m\biggr]
\end{eqnarray}
where $f$ is an unspecified function of curvature invariants
$X\,\equiv\,R$, $Y\,\equiv\,R_{\alpha\beta}R^{\alpha\beta}$ and $Z\,\equiv\,R_{\alpha\beta\gamma\delta}R^{\alpha\beta\gamma\delta}$. The term $\mathcal{L}_m$ is the minimally coupled ordinary matter
contribution. In the metric approach, the field equations are
obtained by varying (\ref{HOGaction}) with respect to
$g_{\mu\nu}$. We get

\begin{eqnarray}\label{fieldequationHOG}
H_{\mu\nu}\,=\,&&f_XR_{\mu\nu}-\frac{f}{2}g_{\mu\nu}-f_{X;\mu\nu}+g_{\mu\nu}\Box
f_X+2f_Y{R_\mu}^\alpha
R_{\alpha\nu}-2[f_Y{R^\alpha}_{(\mu}]_{;\nu)\alpha}+
\nonumber\\\\
&&+\Box[f_YR_{\mu\nu}]+[f_YR_{\alpha\beta}]^{;\alpha\beta}g_{\mu\nu}+
2f_ZR_{\mu\alpha
\beta\gamma}{R_{\nu}}^{\alpha\beta\gamma}-4[f_Z{{R_\mu}^{\alpha\beta}}_\nu]_{;\alpha\beta}\,=\,
\mathcal{X}\,T_{\mu\nu}\nonumber
\end{eqnarray}
where
$T_{\mu\nu}\,=\,-\frac{1}{\sqrt{-g}}\frac{\delta(\sqrt{-g}\mathcal{L}_m)}{\delta
g^{\mu\nu}}$ is the energy-momentum tensor of matter,
$f_X\,=\,\frac{df}{dX}$, $f_Y\,=\,\frac{df}{dY}$,
$f_Z\,=\,\frac{df}{dZ}$, $\Box={{}_{;\sigma}}^{;\sigma}$ and
$\mathcal{X}\,=\,8\pi G$. The convention for Ricci's tensor is
$R_{\mu\nu}={R^\sigma}_{\mu\sigma\nu}$, while for the Riemann
tensor is ${R^\alpha}_{\beta\mu\nu}=\Gamma^\alpha_{\beta\nu,\mu}+...$. The
affinities are the usual Christoffel symbols of the metric. The adopted signature is $(+---)$ (for details, see \cite{landau}). The trace of field equations (\ref{fieldequationHOG}) is the following

\begin{eqnarray}\label{tracefieldequationHOG}
H\,=\,g^{\alpha\beta}H_{\alpha\beta}\,=\,&&f_XR+2f_YR_{\alpha\beta}R^{\alpha\beta}+2f_ZR_{\alpha\beta\gamma\delta}
R^{\alpha\beta\gamma\delta}-2f+
\nonumber\\\\
&&+\Box[3
f_X+f_YR]+2[(f_Y+2f_Z)R^{\alpha\beta}]_{;\alpha\beta}\,=\,\mathcal{X}\,T\nonumber
\end{eqnarray}
where $T\,=\,T^{\sigma}_{\,\,\,\,\,\sigma}$ is the trace of energy-momentum tensor.

Some authors considered a linear Lagrangian containing not only $X$, $Y$ and $Z$ but also the first power of curvature invariants $\Box R$ and ${R^{\alpha\beta}}_{;\alpha\beta}$. Such a choice is justified because all curvature invariants have the same dimension ($L^{-2}$) \cite{santos}. Furthermore, this dependence on these  two last invariants is only formal, since from the contracted Bianchi identity ($2{R^{\alpha\beta}}_{;\alpha\beta}-\Box R\,=\,0$) we have only one independent invariant. In any linear theory of gravity (the function $f$ is linear) the terms $\Box R$ and ${R^{\alpha\beta}}_{;\alpha\beta}$ give us no contribution to the field equations, because they are four-divergences. However if we consider a function of $\Box R$ or ${R^{\alpha\beta}}_{;\alpha\beta}$ by varying the action we still could have the four-divergences but we would have the contributions of sixth order differential. For our aim (we want to stay in framework of fourth order differential equations) the most general theory of fourth order is the action (\ref{HOGaction}).

The Newtonian limit analysis starts from the development (\ref{approx1}) of the
metric tensor. Then we set the metric as

\begin{eqnarray}\label{metric_tensor_PPN_0}
g_{\mu\nu}\,=\,\begin{pmatrix}
1+g^{(2)}_{tt}(t,\mathbf{x}) & 0 \\
0 & -\delta_{ij}+g^{(2)}_{ij}(t,\mathbf{x})+\dots\end{pmatrix}
\end{eqnarray}
where we note that the presence of function $g^{(2)}_{ij}(t,\mathbf{x})$ is superfluous here for our aim but it will be fundamental about the gravitational lensing. The curvature invariants $X$, $Y$, $Z$ become

\begin{eqnarray}
\begin{array}{ll}
X\,\sim\,X^{(2)}+X^{(4)}+\dots\\\\
Y\,\sim\,Y^{(4)}+Y^{(6)}+\dots\\\\
Z\,\sim\,Z^{(4)}+Z^{(6)}\dots
\end{array}
\end{eqnarray}
The function $f$ can be developed as

\begin{eqnarray}
f(X,Y,Z)\,\sim\,f(0)+f_X(0)X^{(2)}+\frac{1}{2}f_{XX}(0){X^{(2)}}^2+f_X(0)X^{(4)}+f_Y(0)Y^{(4)}+f_Z(0)Z^{(4)}+\dots
\end{eqnarray}
and analogous relations for partial derivatives of $f$ are
obtained. From the interpretation of stress-energy tensor components (\ref{PPN-tensor-matter}) as
energy density, momentum density and momentum flux,  we have
$T_{tt}$, $T_{ti}$ and $T_{ij}$ at the various orders

\begin{eqnarray}\label{PPN-matter-density}
\begin{array}{ll}T_{tt}\,=\,T^{(0)}_{tt}+T^{(2)}_{tt}+\mathcal{O}(4)\\\\
T_{ti}\,=\,T^{(1)}_{ti}+\mathcal{O}(3)\\\\T_{ij}=T^{(2)}_{ij}+\mathcal{O}(4)
\end{array}
\end{eqnarray}
where $T^{(N)}_{\mu\nu}$ denotes the term in $T_{\mu\nu}$ of order
$\bar{M}/\bar{r}^3\,\,\bar{v}^{N}$. In particular $T^{(0)}_{tt}$
is the density of rest-mass, while $T^{(2)}_{tt}$ is the
non-relativistic part of the energy density.

From the lowest order of field equations (\ref{fieldequationHOG}) we have

\begin{eqnarray}\label{PPN-field-equation-general-theory-fR-O0}
f(0)\,=\,0
\end{eqnarray}
Not only in $f(R)$-gravity \cite{PRD,PRD1,spher_symm_fR}
but also in $f(X,Y,Z)$-theory a missing cosmological component in
the action (\ref{HOGaction}) implies that the space-time is
asymptotically Minkowskian. The equations (\ref{fieldequationHOG})
and (\ref{tracefieldequationHOG}) at $\mathcal{O}(2)$ - order
become\footnote{We used the properties:
$2{R_{\alpha\beta}}^{;\alpha\beta}-\Box R\,=\,0$ and
${{R_\mu}^{\alpha\beta}}_{\nu;\alpha\beta}\,=\,{{R_\mu}^\alpha}_{;\nu\alpha}-\Box
R_{\mu\nu}$.}

\begin{eqnarray}\label{NL-field-equation}
\begin{array}{ll}
H^{(2)}_{tt}\,=\,f_X(0)R^{(2)}_{tt}-[f_Y(0)+4f_Z(0)]\triangle
R^{(2)}_{tt}
-\frac{f_X(0)}{2}X^{(2)}-[f_{XX}(0)+\frac{f_Y(0)}{2}]\triangle
X^{(2)}\,=\,\mathcal{X}\,T^{(0)}_{tt}\\\\
H^{(2)}_{ij}\,=\,f_X(0)R^{(2)}_{ij}-[f_Y(0)+4f_Z(0)]\triangle
R^{(2)}_{ij}
+\frac{f_X(0)}{2}X^{(2)}\delta_{ij}+[f_{XX}(0)+\frac{f_Y(0)}{2}]\triangle
X^{(2)}\delta_{ij}+\\\\
\,\,\,\,\,\,\,\,\,\,\,\,\,\,\,\,\,\,\,\,\,\,-f_{XX}(0){X^{(2)}}_{,ij}+[f_Y(0)+4f_Z(0)]R^{(2)}_{mi,jm}+f_Y(0)R^{(2)}_{mj,im}\,=\,0\\\\
H^{(2)}\,=\,-f_X(0)X^{(2)}-[3f_{XX}(0)+2f_Y(0)+2f_Z(0)]\triangle
X^{(2)}\,=\,\mathcal{X}\,T^{(0)}
\end{array}
\end{eqnarray}
By introducing the quantities

\begin{eqnarray}\label{mass_definition}
\begin{array}{ll}
{m_1}^2\,\doteq\,-\frac{f_X(0)}{3f_{XX}(0)+2f_Y(0)+2f_Z(0)}\\\\
{m_2}^2\,\doteq\,\frac{f_X(0)}{f_Y(0)+4f_Z(0)}
\end{array}
\end{eqnarray}
we get three differential equations for curvature invariant
$X^{(2)}$, $tt$- and $ij$-component of Ricci tensor
$R^{(2)}_{\mu\nu}$

\begin{eqnarray}\label{NL-field-equation_2}
\begin{array}{ll}
(\triangle-{m_2}^2)R^{(2)}_{tt}+\biggl[\frac{{m_2}^2}{2}-\frac{{m_1}^2+2{m_2}^2}{6{m_1}^2}\triangle\biggr]
X^{(2)}\,=\,-\frac{{m_2}^2\mathcal{X}}{f_X(0)}\,T^{(0)}_{tt}\\\\
(\triangle-{m_2}^2)R^{(2)}_{ij}+\biggl[\frac{{m_1}^2-{m_2}^2}{3{m_1}^2}\,
\partial^2_{ij}-\biggl(\frac{{m_2}^2}{2}-\frac{{m_1}^2+2{m_2}^2}{6{m_1}^2}\triangle\biggr)\delta_{ij}\biggr]
X^{(2)}\,=\,0\\\\
(\triangle-{m_1}^2)X^{(2)}\,=\,\frac{{m_1}^2\mathcal{X}}{f_X(0)}\,T^{(0)}
\end{array}
\end{eqnarray}
The solution for curvature invariant $X^{(2)}$ in third
line of (\ref{NL-field-equation_2}) is

\begin{eqnarray}\label{scalar_invariant_sol_gen}
X^{(2)}(t,\textbf{x})\,=\,\frac{{m_1}^2\mathcal{X}}{f_X(0)}\int
d^3\mathbf{x}'\mathcal{G}_1(\mathbf{x},\mathbf{x}')T^{(0)}(t,\mathbf{x}')
\end{eqnarray}
where $\mathcal{G}_1(\mathbf{x},\mathbf{x}')$ is the Green
function of the field operator $\triangle-{m_1}^2$. The solution for
$g^{(2)}_{tt}$, by remembering
$R^{(2)}_{tt}\,=\,\frac{1}{2}\triangle g^{(2)}_{tt}$, is

\begin{eqnarray}\label{tt_component_sol_gen}
g^{(2)}_{tt}(t,\textbf{x})\,=\,&&\frac{1}{2\pi}\int
d^3\mathbf{x}'d^3\mathbf{x}''\frac{\mathcal{G}_2(\mathbf{x}',\mathbf{x}'')}{|\mathbf{x}-\mathbf{x}'|}\biggl[
\frac{{m_2}^2\mathcal{X}}{f_X(0)}T^{(0)}_{tt}(t,\mathbf{x}'')-\frac{({m_1}^2+2{m_2}^2)\mathcal{X}}{6f_X(0)}\,T^{(0)}
t,\mathbf{x}'')+
\nonumber\\\\
&&+\frac{{m_2}^2-{m_1}^2}{6}X^{(2)}(t,\mathbf{x}'')\biggr]\nonumber
\end{eqnarray}
where $\mathcal{G}_2(\mathbf{x},\mathbf{x}')$ is the Green function of the field operator $\triangle-{m_2}^2$.

The expression (\ref{tt_component_sol_gen}) is the "modified" gravitational
potential (here we have a factor 2) for $f(X,Y,Z)$-gravity. The
solution for the gravitational potential $\Phi\,=\,g^{(2)}_{tt}/2$ has a
Yukawa-like behaviors (\cite{PRD,PRD2}) depending by a
characteristic lengths on whose it evolves.

The $ij$-component of Ricci tensor in terms of metric tensor
(\ref{PPN_ricci}) can be expressed by using the harmonic gauge condition
($g^{\alpha\beta}\Gamma^{\mu}_{\alpha\beta}\,=\,0$) as $R^{(2)}_{ij}|_{HG}\,=\,\frac{1}{2}g^{(2)}_{ij,mm}\,=\,\frac{1}{2}\triangle
g^{(2)}_{ij}$. Then the general solution for $g^{(2)}_{ij}$ from
(\ref{NL-field-equation_2}), in the harmonic gauge, is

\begin{eqnarray}
g^{(2)}_{ij}|_{HG}\,=\,\frac{1}{2\pi}\int
d^3\mathbf{x}'d^3\mathbf{x}''\frac{\mathcal{G}_2(\mathbf{x}',\mathbf{x}'')}{|\mathbf{x}-\mathbf{x}'|}
\biggl[\frac{{m_1}^2-{m_2}^2}{3{m_1}^2}\,
\partial^2_{i''j''}-\biggl(\frac{{m_2}^2}{2}-\frac{{m_1}^2+2{m_2}^2}{6{m_1}^2}\triangle_{\mathbf{x}''}\biggr)\delta_{ij}\biggr]
X^{(2)}(\mathbf{x}'')
\end{eqnarray}

While if we hypothesize
$g^{(2)}_{ij}\,=\,2\,\Psi\,\delta_{ij}$\footnote{We choose a
system of isotropic coordinates. Generally the set of coordinates $(t,r,\theta,\phi)$ are called standard coordinates if the metric is expressed as ${ds}^2\,=\,g_{tt}(t,r)\,dt^2+g_{rr}(t,r){dr}^2-r^2d\Omega$ while if one has ${ds}^2\,=\,g_{tt}(t,\mathbf{x})\,dt^2+g_{ij}(t,\mathbf{x})dx^idx^j$ the set $(t,x^1,x^2,x^3)$ is called isotropic coordinates \cite{weinberg, olmo3}.} we have (without the validity of the harmonic gauge condition)
$R^{(2)}_{ij}\,=\,\triangle\Psi\,\delta_{ij}+(\Psi-\Phi)_{,ij}$
and the second field equation of (\ref{NL-field-equation_2})
becomes

\begin{eqnarray}\label{NL-field-equationij}
\begin{array}{ll} \triangle\Psi\,=\,\int
d^3\mathbf{x}'\mathcal{G}_2(\mathbf{x},\mathbf{x}')\biggl(\frac{{m_2}^2}{2}-\frac{{m_1}^2+2{m_2}^2}{6{m_1}^2}
\triangle_{\mathbf{x}'}\biggr)X^{(2)}(\mathbf{x}')\\\\
(\Phi-\Psi)_{,ij}\,=\, \frac{{m_1}^2-{m_2}^2}{3{m_1}^2}\int
d^3\mathbf{x}'\mathcal{G}_2(\mathbf{x},\mathbf{x}')\,
X^{(2)}_{,i'j'}(\mathbf{x}')
\end{array}
\end{eqnarray}
Then the general solution for $g^{(2)}_{ij}$ from
(\ref{NL-field-equation_2}) is

\begin{eqnarray}\label{solpsi}
g^{(2)}_{ij}\,=\,2\,\Psi\,\delta_{ij}\,=\,-\frac{\delta_{ij}}{2\pi}\int
d^3\mathbf{x}'d^3\mathbf{x}''\frac{\mathcal{G}_2(\mathbf{x}',\mathbf{x}'')}{|\mathbf{x}-\mathbf{x}'|}
\biggl(\frac{{m_2}^2}{2}-\frac{{m_1}^2+2{m_2}^2}{6{m_1}^2}
\triangle_{\mathbf{x}''}\biggr)X^{(2)}(\mathbf{x}'')
\end{eqnarray}
and the second line of (\ref{NL-field-equationij}) is mathematically satisfied.

If we consider the trace of the second line of the set (\ref{NL-field-equationij}) we have a mathematical constraint for the gravitational potentials $\Phi$, $\Psi$

\begin{eqnarray}\label{cond_0}
\triangle(\Phi-\Psi)\,=\,\frac{{m_1}^2-{m_2}^2}{3{m_1}^2}\int
d^3\mathbf{x}'\mathcal{G}_2(\mathbf{x},\mathbf{x}')\,
\triangle_{\mathbf{x}'}X^{(2)}(\mathbf{x}')
\end{eqnarray}
and we can affirm that only in GR the metric potentials are equals (or more generally their difference must be proportional to function $|\mathbf{x}|^{-1}$).


\subsection{The pointlike solution}\label{new_limit_pointlike_sol}


Let us consider a pointlike source with mass $M$. The energy-momentum tensor (\ref{PPN-tensor-matter}) is (we are not interesting to the internal structure)

\begin{eqnarray}\label{emtensor_0}
\begin{array}{ll}T_{\mu\nu}\,=\,\rho(\mathbf{x})u_\mu
u_\nu
\\\\
T\,=\,\rho(\mathbf{x})\end{array}
\end{eqnarray}
where $\rho(\mathbf{x})$ is the mass density and $u_\mu$ satisfies
the condition $g^{tt}{u_t}^2\,=\,1$, $u_i\,=\,0$. Since (\ref{metric_tensor_PPN_0})
the expression (\ref{emtensor_0})
becomes

\begin{eqnarray}\label{emtensorPPN}
\begin{array}{ll}T_{tt}(t,\mathbf{x})\,\sim\,\rho(\mathbf{x})+\rho(\mathbf{x})g^{(2)}_{tt}(t,\mathbf{x})\,
=\,T^{(0)}_{tt}(t,\mathbf{x})+T^{(2)}_
{tt}(t,\mathbf{x})
\\\\
T\,=\,\rho(\mathbf{x})\,=\,T^{(0)}(t,\mathbf{x})\end{array}
\end{eqnarray}
The energy-momentum tensor satisfies the relations $T^{(0)}_{tt}(t,\mathbf{x})\,=\,T^{(0)}(t,\mathbf{x})\,=\,M\delta(\mathbf{x})$, where $\delta(\mathbf{x})$
is the Delta Dirac function. The expressions (\ref{scalar_invariant_sol_gen}), (\ref{tt_component_sol_gen}) and (\ref{solpsi}) are valid
for any values of quantities ${m_1}^2$ and ${m_2}^2$. However, when it wants to calculate the expression of the gravitational potential generated by a given mass distribution it is necessary to determine the algebraic sign of the parameters ${m_1}^2$ and ${m_2}^2$, because from their signs the nature of Green functions $\mathcal{G}_i(\mathbf{x},\mathbf{x}')$ is determined. The possible choices of Green function, for spherically symmetric systems (\emph{i.e.} $\mathcal{G}_i(\mathbf{x},\mathbf{x}')\,=\,\mathcal{G}_i(|\mathbf{x}-\mathbf{x}'|)$), are the following

\begin{eqnarray}\label{green_function}
\mathcal{G}_i(\mathbf{x},\mathbf{x}')\,=\,\left\{\begin{array}{ll}-\frac{1}{4\pi}\frac{e^{-\mu_i|\mathbf{x}-\mathbf{x}'|}}
{|\mathbf{x}-\mathbf{x}'|}\,\,\,\,\,\,\,\,\,\,\,\,\,\,\,\,\,\,\,\,\,\,\,\,\,\,\,\,\,\,\,\,\,\,\,\,\,\,\,\,\,\,\,\,\,
\,\,\,\text{if}\,\,\,\,\,\,\,\,\,\,\,\,\,\,{m_i}^2\,\,>\,0
\\\\
-\frac{1}{4\pi}\frac{\cos
\mu_i|\mathbf{x}-\mathbf{x}'|+ \sin
\mu_i|\mathbf{x}-\mathbf{x}'|}{|\mathbf{x}-\mathbf{x}'|}
\,\,\,\,\,\,\,\,\,\,\,\,\,\,\,\text{if}\,\,\,\,\,\,\,\,\,\,\,\,\,\,{m_i}^2\,<\,0\end{array}\right.
\end{eqnarray}
where $\mu_i\,\doteq\,\sqrt{|{m_i}^2|}$. The first choice in (\ref{green_function}) corresponds to Yukawa-like behavior, while the second one to the oscillating case. 
However, the Green function for the negative squared mass is, in general,  a linear combination of Cos and Sin with coefficients associated to the shift phase. For the sake of simplicity, we assumed here unitary coefficients. 
Both expressions are a generalization and/or an induced correction of the usual gravitational potential ($\propto |\mathbf{x}|^{-1}$), and when ${m_i}^2\,\rightarrow\,\infty$ (\emph{i.e.} $f_{XX}(0)\,,f_Y(0)\,,f_Z(0)\,\rightarrow\,0$ from the (\ref{mass_definition})) we recover the field equations of GR. Independently of algebraic sign of ${m_i}^2$,  one can introduce two scale lengths ${\mu_i}^{-1}$. We note that in the case of $f(R)$-gravity,  we obtain only one scale length (${\mu_1}^{-1}$ with $f_Y(0)\,=\,f_Z(0)\,=\,0$) on the which the Ricci scalar evolves \cite{PRD,PRD1,PRD2}, but in $f(X,Y,Z)$-gravity we have an additional scale length ${\mu_2}^{-1}$ on the which the Ricci tensor evolves.

If we choose ${m_i}^2\,>\,0$ the curvature invariant $X^{(2)}$
(\ref{scalar_invariant_sol_gen}) and the metric potentials $\Phi$
(\ref{tt_component_sol_gen}) and $\Psi$ (\ref{solpsi}) are

\begin{eqnarray}\label{sol_point}
\begin{array}{ll}
X^{(2)}\,=\,
-\frac{r_g\,{\mu_1}^2}{f_X(0)}\frac{e^{-\mu_1|\mathbf{x}|}}{|\mathbf{x}|}
\\\\
\Phi\,=\,-\frac{GM}{f_X(0)}\biggl[\frac{1}{|\textbf{x}|}
+\frac{1}{3}\frac{e^{-\mu_1|\mathbf{x}|}}{|\mathbf{x}|}
-\frac{4}{3}\frac{e^{-\mu_2|\mathbf{x}|}}{|\mathbf{x}|}\biggr]
\\\\
\Psi\,=\,-\frac{GM}{f_X(0)}\biggl[\frac{1}{|\textbf{x}|}
-\frac{1}{3}\frac{e^{-\mu_1|\mathbf{x}|}}{|\mathbf{x}|}
-\frac{2}{3}\frac{e^{-\mu_2|\mathbf{x}|}}{|\mathbf{x}|}\biggr]
\end{array}
\end{eqnarray}
where $r_g\,=\,2GM$ is the Schwarzschild radius and the relativistic invariant is

\begin{eqnarray}\label{metric_new}
ds^2\,=\,(1+2\Phi)dt^2-(1-2\Psi)\delta_{ij}dx^idx^j
\end{eqnarray}

The modified gravitational potential by $f(R)$-gravity is further modified by
the presence of functions of $R_{\alpha\beta}R^{\alpha\beta}$ and
$R_{\alpha\beta\gamma\delta}R^{\alpha\beta\gamma\delta}$. The
curvature invariant $X^{(2)}$ (the Ricci scalar) presents a
massive propagation and when $f(X,Y,Z)\rightarrow f(R)$ we find
the mass definition

\begin{eqnarray}\label{mass_f(R)}
m^2\,=\,-\frac{f'(R=0)}{3f''(R=0)}
\end{eqnarray}
and propagation mode with $m_2$ disappear \cite{PRD,PRD1,PRD2}.  Exponential and oscillating behaviors of  solutions (37) are compatible with respect to the previous obtained outcomes in the Newtonian limit of $f(R)$-gravity \cite{olmo1,olmo2} . Indeed, in the case of a genuine $f(R)$-gravity also the mass definition (\ref{mass_f(R)}) coincides again with the previous results.

The choice of parameters (Yukawa-like case for the Green function)
is made also in \cite{santos} and the constraint conditions on the derivatives of
$f$ obtained here ($f_Y(0)+4f_Z(0)\,>\,0$ and $3f_{XX}(0)+2f_Y(0)+2f_Z(0)\,<\,0$)
are still compatible with respect to paper mentioned.
Obviously the expressions $\Phi$ and $\Psi$ in (\ref{sol_point})
satisfy the constraint condition (\ref{cond_0}).

The gravitational potential $\Phi$, solution of Eqs. (\ref{HOEQ}), has in general
a Yukawa-like behavior depending on a
characteristic length on which it evolves \cite{PRD,mio1}. Then as it is evident
the Gauss theorem is not valid\footnote{It is worth noticing that also if the Gauss theorem does not hold, the Bianchi identities are always valid so the conservation laws are guaranteed. } since the force
law is not $\propto|\mathbf{x}|^{-2}$. The equivalence between a
spherically symmetric distribution and point-like distribution is
not valid and how the matter is distributed in the space is very
important \cite{PRD,PRD1,PRD2,Stabile_Capozziello}. However the field equations (\ref{NL-field-equation_2}) are linear and we can use the superposition principle. If we have a generic density mass $\rho(\mathbf{x})$ the potential $\Phi$ becomes

\begin{eqnarray}\label{superposition_phi}
\Phi(\mathbf{x})\,=\,-\,\frac{G}{f_X(0)}\int
d^3\mathbf{x}'\frac{\rho(\mathbf{x}')}{|\mathbf{x}-\mathbf{x}'|}
\biggl[1+\frac{1}{3}\,e^{-\mu_1|\mathbf{x}-\mathbf{x}'|}-\frac{4}{3}\,e^{-\mu_2|\mathbf{x}-\mathbf{x}'|}\biggr]
\end{eqnarray}
and in the case of a ball source with radius $\xi$ the potential $\Phi$ outside the source is 

\begin{eqnarray}\label{ST_FOG_FE_NL_sol_ball}
\Phi_{ball}(\mathbf{x})\,=\,-\frac{GM}{f_X(0)|\mathbf{x}|}\biggl[1+\frac{F(\mu_1\xi)}{3}\,e^{-\mu_1|\mathbf{x}|}-
\frac{4\,F(\mu_2\xi)}{3}\,e^{-\mu_2|\mathbf{x}|}\biggr]
\end{eqnarray}
where $F(x)\,=\,3\frac{x \cosh x-\sinh x}{x^3}$ is a geometric factor depending on the form of source and satisfying the condition $\lim_{x\rightarrow 0}F(x)\,=\,1$.

Besides the Birkhoff theorem results modified at Newtonian level:
the solution can be only factorized by a space-depending function
and an arbitrary time-depending function \cite{PRD1,mio1}. Furthermore
the correction to the gravitational potential is depending on
the only first two derivatives of $f$ with rispect the $X$ and the first derivatives with rispect the $Y$ and $Z$ in the point $X\,=\,Y\,=\,Z\,=\,0$. This means that different analytical 
theories, from the third derivative perturbation terms on,  admit the same Newtonian
limit \cite{PRD,PRD1,PRD2}.

For a right physical interpretation of $\Phi$ in (\ref{sol_point}) we also impose the condition $\mu_1-4\mu_2\,<\,0$. This is a necessary condition for a correct interpretation of attractive gravitational potential, in addition to the condition $\mu_1\,<\,\mu_2$ for a well with a negative minimum in $|\mathbf{x}|\,=\,0$. These conditions can be summarized in terms of scale lengths. In fact we find that the contribution induced by Ricci and Riemann square goes to zero faster than the contribution induced by Ricci scalar: ${\mu_1}^{-1}\,>\,{\mu_2}^{-1}$. In terms of Lagrangian it means that for any $f(X,Y,Z)$-gravity the correction to Hilbert-Einstein Lagrangian depends more on the contributions of Ricci scalar ($f_{XX}(0)\,\neq\,0$) than ones of Ricci and Riemann square ($f_{Y}(0)\,,\,f_{Z}(0)\,\neq\,0$). Then, if we suppose $f_X(0)\,>\,0$ (for a right Newtonian limit), starting from the condition $\mu_1\,<\,\mu_2$ we get a constraint on the derivatives of $f$ with respect to curvature invariants

\begin{eqnarray}\label{condition}
f_{XX}(0)+f_Y(0)+2f_Z(0)\,<\,0
\end{eqnarray}
In the case of $f(R)$-gravity ($f_Y(0)\,=\,f_Z(0)\,=\,0$) we
reobtain the same condition among the first and second derivatives
of $f$ \cite{PRD,PRD1,PRD2,Stabile_Capozziello}.


\subsection{$f(X,Y,Z)$-gravity and the quadratic Lagrangian}\label{new_limit_properties}


The outcome (\ref{sol_point}) can be obtained also by considering the so-called
\emph{Quadratic Lagrangian} $\mathcal{L}\,=\,\sqrt{-g}(a_1\,R+a_2\,R^2+a_3\,R_{\alpha\beta}R^{\alpha\beta})$
where $a_1$, $a_2$ and $a_3$ are constants. In this case,
\cite{Stabile_Capozziello} we find two characteristic lengths
$\biggl|\frac{2(3a_2+a_3)}{a_1}\biggr|^{1/2}$,
$\biggl|\frac{a_3}{a_1}\biggr|^{1/2}$ and the Newtonian limit of
theory implies as solution Eqs. (\ref{sol_point}). We can state, then, the Newtonian limit of
any $f(X,Y,Z)$-gravity can be reinterpreted by introducing the
\emph{Quadratic Lagrangian} and the coefficients have to satisfy
the following relations

\begin{eqnarray}\label{equivalence}
a_1\,=\,f_X(0),\,\,\,\,\,\,\,\,\,a_2\,=\,\frac{1}{2}f_{XX}(0)-f_Z(0),\,\,\,\,\,\,\,\,\,
a_3\,=\,f_Y(0)+4f_Z(0)\,.
\end{eqnarray}

A first considerations about (\ref{equivalence}) is regarding the
characteristic lengths induced by $f(X,Y,Z)$-gravity. The second
length ${\mu_2}^{-1}$ is originated from the presence, in the
Lagrangian, of Ricci and Riemann tensor square, but also a theory
containing only Ricci tensor square could show the same outcome
(it is successful replacing the coefficients $a_i$ of
\emph{Quadratic Lagrangian} or renaming the function $f$).
Obviously the same is valid also with the Riemann tensor square
alone. Then a such modification of theory enables a massive
propagation of Ricci Tensor and, as it is well known in the
literature, a substitution of Ricci Scalar with any function of
Ricci scalar enables a massive propagation of Ricci scalar. We
can, then, affirm that an hypothesis of Lagrangian containing any
function of only Ricci scalar and Ricci tensor square is not
restrictive and only the experimental constraints can fix the
arbitrary parameters.

A second consideration is starting from the Gauss - Bonnet
invariant defined by the relation $G_{GB}\,=\,X^2-4Y+Z$
\cite{dewitt_book}. In fact the induced field equations satisfy in
four dimensions the following condition

\begin{eqnarray}\label{fieldequationGB}
H^{GB}_{\mu\nu}\,=\,H^{X^2}_{\mu\nu}-4H^{Y}_{\mu\nu}+H^{Z}_{\mu\nu}\,=\,0
\end{eqnarray}
and by substituting them at Newtonian level
($H^Z_{tt}\,\sim\,-4\triangle R^{(2)}_{tt}$) in the equations
(\ref{fieldequationHOG}) we find the field equations (ever at
Newtonian Level) of \emph{Quadratic Lagrangian} which can be recasted as

\begin{eqnarray}\label{FOG_theory_WF}
f(X,Y,Z)\,=\,R-\frac{1}{3}\biggl[\frac{1}{2\,{\mu_1}^2}+\frac{1}{{\mu_2}^2}\biggr]\,R^2
+\frac{R_{\alpha\beta}R^{\alpha\beta}}{{\mu_2}^2}
\end{eqnarray}
The theory of gravity represented by the Lagrangian (\ref{FOG_theory_WF}) is the more general theory considering all invariant curvatures, but we have to take into account  a degeneracy. In fact we  have different $f(X,Y,Z)$-gravity describing  the same Newtonian Limit (see \cite{PRD, PRD1,PRD2,Stabile_Capozziello} for details). The solution of field equations or the observable quantities, as the galactic rotation curve, are parameterized  by the derivatives of $f$, then we  have different functions $f(X,Y,Z)$ which admit the same physics in the weak field limit.


\section{Stellar hydrostatic equilibrium in $f(R)$-gravity}

Extended Theories of Gravity could have important applications at stellar level. Here we outline how stellar structure equations result modified if the underlying theory of gravity is not GR.

\subsection{The Newtonian approach to the hydrostatic equilibrium}\label{hydro_new}


The condition of hydrostatic equilibrium for  stellar structures  in Newtonian dynamics is achieved by considering the  equation \cite{kippe}

\begin{equation}\label{19.1}
\frac{dp}{dr}\,=\,-\frac{d\Phi}{dr}\rho
\end{equation}
Together with  the above equation, the Poisson equation

\begin{equation}\label{19.2}
\frac{1}{r^2}\frac{d}{dr}\left(r^2\frac{d\Phi}{dr}\right)\,=\,4\pi G\rho
\end{equation}
gives the gravitational potential $\Phi$ as solution for a given matter density $\rho$. Since we are taking into account only static and stationary situations, here  we consider only time-independent solutions \footnote{The radius $r$  is assumed as the spatial coordinate. It  ranges from $r\,=\,0$ at the stellar center to  $r\,=\,\xi$ at the surface of the star.}.
In general, the temperature $\tau$ appears in Eqs. (\ref{19.1}) and (\ref{19.2})  the density satisfies an equation of state of the form $\rho\,=\,\rho(p,\tau)$. In any case, we assume that there exists a polytropic relation between $p$ and $\rho$ of the form

\begin{equation}\label{19.3}
p\,=\,K\rho^\gamma
\end{equation}
where $K$ and $\gamma$ are constant. The polytropic constant $K$ is fixed and can be obtained as a combination of fundamental  constants. However there are several realistic  cases where $K$ is not fixed and another equation for its evolution is needed.  The constant $\gamma$  is the {\it polytropic exponent }. Inserting  the polytropic equation of state into  Eq. (\ref{19.1}), we obtain
\begin{equation}\label{19.6}
\frac{d\Phi}{dr}\,=\,-\gamma K \rho^{\gamma-2}\frac{d\rho}{dr}
\end{equation}
For  $\gamma\neq1$,  the above equation can be integrated giving

\begin{eqnarray}\label{densita}
\frac{\gamma K}{\gamma-1}\rho^{\gamma-1}\,=\,-\Phi\,\,\,\,\rightarrow\,\,\,\,\rho\,=\,\biggl[\frac{\gamma-1}{\gamma K}\biggr]^{\frac{1}{\gamma-1}}(-\Phi)^{\frac{1}{\gamma-1}}\,\doteq\,A_n(-\Phi)^n
\end{eqnarray}
where we have chosen the integration constant to give $\Phi=0$ at surface $(\rho=0$). The constant $n$ is called the {\it polytropic index} and is defined as  $n=\frac{1}{\gamma-1}$. Inserting the relation (\ref{densita})  into the Poisson equation, we obtain a differential equation for the gravitational potential
\begin{equation}\label{19.8}
\frac{d^2\Phi}{dr^2}+\frac{2}{r}\frac{d\Phi}{dr}\,=\,4\pi G A_n(-\Phi)^n\,.
\end{equation}
Let us define now the dimensionless variables

\begin{eqnarray}\label{trans1}
\left\{\begin{array}{ll}
z\,=\,|\mathbf{x}|\sqrt{\frac{\mathcal{X}A_n(-\Phi_c)^{n-1}}{2}}\\\\
w(z)\,=\,\frac{\Phi}{\Phi_c}\,=\,(\frac{\rho}{\rho_c})^\frac{1}{n}
\end{array}\right.
\end{eqnarray}
where the subscript $c$ refers to the center of the star and  the relation between $\rho$ and $\Phi$ is given by Eq. (\ref{densita}). At the center  $(r\,=\,0)$,  we have $z\,=\,0$, $\Phi\,=\,\Phi_c$, $\rho\,=\,\rho_c$ and therefore $w\,=\,1$. Then 
Eq. (\ref{19.8}) can be written

\begin{eqnarray}\label{LE}
\frac{d^2w}{dz^2}+\frac{2}{z}\frac{dw}{dz}+w^n\,=\,0
\end{eqnarray}
This is the standard {\it Lan\'{e}-Embden equation} describing the hydrostatic equilibrium of stellar structures in the Newtonian theory \cite{kippe}.


\subsection{Hydrostatic equilibrium in $f(R)$-gravity}\label{hydro_mod}


Let us consider now how hydrostatic equilibrium for stellar structures can change extending the theory to $f(R)$-gravity.
 In this case the action (\ref{HOGaction}) becomes

\begin{eqnarray}\label{f(R)action}
\mathcal{A}\,=\,\int d^{4}x\sqrt{-g}[f(R)+\mathcal{X}\mathcal{L}_m]
\end{eqnarray}
The field equations are obtained by imposing $f(X,Y,Z)\,\rightarrow\,f(R)$ in the equations (\ref{fieldequationHOG}) and (\ref{tracefieldequationHOG}). We get

\begin{eqnarray}\label{fieldequationFOG}
\begin{array}{ll}
f'R_{\mu\nu}-\frac{f}{2}g_{\mu\nu}-f_{;\mu\nu}+g_{\mu\nu}\Box f'=\mathcal{X}\,T_{\mu\nu}\\
\\
3\Box
f'+f'R-2f\,=\,\mathcal{X}\,T
\end{array}
\end{eqnarray}
where $f'\,=\,f_X$. Since we are interested in  analyzing the modification of Lan\'{e}-Embden equation, it is sufficient to solve only the field equation for the gravitational potential $\Phi$. In order to achieve the Newtonian limit of the theory, the metric tensor $g_{\mu\nu}$ (\ref{metric_tensor_PPN_0}) can be approximated as follows 

\begin{eqnarray}\label{metric_tensor_PPN_1}
g_{\mu\nu}\,\sim\,\begin{pmatrix}
1+2\,\Phi(\mathbf{x}) & 0 \\
\\
0 & -\delta_{ij} \end{pmatrix}
\end{eqnarray}
and the field equations (\ref{NL-field-equation}) become

\begin{eqnarray}\label{PPN-field-equation-general-theory-fR-O2}
\begin{array}{ll}
R^{(2)}_{tt}-\frac{R^{(2)}}{2}-f''(0)\triangle
R^{(2)}\,=\,\mathcal{X}\,T^{(0)}_{tt}
\\\\
-3f''(0)\triangle
R^{(2)}-R^{(2)}\,=\,\mathcal{X}\,T^{(0)}
\end{array}
\end{eqnarray}
where for the sake of simplicity we set $f'(0)\,=\,1$. By the definition of mass (\ref{mass_f(R)}) and $R^{(2)}_{tt}\,=\,\triangle\Phi(\mathbf{x})$ we have

\begin{eqnarray}\label{HOEQ}
\begin{array}{ll}
\triangle\Phi-\frac{R^{(2)}}{2}+\frac{\triangle R^{(2)}}{3{m}^2}\,= \,\mathcal{X}\rho
\\\\
(\triangle-m^2)R^{(2)}\,=\,m^2\mathcal{X}\rho
\end{array}
\end{eqnarray}
We note that  for $f''(0)\,\rightarrow\,0$ (or $m\,\rightarrow\,\infty$) we have the  standard Poisson equation: $\triangle\Phi\,=\,4\pi G\rho$. This means that as soon as the second derivative of $f$ is different from zero, deviations from the Newtonian limit of GR emerge. The equations (\ref{HOEQ}) can be considered the \emph{modified Poisson equation}
for $f(R)$-gravity. They  do not depend on gauge condition choice \cite{Stabile_Capozziello}. From the Bianchi identity, satisfied by the field equations (\ref{fieldequationHOG}), we have

\begin{eqnarray}\label{equidr}
{T^{\mu\nu}}_{;\mu}\,=\,0\,\,\,\,\rightarrow\,\,\,\,\frac{\partial p}{\partial x^k}\,=\,-\frac{1}{2}(p+\rho)\frac{\partial \ln g_{tt}}{\partial x^k}
\end{eqnarray}
If the dependence on the temperature  $\tau$ is negligible, \emph{i.e.} $\rho\,=\,\rho(p)$,  this relation can be introduced into Eqs. (\ref{HOEQ}), which become a system of three equations for $p$, $\Phi$ and $R^{(2)}$ and can be solved without the other structure equations.

Let us suppose that matter satisfies still a polytropic equation $p\,=\,K\,\rho^\gamma$. If we introduce the relation (\ref{densita}) into field equations (\ref{HOEQ}) we obtain an integro-differential equation for the gravitational potential $\Phi$, that is 

\begin{eqnarray}\label{deltafi}
\triangle\Phi(\mathbf{x})-\frac{2\mathcal{X}A_n}{3}(-\Phi(\mathbf{x}))^n\,=\,\frac{m^2\mathcal{X}A_n}{6} \int d^3\mathbf{x}'\mathcal{G}_m(\mathbf{x},\mathbf{x}')(-\Phi(\mathbf{x}'))^n
\end{eqnarray}
The integro-differential nature of Eq.(\ref{deltafi}) is the proof of the non-viability of Gauss theorem for $f(R)$-gravity. Adopting again the dimensionless variables

\begin{eqnarray}\label{trans}
z\,=\,\frac{|\mathbf{x}|}{\xi_0},\qquad\qquad w(z)\,=\,\frac{\Phi}{\Phi_c}
\end{eqnarray}
where

\begin{eqnarray}\label{charpar}
\xi_0\,\doteq\,\sqrt{\frac{3}{2\mathcal{X}A_n(-\Phi_c)^{n-1}}}
\end{eqnarray}
is a characteristic length linked to stellar radius $\xi$, Eq. (\ref{deltafi}) becomes

\begin{eqnarray}\label{LEmod}
\frac{d^2w(z)}{dz^2}+\frac{2}{z}\frac{d w(z)}{dz}+w(z)^n\,=\,\frac{m\xi_0}{8}\frac{1}{z}\int_0^{\xi/\xi_0}
dz'\,z'\,\biggl\{e^{-m\xi_0|z-z'|}-e^{-m\xi_0|z+z'|}\biggr\}\,w(z')^n
\end{eqnarray}
which is the \emph{modified Lan\'{e}-Emden equation} deduced from $f(R)$-gravity. Clearly  the particular $f(R)$-model is specified by the parameters $m$ and $\xi_0$. 
If $m\,\rightarrow\,\infty$ (\emph{i.e.} $f(R)\,\rightarrow\,R$),  Eq. (\ref{LEmod}) becomes Eq. (\ref{LE}). We are only interested in solutions of Eq. (\ref{LEmod}) that are finite at the center, that is for $z\,=\,0$. Since the center must be an equilibrium point,  the gravitational acceleration $|\mathbf{g}|\,=\,d\Phi/dr\,\propto\,dw/dz$ must vanish for $w'(0)\,=\,0$. Let us assume we have solutions $w(z)$ of Eq.(\ref{LEmod}) that fulfill the  boundary conditions $w(0)\,=\,1$ and $w(\xi/\xi_0)\,=\,0$; then according to the choice (\ref{trans}), the radial distribution of  density is given by

\begin{eqnarray}
\rho(|\mathbf{x}|)\,=\,\rho_cw^n\,,\,\,\,\,\,\,\,\,\rho_c\,=\,A_n{\Phi_c}^n
\end{eqnarray}
and the pressure by

\begin{eqnarray}
p(|\mathbf{x}|)\,=\,p_cw^{n+1}\,,\,\,\,\,\,\,\,\,p_c\,=\,K{\rho_c}^\gamma
\end{eqnarray}

For $\gamma\,=\,1$ (or $n\,=\,\infty$) the integro-differential Eq. (\ref{LEmod}) is not correct. This means that the theory does not contain the case of isothermal sphere of ideal gas. 
In this case, the polytropic relation is $p\,=\,K\,\rho$. Putting this relation into Eq.(\ref{equidr}) we have
\begin{eqnarray}\label{densitaiso}
\frac{-\Phi}{K}\,=\,\ln\rho-\ln\rho_c\,\,\,\,\rightarrow\,\,\,\,\rho\,=\,\rho_c\,e^{-\Phi/K}
\end{eqnarray}
where  the constant of integration is chosen in such a way that the gravitational potential is zero at the center. If we introduce Eq.(\ref{densitaiso}) into Eqs. (\ref{HOEQ}), we have

\begin{eqnarray}\label{deltafiiso}
\triangle\Phi(\mathbf{x})-\frac{2\mathcal{X}\rho_c}{3}e^{-\Phi(\mathbf{x})/K}\,=\,\frac{m^2\mathcal{X}\rho_c}{6} \int d^3\mathbf{x}'\mathcal{G}_m(\mathbf{x},\mathbf{x}')e^{-\Phi(\mathbf{x}')/K}
\end{eqnarray}
Assuming the dimensionless variables $z\,=\,\frac{|\mathbf{x}|}{\xi_1}$ and $w(z)\,=\,\frac{-\Phi}{K}$ where $\xi_1\,\doteq\,\sqrt{\frac{3K}{2\mathcal{X}\rho_c}}$, Eq. (\ref{deltafiiso}) becomes

\begin{eqnarray}\label{LEmodiso}
\frac{d^2w(z)}{dz^2}+\frac{2}{z}\frac{d w(z)}{dz}+e^{w(z)}\,=\,\frac{m\xi_1}{8}\frac{1}{z}\int_0^{\xi/\xi_1}
dz'\,z'\,\biggl\{e^{-m\xi_1|z-z'|}-e^{-m\xi_1|z+z'|}\biggr\}\,e^{w(z')}
\end{eqnarray}
which is the \emph{modified "isothermal" Lan\'{e}-Emden equation} derived $f(R)$-gravity.


\subsection{Solutions for the standard and modified Lan\'{e}-Emden Equations}\label{hydro_solutions}

\begin{figure}[htbp]
\centering
\includegraphics[scale=.5]{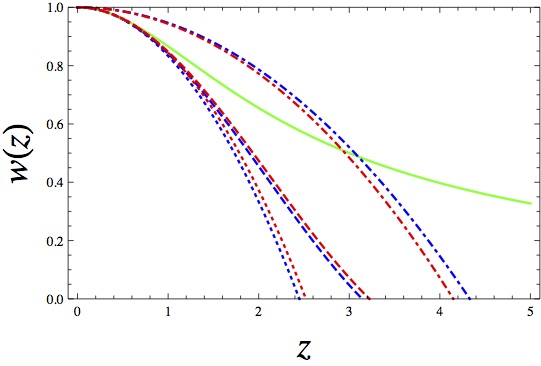}\\
\caption{Plot of solutions (blue lines) of standard Lan\'{e}-Emden Eq. (\ref{LE}): $w^{(0)}_{GR}(z)$ (dotted line) and $w^{(1)}_{GR}(z)$ (dashed line). The green line corresponds to $w^{(5)}_{GR}(z)$. The red lines are the solutions of modified Lan\'{e}-Emden Eq. (\ref{LEmod}): $w^{(0)}_{_{f(R)}}(z)$ (dotted line) and $w^{(1)}_{_{f(R)}}(z)$ (dashed line). The blue dashed-dotted line is the  potential  derived from GR ($w_{GR}(z)$) and the red dashed-dotted line the  potential derived from  $f(R)$-gravity ($w_{_{f(R)}}(z)$) for a uniform spherically symmetric mass distribution. The assumed values are $m\xi\,=\,1$ and $m\xi_0\,=\,.4$. From a rapid inspection of these plots, the differences between GR and $f(R)$ gravitational potentials are clear and the tendency is that at larger radius $z$ they become more evident.}
\label{fig}
\end{figure}

The task is now to solve the modified Lan\'{e}-Emden equation and compare its solutions to those of standard Newtonian theory.
Only for three values of $n$,  the solutions of Eq.(\ref{LE}) have analytical expressions \cite{kippe}

\begin{eqnarray}\label{LEsol}
&&n\,=\,0\,\,\,\,\rightarrow\,\,\,\,w^{(0)}_{GR}(z)\,=\,1-\frac{z^2}{6}\nonumber\\
&&n\,=\,1\,\,\,\,\rightarrow\,\,\,\,w^{(1)}_{GR}(z)\,=\,\frac{\sin z}{z}\\
&&n\,=\,5\,\,\,\,\rightarrow\,\,\,\,w^{(5)}_{GR}(z)\,=\,\frac{1}{\sqrt{1+\frac{z^2}{3}}}\nonumber
\end{eqnarray}
We label these solution with $_{GR}$ since they agree with the Newtonian limit of GR.
The surface of the polytrope of index $n$ is defined by the value $z\,=\,z^{(n)}$, where $\rho\,=\,0$ and thus $w\,=\,0$. For $n\,=\,0$ and $n\,=\,1$ the surface is reached for a finite value of $z^{(n)}$. The case $n\,=\,5$ yields a model of infinite radius. It can be shown that for $n\,<\,5$ the radius of polytropic models is finite; for $n\,>\,5$ they have infinite radius. From Eqs.(\ref{LEsol}) one finds $z^{(0)}_{GR}\,=\,\sqrt{6}$, $z^{(1)}_{GR}\,=\,\pi$ and $z^{(5)}_{GR}\,=\,\infty$. A general property of the solutions is that $z^{(n)}$ grows monotonically with the polytropic index $n$. In Fig. \ref{fig} we show the behavior of solutions $w^{(n)}_{GR}$ for $n\,=\,0,\,1,\,5$.
Apart from the three cases where analytic solutions are known, the classical Lan\'{e}-Emden Eq. (\ref{LE}) has to be be solved numerically, considering with the expression

\begin{eqnarray}\label{taydev}
w^{(n)}_{GR}(z)\,=\,\sum_{i\,=\,0}^\infty a^{(n)}_iz^i
\end{eqnarray}
for the neighborhood of the center. Inserting Eq.(\ref{taydev}) into Eq. (\ref{LE}) and by comparing coefficients one finds, at lowest orders, a classification of solutions by the index $n$, that is

\begin{eqnarray}\label{gensolGR}
w^{(n)}_{GR}(z)\,=\,1-\frac{z^2}{6}+\frac{n}{120}z^4+\dots
\end{eqnarray}
The case  $\gamma\,=\,5/3$ and $n\,=\,3/2$ is the non-relativistic limit  while the case  $\gamma\,=\,4/3$ and $n\,=\,3$ is the relativistic limit of a completely degenerate gas.

Also for modified Lan\'{e}-Emden Eq. (\ref{LEmod}),  we have an  exact solution for $n\,=\,0$. In fact, it is straightforward to find out 

\begin{eqnarray}\label{LEmodsol0}
w^{(0)}_{_{f(R)}}(z)\,=\,1-\frac{z^2}{8}+\frac{(1+m\xi)e^{-m\xi}}{4m^2{\xi_0}^2}\biggl[1-\frac{\sinh m\xi_0 z}{m\xi_0 z}\biggr]
\end{eqnarray}
where  the boundary conditions $w(0)\,=\,1$ and $w'(0)\,=\,0$ are satisfied. A comment on the GR limit (that is  $f(R)\rightarrow R$) of solution (\ref{LEmodsol0}) is necessary. In fact when we perform the limit $m\,\rightarrow\,\infty$, we do not recover  exactly $w^{(0)}_{GR}(z)$. The difference is in the definition of quantity $\xi_0$. In $f(R)$-gravity we have the definition (\ref{charpar}) while in GR it is ${\displaystyle \xi_0\,=\,\sqrt{\frac{2}{\mathcal{X}A_n\Phi_c^{n-1}}}}$, since in the first equation of (\ref{HOEQ}), when we perform $f(R)\rightarrow R$, we have to eliminate the trace equation condition. In general, this means that the Newtonian limit and the Eddington parameterization of different relativistic theories of gravity cannot coincide with those of GR (see \cite{edd} for further details on this point).

The point $z_{_{f(R)}}^{(0)}$ is calculated by imposing $w^{(0)}_{_{f(R)}}(z_{_{f(R)}}^{(0)})\,=\,0$ and by considering the Taylor expansion 

\begin{equation}
\frac{\sinh m\xi_0z}{m\xi_0z}\,\sim\,1+\frac{1}{6}(m\xi_0z)^2+\mathcal{O}(m\xi_0z)^4
\end{equation}
we obtain ${\displaystyle z_{_{f(R)}}^{(0)}\,=\,\frac{2\sqrt{6}}{\sqrt{3+(1+m\xi)e^{-m\xi}}}}$.
Since the stellar radius $\xi$ is given by definition $\xi\,=\,\xi_0\,z_{_{f(R)}}^{(0)}$, we obtain the  constraint

\begin{eqnarray}\label{radius_constraint}
\xi\,=\,\sqrt{\frac{3\Phi_c}{2\pi G}}\frac{1}{\sqrt{1+\frac{1+m\xi}{3}e^{-m\xi}}}
\end{eqnarray}
By solving numerically the constraint\footnote{In principle, there is a solution for any value of $m$.} Eq.(\ref{radius_constraint}), we find the modified expression of the radius $\xi$. If $m\,\rightarrow\,\infty$ we have the standard expression $\xi\,=\,\sqrt{\frac{3\Phi_c}{2\pi G}}$ valid for the Newtonian limit of GR. Besides, it is worth noticing  that in the  $f(R)$-gravity case, for $n=0$, the radius is smaller than in GR. On the other hand,  the gravitational potential $-\Phi$ gives rise to a deeper potential well than the corresponding Newtonian potential derived from GR \cite{Stabile_Capozziello}. 

In the case $n\,=\,1$, Eq. (\ref{LEmod}) can be recast as follows

\begin{eqnarray}\label{LEmod2}
\frac{d^2\tilde{w}(z)}{dz^2}+\tilde{w}(z)\,=\,\frac{m\xi_0}{8}\int_0^{\xi/\xi_0}
dz'\,\biggl\{e^{-m\xi_0|z-z'|}-e^{-m\xi_0|z+z'|}\biggr\}\,\tilde{w}(z')
\end{eqnarray}
where $\tilde{w}\,=\,z\,w$. If we consider the solution of (\ref{LEmod2}) as a small perturbation to the one of GR, we have

\begin{eqnarray}\label{hypsol}
\tilde{w}^{(1)}_{_{f(R)}}(z)\,\sim\,\tilde{w}^{(1)}_{GR}(z)+e^{-m\xi}\Delta\tilde{w}^{(1)}_{_{f(R)}}(z)
\end{eqnarray}
The coefficient $e^{-m\xi}\,<\,1$ is the parameter with respect to which we perturb Eq. (\ref{LEmod2}). Besides these position  ensure us that when $m\,\rightarrow\,\infty$ the solution converge to something like 
$\tilde{w}^{(1)}_{GR}(z)$. Substituting Eq.(\ref{hypsol}) in Eq.(\ref{LEmod2}),  we have

\begin{eqnarray}
\frac{d^2\Delta\tilde{w}^{(1)}_{_{f(R)}}(z)}{dz^2}+\Delta\tilde{w}^{(1)}_{_{f(R)}}(z)\,=\,\frac{m\xi_0\,e^{m\xi}}{8}\int_0^{\xi/\xi_0}
dz'\,\biggl\{e^{-m\xi_0|z-z'|}-e^{-m\xi_0|z+z'|}\biggr\}\,\tilde{w}^{(1)}_{GR}(z')
\end{eqnarray}
and the solution is easily found

\begin{eqnarray}
w^{(1)}_{_{f(R)}}(z)\,\sim\,&&\frac{\sin z}{z}\biggl\{1+\frac{m^2{\xi_0}^2}{8(1+m^2{\xi_0}^2)}\biggl[1+\frac{2\,e^{-m\xi}}{1+m^2{\xi_0}^2}(\cos\xi/\xi_0+m\xi_0
\sin\xi/\xi_0)
\biggr]\biggr\}
\nonumber\\\nonumber\\&&
-\frac{m^2{\xi_0}^2}{8(1+m^2{\xi_0}^2)}\biggl[\frac{2\,e^{-m\xi}}{1+m^2{\xi_0}^2}(\cos\xi/\xi_0+m\xi_0\sin\xi/\xi_0)
\frac{\sinh m\xi_0z}{m\xi_0z}+\cos z\biggr]
\end{eqnarray}
Also in this case,  for $m\,\rightarrow\,\infty$, we do not  recover exactly $w^{(1)}_{GR}(z)$. The reason is the same of previous $n\,=\,0$ case \cite{edd}. Analytical solutions for other values of $n$ are not available.

To conclude this section,  we report  the  gravitational potential profile generated by a spherically symmetric source of  uniform mass with radius $\xi$. By solving field Eqs. (\ref{HOEQ}) {\it inside the star}  and considering the boundary conditions $w(0)\,=\,1$ and $w'(0)\,=\,0$, we get
\begin{eqnarray}\label{sol_pot}
w_{_{f(R)}}(z)\,=\,\biggl[\frac{3}{2\xi}+\frac{1}{m^2\xi^3}-\frac{e^{-m\xi}(1+m\,\xi)}{m^2\xi^3}\biggr]^{-1}
\biggl[\frac{3}{2\xi}+\frac{1}{m^2\xi^3}-\frac{{\xi_0}^2z^2}{2\xi^3}-\frac{e^{-m\xi}(1+m\,\xi)}{m^2\xi^3}\frac{\sinh m\xi_0z}{m\xi_0z}\biggr]
\end{eqnarray}
In the limit $m\,\rightarrow\,\infty$, we recover the GR case $w_{GR}(z)\,=\,1-\frac{{\xi_0}^2z^2}{3\xi^2}$. In Fig. \ref{fig} we show the behaviors of $w^{(0)}_{_{f(R)}}(z)$ and $w^{(1)}_{_{f(R)}}(z)$ with respect to the corresponding GR cases. Furthermore, we plot the potential generated by  a uniform spherically symmetric  mass distribution in GR and $f(R)$-gravity and the case $w^{(5)}_{GR}(z)$.

An important remark is in order at this point.  In a general discussion of the problem, also time-dependent perturbations should be taken into account.  In particular, quadratic gravity possess gosth-like instability related to the sign of coefficient in front of the term $R^2$. In such  cases, unstable or unphysical solutions come out and have to be discriminated with respect to physical cases. A discussion in this sense is in \cite{farinelli,formisano}.   

\section{Rotation curves of galaxies}\label{curve}\label{gal_rotation}

Another important astrophysical application of the above Newtonian limit of Fourth Order Gravity is related to the  rotation curves of spiral galaxies.
The approach can be considered also an alternative test for  DM behavior.


In general, the motion of body embedded in the gravitational field is given by geodesic equation (\ref{geodesic}) and in the Newtonian limit, formally, we obtain the classical structure of the motion equation (\ref{geodesic_new})

\begin{eqnarray}
\frac{d^2\,\mathbf{x}}{dt^2}\,=\,-\nabla\Phi(\mathbf{x})
\end{eqnarray}
but the gravitational potential is given by (\ref{sol_point}) or generally by (\ref{superposition_phi}). The study of motion is very simple if we consider a particular symmetry of mass distribution $\rho$, otherwise the analytical solutions are not available. Our aim is to evaluate the corrections to the classical motion in the easiest situation: the circular motion. In this case we do not consider the radial and vertical motion. The condition of stationary motion on the circular orbit is

\begin{eqnarray}\label{stazionary_motion}
\frac{{v_c(|\mathbf{x}|)}^2}{|\mathbf{x}|}\,=\,\frac{\partial\Phi(\mathbf{x})}{\partial|\mathbf{x}|}
\end{eqnarray}
where $v_c$ is the velocity.

The distribution of mass can be modeled simply by introducing two sets of coordinates: the spherical coordinates $(r,\theta,\phi)$ and the cylindrical coordinates $(R,\theta,z)$. An useful mathematical tool is the Gauss flux theorem for gravity: \emph{The gravitational flux through any closed surface is proportional to the enclosed mass}. The law is expressed in terms of the gravitational field. The gravitational field $\mathbf{g}$ is defined so that the gravitational force experienced by a particle with mass $m$ is $\mathbf{F}_{grav}\,=\,m\,\mathbf{g}$. Since the Newtonian mechanics satisfies this theorem and, by thinking to a spherical system of mass distribution, we get, from (\ref{stazionary_motion}), the equation

\begin{eqnarray}\label{circular_velocity}
{v_c(r)}^2\,=\,\frac{G\,M(r)}{r}\,=\,\frac{4\pi G}{r}\int_0^rdy\,y^2\,\rho(y)
\end{eqnarray}
where $M(r)$ is the only mass enclosed in the sphere with radius $r$. The Green function of the $f(X,Y,Z)$-gravity ($\neq\,|\mathbf{x}-\mathbf{x}'|^{-1}$), instead, does not satisfy the theorem \cite{Stabile_Capozziello}. In this case we must consider directly the gravitational potential (\ref{superposition_phi}) generated by using the superposition principle.

Apart the mathematical difficulties incoming from the research of gravitational potential for a given mass distribution, the non-validity of Gauss theorem implies, for example, that a sphere can not be reduced to a point (see solution (\ref{ST_FOG_FE_NL_sol_ball})). In fact the gravitational potential generated by a ball (also with constant density) is depending also on the Fourier transform of ball \cite{Stabile_Capozziello}. However our aim is to consider not the simple case of motion of body in the vacuum, but the more interesting case of motion in the matter. So we must leave any possibility of idealization and consider directly the calculation of the potential (\ref{superposition_phi}).

Two remarks on the (\ref{superposition_phi}) are needed. The two corrections have different algebraic sign, and in particular the Yukawa correction with $\mu_1$ implies a stronger gravitational force, while the second one (purely induced by Ricci and Riemann square) contributes with a repulsive force. By remembering that the motivations of extending the outcome of GR to new theories is supported by missing matter justifying the flat rotation curves of galaxies, the first correction is a nice candidate. A crucial point is given by the spatial range of correction. In fact the Yukawa corrections imply a massive propagation; then, more massive is the particle, shorter is the spatial range.

At last in Newtonian Mechanics the Gauss theorem gives us a spherically symmetric gravitational potential even if the spherically symmetric source is rotating. In GR as well as in Fourth Order Gravity, however, the rotating spherically symmetric source generates an axially symmetric space-time (the well known Kerr metric) and only if the source is at rest one has the space-time with the same symmetry (the well known Schwarzschild metric). Then the galaxy being a rotating system will generate an axially symmetric space-time while we are using the solution (\ref{superposition_phi}). This aspect is not contradictory because the solutions are calculated in the Newtonian limit (\emph{i.e.} $v^2\,\ll\,1$) and under this assumption the Kerr metric collapses into Schwarzschild metric. In fact we have

\begin{eqnarray}\label{Kerr-Scwh}
g^{Kerr}_{\mu\nu}\,=\,&&\begin{pmatrix}
1-\frac{r_g r}{\Sigma^2}& 0 & 0 & \frac{r_g r \eta}{\Sigma^2}\sin^2\theta\\
\\
0 & -\frac{\Sigma^2}{H^2} & 0 & 0\\
\\
0 & 0 & -\Sigma^2 & 0\\
\\
\frac{r_g r \eta}{\Sigma^2}\sin^2\theta & 0 & 0 & -\biggl(r^2+\eta^2+\frac{r_g r \eta^2}{\Sigma^2}\sin^2\theta\biggr)\sin^2\theta\\
\end{pmatrix}\rightarrow\nonumber\\\\
&& \rightarrow g^{Schw}_{\mu\nu}\,=\,\begin{pmatrix}
1-\frac{r_g}{r}& 0 & 0 & 0\\
\\
0 & -1-\frac{r_g}{r} & 0 & 0\\
\\
0 & 0 & -r^2 & 0\\
\\
0 & 0 & 0 & -r^2\sin^2\theta\\
\end{pmatrix}\nonumber
\end{eqnarray}
where $\Sigma^2\,=\,r^2+\eta^2\cos^2\theta$, $H^2\,=\,r^2-r_g r+\eta^2$, $\eta\,=\,L/M$ and $L$ is the angular momentum along the z-axis.


\subsection{Mass models  for galaxies}\label{mass_model}


From the point of view of morphology, a galaxy can be modeled by considering at least two
components: the bulge and the disk. Obviously the galaxy is a more complicated structure and there are others components, but for our aim this idealization is satisfactory. The bulge, generally, can be represented easily with cylindrical coordinates (but in a more crude idealization it is like a ball), while the disk has a radius bigger than the thickness. However we find that the rotation curve does not present the Keplerian behavior outside the matter, but the curve remain constant for any distance. Then we must formulate the existence of exotic matter that can justify the experimental observation. A simple discussion about the distribution of DM can be formulated by imposing the constant value of velocity in (\ref{circular_velocity}) for large distances. In fact we find

\begin{eqnarray}\label{density_DM}
v_c(r)\,\sim\,\text{constant}\,\rightarrow\,\rho_{DM}(r)\,\sim\,r^{-2}
\end{eqnarray}

A matter distribution as (\ref{density_DM}) has a problem when we want to calculate the total mass. In fact if we have $\rho\,\sim\,r^{-2}$, the mass diverges. A such exotic behavior seems no-physical, but this outcome is only the consequence of constant rotation curve. In fact by increasing the distance also the mass must increase with the power law for any distance (\ref{circular_velocity}). However, since the Gauss theorem holds in GR, the matter outside the sphere of integration does not contribute to the gravitational flux and then we do not have difference with respect to the ordinary matter.

The same arguments are not valid in $f(X,Y,Z)$-gravity: the non-viability of Gauss theorem implies that the range of integration of DM could cover all range and also the matter outside is considered. A cut off is needed now. Here then for completeness we consider that the galaxy is composed by three components: the bulge, the disk and an alone of DM.

It should be noted that the spatial behaviors of DM (generally spherically symmetric) are made only \emph{a posteriori}: the cornerstones  of rotation curves study  are the GR and the distribution of ordinary matter. Only after this assumption the distribution of DM is such as to justify the gap between the theoretical prediction and the experimental observation.

Before jumping to analysis of rotation curves we want to resume the principal spatial distributions of mass in the three galactic components. In literature there are many forms of density, but it is possible to resume them as follows.

More realistic models are the ones with mass density depending also on the $z$-coordinate for bulge and disk. Particularly one can consider the following choice

\begin{eqnarray}\label{density_1}
\left\{\begin{array}{ll}
\rho_{bulge}(R,z)\,=\,\frac{M_b}{4\pi\,V_0}\frac{{\xi_b}^{\gamma}}{[R^2+z^2/q^2]
^{\gamma/2}}\biggl[\frac{\xi_b+\sqrt{R^2+z^2/q^2}}
{\xi_b}\biggr]^{\gamma-\beta}\,e^{-\frac{R^2+z^2/q^2}{{\xi_t}^2}}\\\\
\rho_{disk}(R,z)\,=\,\frac{M_d}{4\pi\,{\xi_d}^2z_d}\,e^{-\frac{R}{\xi_d}
-\frac{|z|}{z_d}}\\\\
\rho_{DM}(r)\,=\,\frac{M_{DM}^{vir}}{4\pi{\xi_s}^3\,g(\xi_{DM}^{vir}/\xi_s)}\frac{\xi_s}{r}\frac{1}{[1+r/\xi_s]^2}
\end{array}\right.
\end{eqnarray}
In this choice, suggested by Dehnen \& Binney \cite{deh}, the bulge is described as a truncated power - law model (first line of
(\ref{density_1})) where $V_0\,=\,\int_{0}^{\infty}
dR'\,R'\,\int_{0}^{\infty}dz'\tilde{\rho}_{bulge}(R',z')$.
$\beta$, $\gamma$, $q$ and $\xi_t$ are the parameters. While for
the disk, one adopted a double exponential where the total
mass is $M_d\,=\,2\pi{\xi_d}^2\,\Sigma_\odot\,e^{\frac{\xi_0}{\xi_d}}$
with $\Sigma_\odot\,=\,(48\,\pm\,8)\,M_\odot/\text{pc}^2$ and
$\xi_0\,=\,8.5\,\text{Kpc}$ \cite{kui}. Finally in the case of DM the density profile is the NFW model \cite{navarro, NFW} where $g(x)\,=\,\ln(1+x)-\frac{x}{x+1}$, $M_{DM}^{vir}$ and
$\xi_{DM}^{vir}$ are the virial mass and virial radius and $\xi_s$
is a characteristic length. For the Milky Way
$\xi_{DM}^{vir}/\xi_s\,=\,10\,\div\,15$.

Leaving the axis-symmetry one can consider a more simple model:
the spherical symmetry model. With this approach one has \cite{wyse, noo, bin}

\begin{eqnarray}\label{density_2}
\left\{\begin{array}{ll}
\rho_{bulge}(r)\,=\,\frac{k\,M_b}{4\eta\,\pi\,{\xi_b}^{9/4}}\int_r^\infty dx\,\frac{e^{-k[(x/{\xi_b})^{1/4}-1]}}{x^{3/4}\sqrt{x^2-r^2}}\\\\
\sigma_{disk}(R)\,=\,\frac{M_d}{2\pi\,{\xi_d}^2}\,
e^{-\frac{R}{\xi_d}}\\\\
\rho_{DM}(r)\,=\,\frac{M_{DM}}{2(4-\pi)\pi{\xi_{DM}}^3}\,\frac{1}{1+\frac{r^2}{{\xi_{DM}}^2}}
\end{array}\right.
\end{eqnarray}
where $\xi_b$, $\xi_d$, $\xi_{DM}$, $M_b$, $M_d$, $M_{DM}$, are
the radii and the masses of bulge, disk and DM.
$k\,=\,7.6695$ and $\eta\,=\,22.665$ are dimensionless constants.
The density profile of bulge considered is the well-known
formula of de Vaucouleurs \cite{vau}.

A  simpler model, resuming the previous ones, can be

\begin{eqnarray}\label{density_3}
\left\{\begin{array}{ll}
\rho_{bulge}(r)\,=\,\frac{M_b}{2\,\pi\,{\xi_b}^{3-\gamma}\,\Gamma(\frac{3-\gamma}{2})}\frac{e^{-\frac{r^2}{{\xi_b}^2}}}{r^\gamma}\\\\
\sigma_{disk}(R)\,=\,\frac{M_d}{2\pi\,{\xi_d}^2}\,
e^{-\frac{R}{\xi_d}}\\\\
\rho_{DM}(r)\,=\,\frac{\alpha\,M_{DM}}{\pi\,(4-\pi){\xi_{DM}}^3}\,\frac{1}{1+\frac{r^2}{{\xi_{DM}}^2}}
\end{array}\right.
\end{eqnarray}
where $\Gamma(x)$ is the Gamma function, $0\,\leq\,\gamma\,<\,3$
is a free parameter and $0\,\leq\,\alpha\,<\,1$ is the ratio of DM
inside the sphere with radius $\xi_{DM}$ with respect to the total DM. The radius $\xi_{DM}$ and the mass $M_{DM}$ play conceptually the same role respectively of $\xi_{DM}^{vir}$ and $M_{DM}^{vir}$. However, as before claimed, the hot point is the choice of DM model. Given all these models it is almost normal that there are many different estimates of DM. Therefore, the parameters of DM model may not be unique \cite{bur}.


\subsection{Rotation curves in $f(X,Y,Z)$-gravity}\label{curve_FOG}


We are interested to evaluate the circular velocity (\ref{stazionary_motion}) adopting the mass models (\ref{density_3}). So the potential (\ref{superposition_phi}), by setting $f_X(0)\,=\,\,1$, becomes

\begin{eqnarray}
\Phi(r,R,z)\,=&&G\,\biggl\{\frac{2\,M_b}{3\,{\xi_b}^{3-\gamma}\,\Gamma(\frac{3-\gamma}{2})}\frac{1}{r}\int_0^{\infty}
dr'\,r'^{1-\gamma}\,e^{-\frac{r'^2}{{\xi_b}^2}}\biggl[3\,\frac{|r-r'|-r-r'}{2}
\nonumber\\\nonumber\\\nonumber\\
&-&\frac{e^{-\mu_1a|r-r'|}-e^{-\mu_1a(r+r')}}{2\,\mu_1\,a}
+2\,\frac{e^{-\mu_2a|r-r'|}-e^{-\mu_2a(r+r')} }{\mu_2\,a}\biggr]
\nonumber\\\nonumber\\\nonumber\\
&+&\frac{4\,\alpha\,M_{DM}}{3(4-\pi)\,{\xi_{DM}}^3}\frac{1}{r}\int_0^{\Xi/a}
dr'\,\frac{r'}{1+\frac{r'^2}{{\xi_{DM}}^2}}\biggl[3\,\frac{|r-r'|-r-r'}{2}
\nonumber
\end{eqnarray}
\begin{eqnarray}\label{sol_gen_3}
&-&\frac{e^{-\mu_1a|r-r'|}-e^{-\mu_1a(r+r')}}{2\,\mu_1\,a}
+2\,\frac{e^{-\mu_2a|r-r'|}-e^{-\mu_2a(r+r')} }{\mu_2\,a}
\biggr]
\nonumber\\\nonumber\\\nonumber\\
&-&\,\frac{M_d}{\pi\,{\xi_d}^2}\,\biggr[\int_0^\infty
dR'\,e^{-\frac{R'}{\xi_d}}\,R'\,\biggl(\frac{\mathfrak{K}(\frac{4RR'}{(R+R')^2+z^2})}{\sqrt{(R+R')^2+z^2}}+
\frac{\mathfrak{K}(\frac{-4RR'}{(R-R')^2+z^2})}{\sqrt{(R-R')^2+z^2}}\biggr)
\nonumber\\\nonumber\\\nonumber\\
&+&\int_0^\infty dR'\,e^{-\frac{R'}{\xi_d}}\,R'\,\int_0^{\pi}
d\theta'\frac{e^{-\mu_1a\,\Delta(R,R',z,0,\theta')}
-4\,e^{-\mu_2a\,\Delta(R,R',z,0,\theta')}}{3\,\Delta(R,R',z,0,\theta')}\biggr]\biggl\}
\end{eqnarray}
where $\Xi$ is the distance on which we observe the rotation
curve, $\mathfrak{K}$ is the elliptic function and the modulus of
distance is given by

\begin{eqnarray}\label{distance}
\Delta(R,R',z,z',\theta')\,\doteq\,|\mathbf{x}-\mathbf{x}'|\,=\,\sqrt{(R+R')^2+(z-z')^2-4RR'\cos^2\theta'}\
\end{eqnarray}
The constant $a$ is a scale factor defined by the
substitution $R\,, r\,\rightarrow\,a\,r\,,a\,R$ so all quantities are
dimensionless. At last, by remembering $r\,=\,\sqrt{R^2+z^2}$ the circular speed (\ref{stazionary_motion}) in the galactic plan is given by

\begin{eqnarray}\label{velocity}
v_c(R)\,=\,\sqrt{R\,\frac{\partial}{\partial R}\,\Phi(R,R,0)}
\end{eqnarray}

\begin{figure}[htbp]
\centering
\includegraphics[scale=.3]{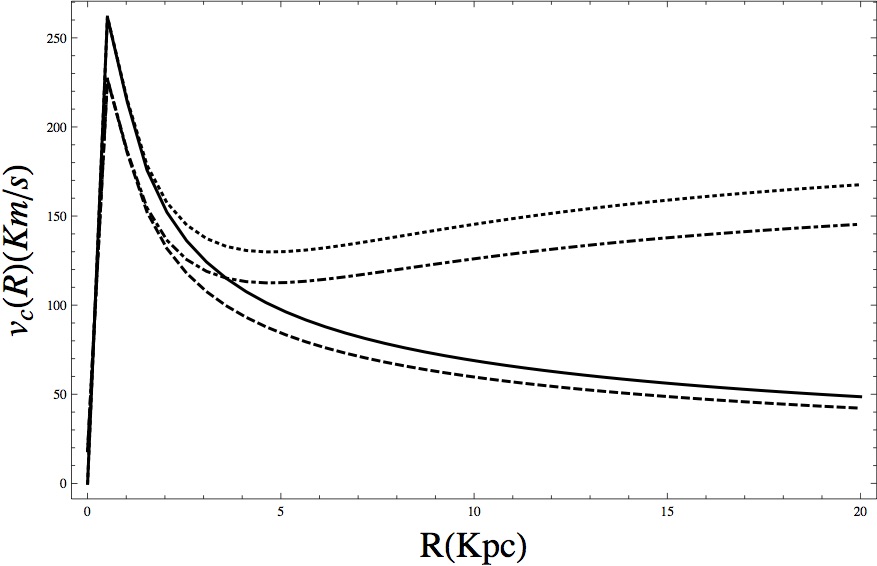}\\
\caption{The rotation curve induced by bulge component (first line of (\ref{density_3})): GR (dashed line), GR + DM (dashed and dotted line), FOG (solid line) and FOG + DM (dotted line).}
\label{vel_1}
\end{figure}

\begin{figure}[htbp]
\centering
\includegraphics[scale=.3]{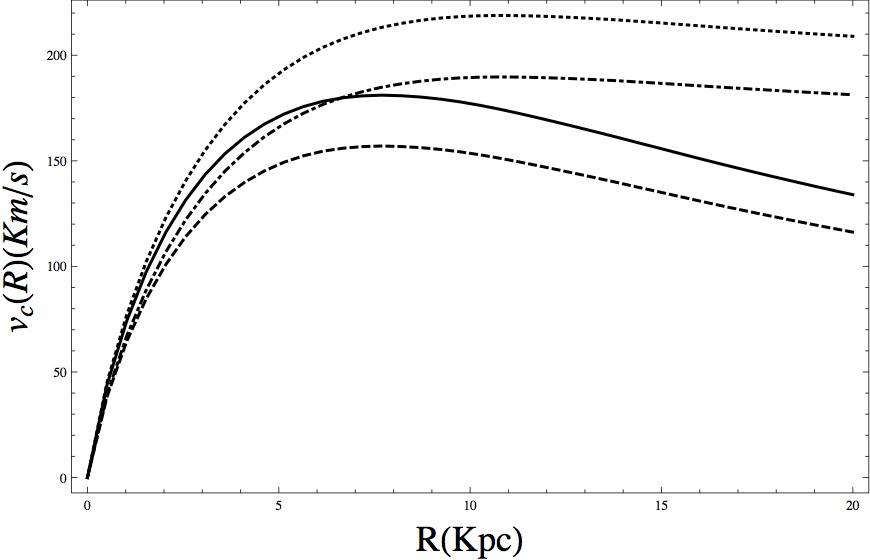}\\
\caption{The rotation curve induced by disk component (second line of (\ref{density_3})): GR (dashed line), GR + DM (dashed and dotted line), FOG (solid line) and FOG + DM (dotted line).}
\label{vel_2}
\end{figure}

In the Figs. \ref{vel_1}, \ref{vel_2}, we report the spatial behaviors of rotation curve induced by the bulge and disk component. The behavior for any component is compared in the framework of GR, FOG, GR + DM and FOG + DM. The values of free parameters of model are in the first line in Table \ref{tab_1} and refereing to Milky Way. The values of scale lengths $\mu_1$, $\mu_2$ are set at $10^{-2}\,a^{-1}$, $10^2\,a^{-1}$. In both components we note for $R\,>>\,\xi_b, \xi_d$ the Keplerian behavior, while it is missing only when we consider also the DM component. The shape of the rotation curve is similar to ones obtained by varying the total mass and scale radius. For a given scale radius, the peak velocity varies proportionally to a square root of the mass. For a fixed total mass, the peak-velocity position moves inversely proportionally to the scale radius, or along a Keplerian line.

\begin{figure}[htbp]
\centering
\includegraphics[scale=.3]{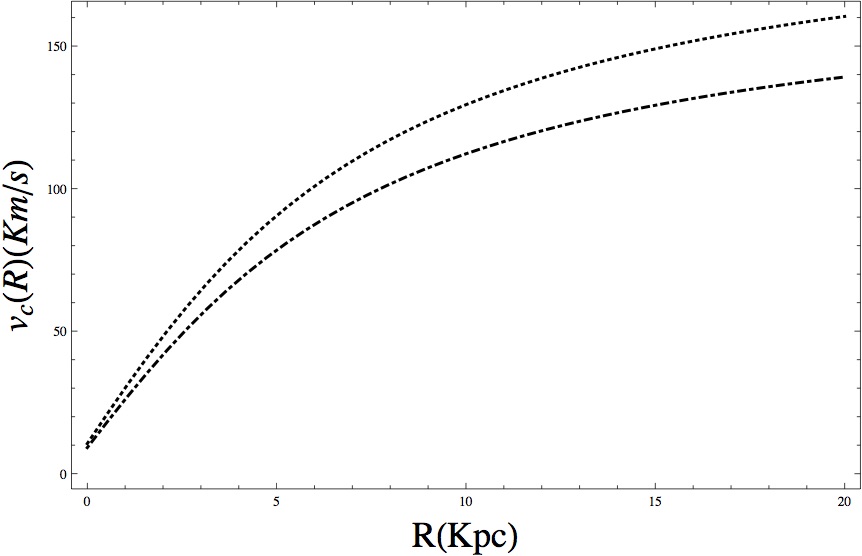}\\
\caption{The rotation curve induced by DM component (third line of (\ref{density_3})): GR + DM (dashed and dotted line) and FOG + DM (dotted line).}
\label{vel_3}
\end{figure}

As it is known in literature $f(X,Y,Z)$-gravity, and in particular $f(R)$-gravity, mimics a partial contribution of DM. In fact the corrective term $\propto\,e^{-\mu_1|\mathbf{x}|}/|\textbf{x}|$ contributes to enhance the attraction and thus the rotation curve must increase to balance the force. In the case of the other term, we have a correction $\propto\,-e^{-\mu_2|\mathbf{x}|}/|\textbf{x}|$ that contributes, being repulsive, to decrease the velocity. However in both cases these terms are asymptotically null and $f(X,Y,Z)$-gravity and GR must lead to the same result. Only with the addition of DM it is possible to raise the curve and have almost constant values. In the Fig. \ref{vel_3} we report the component of rotation curve induced by only auto-gravitating DM.

\begin{table}[hbt]
\begin{center}
\caption{Parameters of models (\ref{density_3}). The unity of mass is $10^{10}\,M_\odot$ and $a\,=\,1\,\text{Kpc}$.}
\begin{tabular}{lcccccccccc}
\hline
\hline
Galaxy & $M_b$ & $\xi_b$ & $\gamma$ & $M_d$ & $\xi_d$ & $M_{DM}$ & $\xi_{DM}$ & $\alpha$ & $\Xi$ \\
\hline
Milky Way & 0.77 & 0.5 & 1.5 & 5.20 & 3.5 & 1.68 & 5.5 & 0.50 & 20 \\
NGC 3198 & 0 & / & / & 2.60 & 3.5 & 0.84 & 5.5 & 0.53 & 20 \\
\hline
\hline
\end{tabular}
\label{tab_1}
\end{center}
\end{table}

\begin{figure}[htbp]
\centering
\includegraphics[scale=.3]{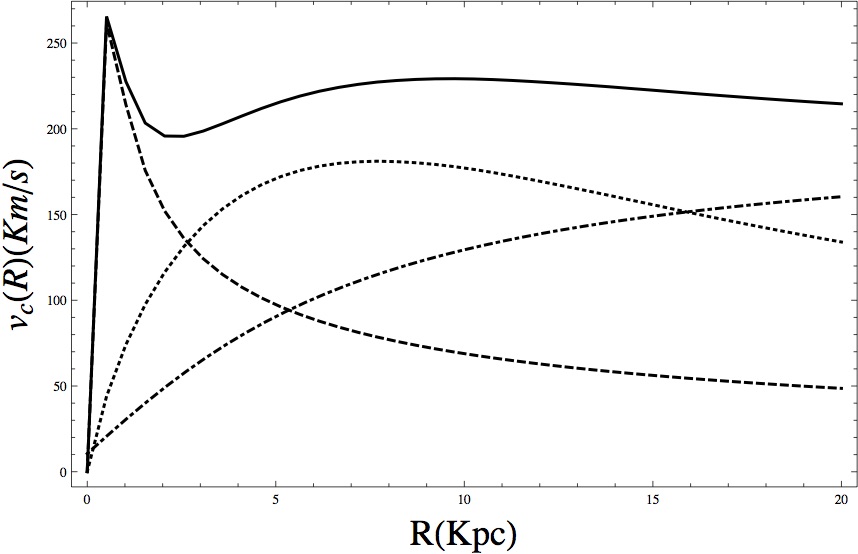}\\
\caption{Comparison between the rotation curves of galactic components: bulge (dashed line), disk (dotted line), DM (dotted and dashed line) and the global galactic rotation curve (solid line). All curves have been valuated in the framework of FOG + DM.}
\label{vel_4}
\end{figure}

\begin{figure}[htbp]
\centering
\includegraphics[scale=.3]{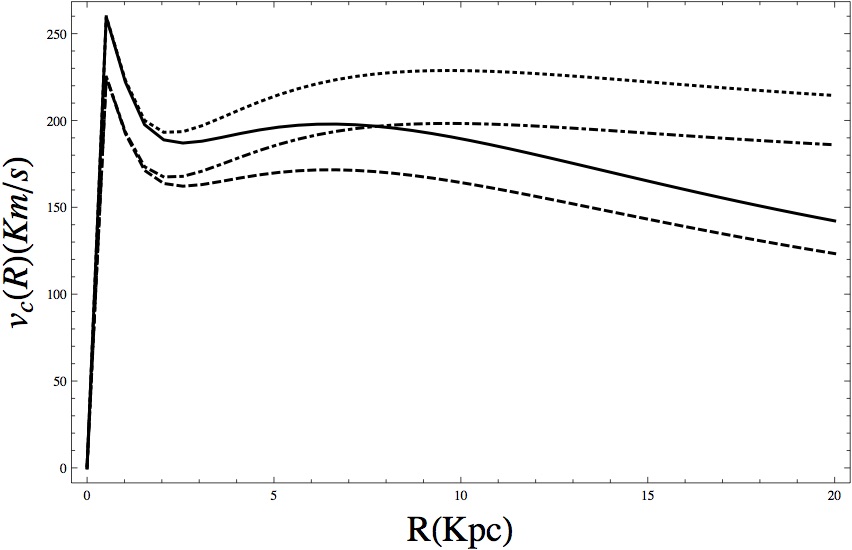}\\
\caption{The global rotation curve in the framework of GR (dashed line), GR + DM (dashed and dotted line), FOG (solid line) and FOG + DM (dotted line). The values of "masses" are $\mu_1\,=\,10^{-2}\,a^{-1}$ and $\mu_2\,=\,10^2\,a^{-1}$.}
\label{vel_5}
\end{figure}

\begin{figure}[htbp]
\centering
\includegraphics[scale=.3]{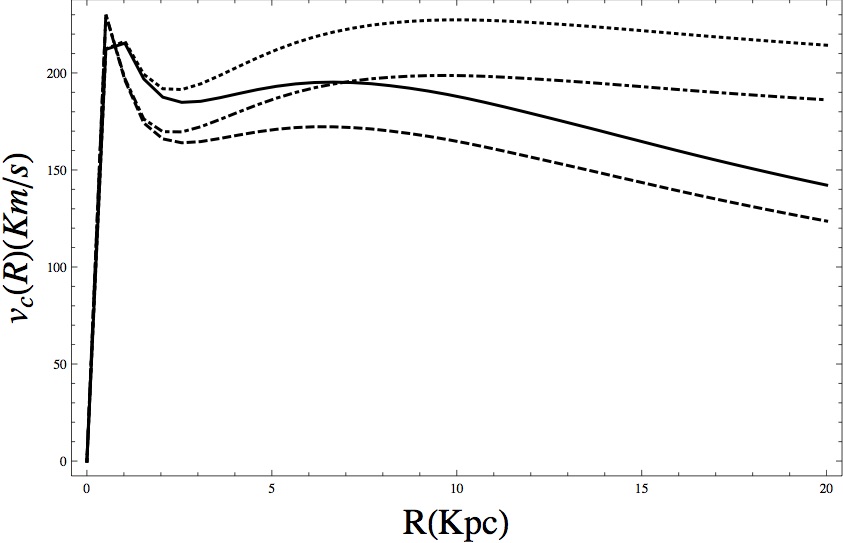}\\
\caption{The global rotation curve in the framework of GR (dashed line), GR + DM (dashed and dotted line), FOG (solid line) and FOG + DM (dotted line). The values of "masses" are $\mu_1\,=\,10^{-2}\,a^{-1}$ and $\mu_2\,=\,5\,a^{-1}$.}
\label{vel_6}
\end{figure}

In Fig. \ref{vel_4} we show the global behavior (experimentally expected) of rotation curve compared with respect to the bulge, disk and DM component for the $f(X,Y,Z)$-gravity. While in Fig. \ref{vel_5} there is the global rotation curve in the framework of GR, FOG, GR + DM and FOG + DM. At last in Fig. \ref{vel_6} we replicate the outcome of Fig. \ref{vel_5} but we inserted the value $\mu_2\,=\,5\,a^{-1}$. In this case the rotation curve induced by $f(X,Y,Z)$-gravity allows lower values as previously we claimed.


\begin{figure}[htbp]
\centering
\includegraphics[scale=.3]{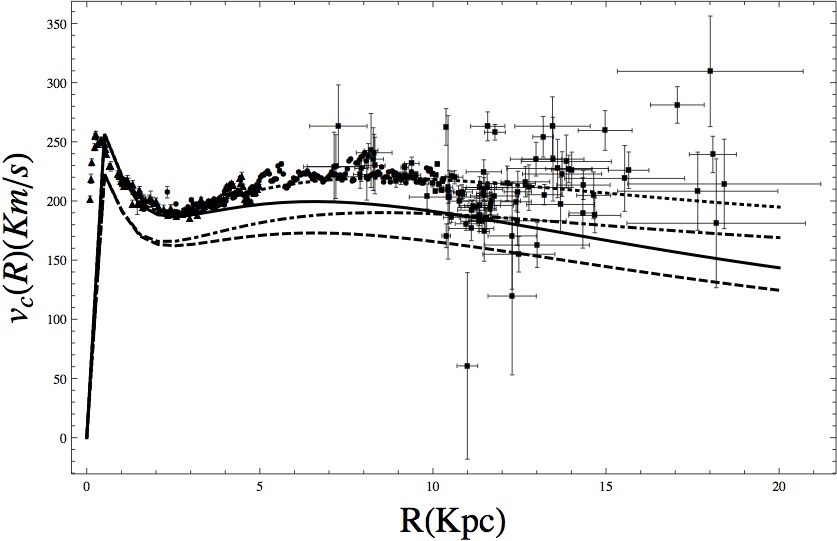}\\
\caption{Superposition of theoretical behaviors (GR (dashed line), GR + DM (dashed and dotted line), FOG (solid line), FOG + DM (dotted line)) on the experimental data for Milky Way. The mass model used is shown in (\ref{density_3}) and the values of parameters are in Table \ref{tab_1}. The values of "masses" are $\mu_1\,=\,10^{-2}\,\text{Kpc}^{-1}$ and $\mu_2\,=\,10^2\,\text{Kpc}^{-1}$.}
\label{vel_7}
\end{figure}

From the experimental point of view we used an updated rotation curve of Milky Way by integrating the existing data from the literature, and plot them in the same scale \cite{sofue}. The data used are available in a digitized from the URL http://www.ioa.s.u-tokyo.ac.jp/~sofue/mw/rc2009/. The unified rotation curve shows clearly the three dominant components: bulge, disk, and flat rotation due to the DM \cite{BG, CL, FBS, BFS, DB, sofue_2, sofue_3, HBCHI}. These data, finally, have been updated further by \cite{data_MW}. The whole set of data are plotted in Fig. \ref{vel_7} and on them the theoretical rotation curve induced by $f(X,Y,Z)$-gravity with DM has been superimposed. The values of best fit are shown in Table \ref{tab_1} with $\mu_1\,=\,10^{-2}\,\text{Kpc}^{-1}$ and $\mu_2\,=\,10^2\,\text{Kpc}^{-1}$.

\begin{figure}[htbp]
\centering
\includegraphics[scale=.3]{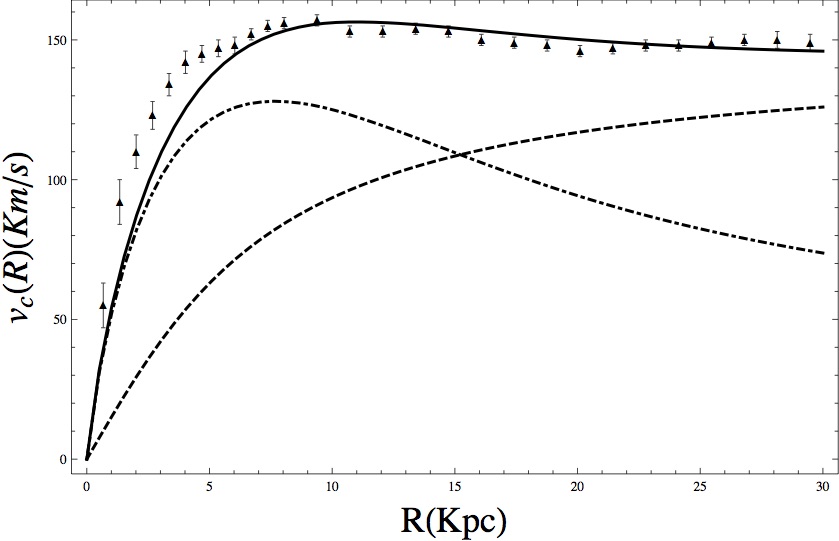}\\
\caption{Superposition of theoretical behaviors (GR (dashed line), GR + DM (dashed and dotted line), FOG (solid line), FOG + DM (dotted line)) on the experimental data for NGC 3198. The mass model used is shown in (\ref{density_3}) and the values of parameters are in Table \ref{tab_1}. The values of "masses" are $\mu_1\,=\,10^{-2}\,\text{Kpc}^{-1}$ and $\mu_2\,=\,10^2\,\text{Kpc}^{-1}$.}
\label{vel_8}
\end{figure}

The same mass model (\ref{density_3}) has been considered also for the galaxy NGC 3198. This galaxy has been chosen since the bulge is missing. Then we set $M_b\,=\,0$ in the (\ref{sol_gen_3}). In Fig. \ref{vel_8} we show the experimental data \cite{data_3198} and the superposition of theoretical behavior. Also in this case we find a nice outcome for a new set of parameters shown in Table \ref{tab_1}, while the values of $\mu_1$ and $\mu_2$ are the same of Milky Way.


The initial aim, \emph{i.e.} to extend the GR to a new class of theories, as we claimed in the introduction, is to justify the rotation curve without the DM component. From the previous outcomes, we see that even if the $f(X,Y,Z)$-gravity, or better a $f(R)$-gravity, admits a stronger attractive force, it is unable to realize our aim. \emph{Also in this framework we need Dark Matter}. Obviously we need a smaller amount of DM on the middle distances, but for large distances we have the same problems of GR.


\subsection{Rotation curves in $R^n$-gravity and $f(X,Y,Z)$-gravity}\label{CCT_theory}


The problem of DM seems to have been solved in literature, in the framework of $f(R)$-gravity, by considering the Lagrangian $\mathcal{L}\,=\,R^n$ with $n\,\epsilon\,\mathrm{Q}$ \cite{CCCT, CCT}. In these papers the gravitational potential for a point-like source can be

\begin{eqnarray}\label{pot_CCCT}
\Phi_{R^n}(r)\,=\,-\frac{GM}{r}\biggl[1+\frac{(r/{r_c})^\beta-1}{2}\biggr]
\end{eqnarray}
where $r_c$ is a characteristic length and $\beta$ is a dimensionless parameter. To recover the condition $\lim_{r\rightarrow\infty}\Phi_{R^n}(r)\,=\,0$ one must have $0\,\leq\,\beta\,<\,1$. In the case $\beta\,=\,0$ the GR is found.

\begin{figure}[htbp]
\centering
\includegraphics[scale=.4]{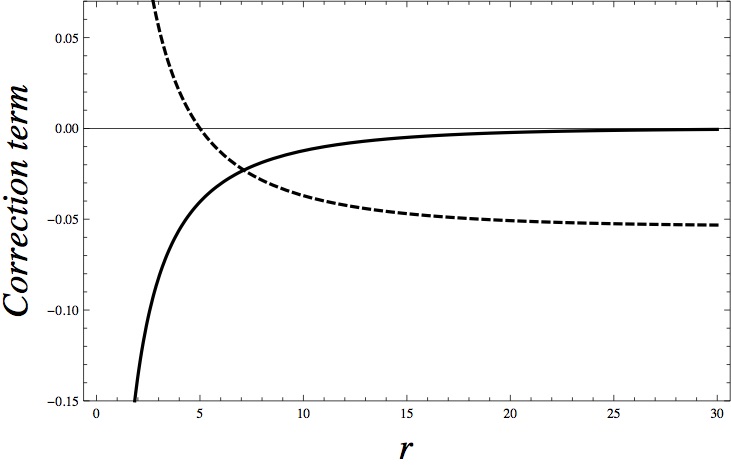}\\
\caption{Comparison between the corrective terms induced by $f(R)$-gravity ($-\frac{e^{-\mu_1\,r}}{3\,r}$, solid line) and $R^n$-gravity ($\frac{1-(r/{r_c})^\beta}{2\,r}$, dashed line). $\mu_1\,=\,0.1$, $\beta\,=\,0.8$ and $r_c\,=\,5$. The unities for $r_c$ and $\mu_1$ are arbitrary. The dashed curve shows a very slow ascent.}
\label{Y_CCCT}
\end{figure}

We comment about the physical behavior of potential (\ref{pot_CCCT}) and we want to add some reflections considering the result of the rotation curve shown above. Before to analyze the mathematical properties of metric linked to potential (\ref{pot_CCCT}), we want to show the different values of correction to the Newtonian potential. In Fig. \ref{Y_CCCT} we report the radial behavior of the corrections to $1/r$ for the potentials (\ref{sol_point}) and (\ref{pot_CCCT}) (to minimize the difference we considered only $f(R)$-gravity). From the plot we note a discrepancy between the two corrections. The correction by $\biggl(R-\frac{R^2}{6\,{\mu_1}^2}\biggr)$-gravity acts over distances much smaller, while the correction induced by $R^n$-gravity provides a potential nearly constant over large intervals and slowly goes to zero ($\sim\,r^{\beta-1}$). For this aspect the potential (\ref{pot_CCCT}) does not need the DM component. Then with a procedure of fine tuning of $r_c$ and $\beta$ it was possible to justify the experimental rotation curve for a wide class of galaxies \cite{CCT} when $n\,=\,3.5$. This choice was possible because there must be a relationship $\beta\,=\,\beta(n)$ so that the potential (\ref{pot_CCCT}) was compatible with respect to the field equations. These are the positive aspects of the potential (\ref{pot_CCCT}) used in \cite{CCCT, CCT}.

We conclude this section by reviewing the fundamental weaknesses of $R^n$-gravity.

\begin{itemize}
\item The potential (\ref{pot_CCCT}) presents an analogous behavior of potential (\ref{sol_point}). In fact for $r\,<\,r_c$ one has $\frac{1-(r/{r_c})^\beta}{2\,r}\,>\,0$ then the correction is "repulsive" like one induced by Ricci tensor square, while for $r\,>\,r_c$ one has an attractive correction. Now by remembering the reason of extension of GR, now we have unlike a repulsive contribution for $r\,<\,r_c$. If in $f(X,Y,Z)$-gravity we can delete the Ricci tensor square contribution and we have only the $f(R)$-gravity, in $R^n$-gravity we must collapse only in GR.

\item The potential (\ref{pot_CCCT}) belongs to general class of solutions for $R^n$-gravity classified by a perturbative method \cite{Stabile_Capozziello}, but the solutions are $n$-independent. Obviously the general solutions (it would be hard challenge to find them) are $n$-dependent, but at first order with respect to the perturbative parameter and in the vacuum\footnote{This parameter is generally $c^{-2}$, but the analysis is the same if we consider the dimensionless ratio $r_g/r$.} the field equations are identically vanishing. So we say that the presence of matter has been not considered and the choice of arbitrary constant has been evaluated only by matching $R^n$-gravity with GR in the limit $\beta\,\rightarrow\,0$. In fact by solving the field equations correctly in presence of matter (also with the point-like source) we would obtain solutions depending on the perturbative parameter and the technique is misplaced.
\end{itemize}

For these two aspects, but especially for the second point, $R^n$-gravity does not admit the Newtonian limit if $n\,\neq\,1$. The potential (\ref{pot_CCCT}) does not follow a correct framework when extending the GR to the new theories we want to generalize the Newtonian potential. Generally all theories without Ricci scalar in the Lagrangian suffer from the same problem. For example also $R^2$-gravity is in the same situation: is not possible to extend the solution in the matter \cite{Stabile_Capozziello}. \emph{Although we have solutions as $1/r$ with additional asymptotically flat terms, it is not automatic the assertion that these solutions are the Newtonian limit of theory}.

Let us analyze now the mathematical properties of the metric trying to justify the difference of spatial behaviors in Fig. \ref{Y_CCCT}. To simplify the calculation we choose the set of the standard coordinates. The metric (\ref{metric_tensor_PPN_0}), from the expressions (\ref{sol_point}), becomes\footnote{The set of standard coordinates is defined by the condition to obtain the standard definition of the circumference with radius $r$. From the metric tensor (\ref{metric_tensor_PPN_0}) we must impose the condition $\biggl[1-2\Psi(r)\biggr]r^2\,=\,\tilde{r}^2$ for the new radial coordinate.}

\begin{eqnarray}\label{standard_metric}
ds^2\,=&&\biggl[1-\frac{r_g}{r}\biggl(1+\frac{1}{3}\,e^{-\mu_1r}-\frac{4}{3}\,e^{-\mu_2r}\biggr)\biggr]dt^2+
\nonumber\\\\
&&\qquad\qquad-\biggl[1+\frac{r_g}{r}\biggl(1-\frac{\mu_1r+1}{3}\,e^{-\mu_1r}-\frac{2(\mu_2r+1)}{3}\,e^{-\mu_2r}\biggr)\biggr]dr^2-r^2d\Omega\nonumber
\end{eqnarray}
where $d\Omega\,=\,d\theta^2+\sin^2\theta d\phi^2$ is the solid
angle, while the element of distance linked to potential (\ref{pot_CCCT}) can be written as follows

\begin{eqnarray}\label{me_stand}
ds^2\,=\,\biggl[1+2\Phi^{SC}_{R^n}(r)\biggr]\,dt^2-\biggl[1-2\Psi^{SC}_{R^n}(r)\biggr]\,dr^2-r^2d\Omega
\end{eqnarray}
where $\Phi^{SC}_{R^n}(r)$ is the the potential (\ref{pot_CCCT}) and $\Psi^{SC}_{R^n}(r)$ is the other potential missing in the paper \cite{CCT}. The metrics (\ref{standard_metric}) and (\ref{sol_point}) represent the same space-time at first order of $r_g/r$. However in their analysis the knowledge of last potential is useless because its contribution in the geodesic motion is at fourth order. By following the paradigm of weak field limit at small velocity \cite{PRD, PRD1, PRD2} for the $R^n$-gravity we find

\begin{eqnarray}\label{me}
\Psi^{SC}_{R^n}(r)\,=\,-\frac{GM+K(\beta)+K_X}{r}+\frac{\beta-1}{4}\frac{GM}{r}\biggl(\frac{r}{r_c}\biggr)^\beta+
\frac{1}{4\,r}\int dr\,r^2 R(r)
\end{eqnarray}
where $K(\beta)$ and $K_X$ are constants depending, respectively, on the value of $\beta$ and on the integral operation, while the Ricci scalar $R$ could be an arbitrary function. In fact it needs some comment about the index $n$ in the Ricci scalar. If $n$ is a integer number, then the Ricci scalar can assume any value and can be also a generic space depending function. More attention is needed if $n$ is a rational number. The field equations (\ref{fieldequationFOG}) take into account up to third derivatives with respect to the Ricci scalar, then we must ensure that the function $f(R)$ and its derivatives are always well defined \cite{Stabile_Capozziello}. In this case for $n\,<\,3$ the solution Ricci flat ($R\,=\,0$) or space depending and asymptotically vanishing are excluded. Only solutions with constant values are allowed, but the algebraic sign is crucial. A such behavior is expected any time we have the condition $\lim_{R\rightarrow 0}\,f(R)\,=\,\text{constant}$ \cite{Stabile_Capozziello}. Now in all these considerations we do not recover the condition $\lim_{r\rightarrow\infty}\Psi^{SC}_{R^n}(r)\,=\,0$: then we have a theory which does not provide  the Minkowskian limit. It is using the first perturbative contribution of a metric component (providing the flatness at infinity), while other contributions in the remaining metric components (negligible in the Newtonian limit) do not cover the same asymptotic limit. Then only for $n\,>\,3$ we can have the flatness at infinity.

Then we could say that for $n\,>\,3$ the Minkowskian limit is recovered but a such perturbative approach can be performed only in the vacuum. The objection previously shown comes back. Up to third order ($c^{-6}$ or ${r_g}^3$) the geometrical side of field equation is identically null, but the matter side could not be null at first order ($c^{-2}$ or $r_g$).

The class of $R^n$-gravity are examples of theories where the weak field limit procedure does not generate automatically the Minkowskian limit. In fact only if we consider theories satisfying the condition $\lim_{X\rightarrow 0}\,f(R)\,=\,0$ \cite{Stabile_Capozziello}, their weak field limit is compatible with the request of asymptotically flatness. Moreover $f(R)$-gravity mimicking an additional source due to its scalar curvature \cite{PRD,PRD1,minko} we would have a constant matter that pervades all space giving us a justification of more intense gravitational potential. In addition if $\lim_{R\rightarrow 0}\,f(R)\,=\,\text{costant}$ we do not have the Minkowskian limit, but we can interpret the apparent mass, only from the experimental point of view, as DM. These aspects, then, can be a mathematical motivation for different shape of point-like gravitational potential, but also source of further attention.


\section{The Gravitational Lensing in $f(X,Y,Z)$-gravity}

Gravitational lensing can be  considered as one of the main probes for DM at galactic and extragalactic level. However, modified gravity could affect the lensing quantities vs DM. Let us develop this issue in the framework of Fourth Order Gravity. 

\subsection{The Point-Like Source}\label{GL_pointlike}


Let us start by defining the  Lagrangian of a photon moving in the gravitational field with metric (\ref{standard_metric}). It is

\begin{eqnarray}\label{phot_lagr}
\mathcal{L}\,=\,\frac{1}{2}\biggl[\biggl(1-\frac{r_g}{r}\,\Xi(r)\biggr)
\dot{t}^2-\biggl(1+\frac{r_g}{r}\,\Lambda(r)\biggr)\dot{r}^2
-r^2\dot{\theta}^2-r^2\sin^2\theta\dot{\phi}^2\biggr]
\end{eqnarray}
where
$\Xi(r)\,\doteq\,1+\hat{\Xi}(r)\,\doteq\,1+\frac{1}{3}\,e^{-\mu_1r}-\frac{4}{3}\,e^{-\mu_2r}$,
$\Lambda(r)\,\doteq\,1+\hat{\Lambda}(r)\,\doteq\,1-\frac{\mu_1r+1}{3}\,e^{-\mu_1r}-\frac{2(\mu_2r+1)}{3}\,e^{-\mu_2r}$
and the dot represents the derivatives with respect to the affine
parameter $\lambda$. Since the variable $\theta$ does not have dynamics
($\ddot{\theta}\,=\,0$) we can choose for simplicity
$\theta\,=\,\pi/2$. By applying the Euler-Lagrangian equation to
Lagrangian (\ref{phot_lagr}) for the cyclic variables $t$, $\phi$
we find two motion constants

\begin{eqnarray}\label{motion_constant}
\begin{array}{ll}
\frac{\partial\mathcal{L}}{\partial\dot{t}}\,=\,\biggl(1-\frac{r_g}{r}\,\Xi(r)\biggr)\dot{t}\,\doteq\,\mathcal{T}
\\\\
\frac{\partial\mathcal{L}}{\partial\dot{\phi}}\,=\,-r^2\dot{\phi}\,\doteq\,-J
\end{array}
\end{eqnarray}
and respect to $\lambda$ we find the "energy" of Lagrangian\footnote{The (\ref{phot_lagr}) is a quadratic form, so
it corresponds to its Hamiltonian.}

\begin{eqnarray}\label{energy}
\mathcal{L}\,=\,0
\end{eqnarray}
By inserting the equations (\ref{motion_constant}) into (\ref{energy}) we find a differential equation for $\dot{r}$

\begin{eqnarray}\label{diffequ}
\dot{r}_{\pm}\,=\,\pm\,\mathcal{T}\,\sqrt{\frac{1}{1+\frac{r_g}{r}\,\Lambda(r)}\biggl[\frac{1}{1-\frac{r_g}{r}\,\Xi(r)}-
\frac{J^2}{r^2}\bigg]}
\end{eqnarray}
$\dot{r}_+$ is the solution for leaving photon, while $\dot{r}_-$
is one for incoming photon. Let $r_0$ be a minimal distance from
the lens center (Fig. \ref{deflection}). We must impose the condition
$\dot{r}_{\pm}(r_0)\,=\,0$ from the which we find

\begin{eqnarray}\label{Jcond}
J^2\,=\,\frac{{r_0}^2\mathcal{T}^2}{1-\frac{r_g}{r_0}\,\Xi(r_0)}
\end{eqnarray}
Now the deflection angle $\alpha$ (Fig. \ref{deflection}) is defined by following relation

\begin{eqnarray}\label{angle_deflec}
\alpha\,=\,&&-\pi+\phi_{fin}\,=\,-\pi+\int_0^{\phi_{fin}}d\phi\,=\,-\pi+\int_{\lambda_{in}}^{\lambda_{fin}}
\dot{\phi}\,d\lambda\,=\,-\pi+
\int_{\lambda_{in}}^{\lambda_0}\dot{\phi}\,d\lambda+\int_{\lambda_0}^{\lambda_{fin}}\dot{\phi}\,d\lambda\,=\nonumber\\
\nonumber\\
&&-\pi+\int_{\infty}^{r_0}\frac{\dot{\phi}}{\dot{r}_-}\,dr+\int_{r_0}^{\infty}\frac{\dot{\phi}}{\dot{r}_+}\,dr\,=\,
-\pi+2\int_{r_0}^{\infty}\frac{\dot{\phi}}{\dot{r}_+}\,dr
\end{eqnarray}
where $\lambda_0$ is the value of $\lambda$ corresponding to the minimal value ($r_0$) of radial coordinate $r$. By putting the expressions of $J$, $\dot{\phi}$ and $\dot{r}_+$ into (\ref{angle_deflec}) we get the deflection angle

\begin{eqnarray}\label{angle_deflec_1}
\alpha\,=\,-\pi+2\int_{r_0}^\infty\frac{dr}{r\sqrt{\frac{1}{1+\frac{r_g}{r}\,\Lambda(r)}\biggl[\frac{1-\frac{r_g}
{r}\,\Xi(r_0)}{1-\frac{r_g}{r}\,\Xi(r)}\frac{r^2}{{r_0}^2}-1\bigg]}}
\end{eqnarray}
which in the case $r_g/r\,\ll\,1$\footnote{We do not consider the GL generated by a black hole.} becomes

\begin{eqnarray}\label{angle_deflec_2}
\alpha\,=\,2\,r_g\biggl[\frac{1}{r_0}+\mathcal{F}_{\mu_1,\,\mu_2}(r_0)\biggr]
\end{eqnarray}
where

\begin{eqnarray}\label{fun1}
\mathcal{F}_{\mu_1,\,\mu_2}(r_0)\,\doteq\,\frac{1}{2}\int_{r_0}^\infty
\frac{r_0r^2[\hat{\Lambda}(r)-\hat{\Xi}(r)]+r^3\hat{\Xi}(r_0)-{r_0}^3\hat{\Lambda}(r)}{r^3(r^2-{r_0}^2)
\sqrt{1-\frac{{r_0}^2}{r^2}}}\,dr
\end{eqnarray}
From the definition of $\hat{\Xi}$ and $\hat{\Lambda}$ we note that in the case $f(X,Y,Z)\,\rightarrow\,R$ we obtain $\mathcal{F}_{\mu_1,\,\mu_2}(r_0)\,\rightarrow\,0$. In a such way we extended and contemporarily recovered the outcome of GR.

\begin{figure}[htbp]
\centering
\includegraphics[scale=.5]{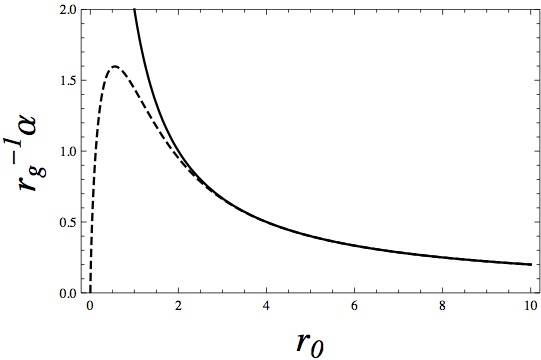}\\
\caption{Comparison between the deflection angle of GR (solid line) and one of $f(X,Y,Z)$-gravity (dashed line) (\ref{angle_deflec_2}) for a fixed value $\mu_2\,=\,2$ and any $\mu_1$.}
\label{plot_1}
\end{figure}

The analytical dependence of function $\mathcal{F}_{\mu_1,\,\mu_2}(r_0)$ from the parameters $\mu_1$ and $\mu_2$ is given by evaluating the integral (\ref{fun1}). A such as integral is not easily evaluable from the analytical point of view. However this aspect is not fundamental, since we can numerically appreciate the deviation from the outcome of GR. In fact in Fig. \ref{plot_1} we show the plot of deflection angle (\ref{angle_deflec_2}) by $f(X,Y,Z)$-gravity for a given set of values for $\mu_1$ and $\mu_2$. The spatial behavior of $\alpha$ is ever the same if we do not modify $\mu_2$. This outcome is really a surprise: by the numerical evaluation of the function $\mathcal{F}_{\mu_1,\,\mu_2}(r_0)$ one notes that the dependence of $\mu_1$ is only formal. If we solve analytically the integral we must find a $\mu_1$ independent function. However, this statement should not be justified only by numerical evaluation but it needs an analytical proof. For these reasons in the next section we reformulate the theory of Gravitational  Lensing (GL) generated by a generic matter distribution and demonstrate that for $f(R)$-gravity one has the same outcome of GR.

\subsection{Extended  Sources}\label{GL_extended}


In this section we want to recast the framework of GL for a generic matter source distribution $\rho(\mathbf{x})$ so the photon can undergo many deviations. In this case we leave the hypothesis that the flight of photon belongs always to the same plane, but we consider only the deflection angle as the angle between the directions of incoming and leaving photon. Finally we find the generalization of GL in $f(X,Y,Z)$-gravity including the previous outcome of deflecting point-like source (and resolving the integral (\ref{fun1})).

The starting point is the relativistic distance (\ref{metric_new}) where $\Phi$ is given by the superposition principle (\ref{superposition_phi}) and an analogous relation is valid also for $\Psi$. By introducing the four velocity $u^\mu\,=\,\dot{x}^\mu\,=\,(u^0, \mathbf{u})$ the flight of photon is regulated by the condition

\begin{eqnarray}\label{travel_photon}
ds^2\,=\,g_{\alpha\beta}u^\alpha u^\beta\,=\,(1+2\Phi){u^0}^2-(1-2\Psi)|\mathbf{u}|^2\,=\,0
\end{eqnarray}
then $u^\mu$ is given by

\begin{eqnarray}\label{four_velocity}
u^\mu\,=\,\biggl(\sqrt{\frac{1-2\Psi}{1+2\Phi}}\,|\mathbf{u}|,\mathbf{u}\biggr)
\end{eqnarray}
In the Newtonian limit we find that the geodesic motion equation becomes

\begin{eqnarray}\label{geodesic_motion}
\dot{u}^\mu+\Gamma^\mu_{\alpha\beta}u^\alpha u^\beta\,=\,0\,\rightarrow\,\dot{\mathbf{u}}+|\mathbf{u}|^2\nabla(\Phi+\Psi)-2\mathbf{u}\nabla\Psi\cdot\mathbf{u}\,=\,0
\end{eqnarray}
and by supposing $|\mathbf{u}|^2\,=\,1$ we can recast the equation in a more known aspect

\begin{eqnarray}\label{geodesic_motion_1}
\dot{\mathbf{u}}\,=\,-2\biggl[\nabla_\bot\Psi+\frac{1}{2}\nabla(\Phi-\Psi)\biggr]
\end{eqnarray}
where $\nabla_\bot\,=\,\nabla-\biggl(\frac{\mathbf{u}}{|\mathbf{u}|}\cdot\nabla\biggr)\frac{\mathbf{u}}{|\mathbf{u}|}$ is the two dimensions nabla operator orthogonal to direction of vector $\mathbf{u}$. In GR we would had only $\dot{\mathbf{u}}\,=\,-2\,\nabla_\bot\Phi$ since we have $\Psi\,=\,\Phi$. In fact the field equations (\ref{NL-field-equation_2}) admit the pointlike solutions (\ref{sol_point}) and the condition (\ref{cond_0}) is satisfied \cite{PRD2}. We can affirm, then, that only in GR the metric potentials are equals (or more generally their difference must be proportional to function $|\mathbf{x}|^{-1}$). The constraint (\ref{cond_0}) has been found also many times in the context of cosmological
perturbation theory \cite{sugg1,sugg2,sugg3,sugg4,sugg5}.

The deflection angle (\ref{angle_deflec}) is now defined by equation

\begin{eqnarray}\label{angle_vect}
\vec{\alpha}\,=\,-\int_{\lambda_i}^{\lambda_f}\frac{d\mathbf{u}}{d\lambda}d\lambda
\end{eqnarray}
where $\lambda_i$ and $\lambda_f$ are the initial and final value of affine parameter \cite{schneider}. For a generic matter distribution we can not \emph{a priori} claim that the deflection angle belongs to lens plane (as pointlike source), but we can only link the deflection angle to the difference between the initial and final velocity $\mathbf{u}$. So we only analyze the directions of photon before and after the interaction with the gravitational mass. Then the (\ref{angle_vect}) is placed by assuming $\vec{\alpha}\,=\,\Delta\mathbf{u}\,=\,\mathbf{u}_i-\mathbf{u}_f$. From the geodesic equation (\ref{geodesic_motion_1}) the deflection angle becomes

\begin{eqnarray}\label{angle_vect_1}
\vec{\alpha}\,=\,2\,\int_{\lambda_i}^{\lambda_f}\biggl[\nabla_\bot\Psi+\frac{1}{2}\nabla(\Phi-\Psi)\biggr]d\lambda
\end{eqnarray}
The formula (\ref{angle_vect_1}) represents the generalization of deflection angle in the framework of GR. By considering the photon incoming along the z-axes we can set $\mathbf{u}_i\,=\,(0,0,1)$. Moreover we decompose the general vector $\mathbf{x}\,\epsilon\,\mathbb{R}^3$ in two components: $\vec{\xi}\,\epsilon\,\mathbb{R}^2$ and $z\,\epsilon\,\mathbb{R}$. The differential operator now can be decomposed as follows $\nabla\,=\,\nabla_\bot+\hat{z}\,\partial_z\,=\,\nabla_{\vec{\xi}}+\hat{z}\,\partial_z$, while the modulus of distance is $|\mathbf{x}-\mathbf{x}'|\,=\,\sqrt{|\vec{\xi}-\vec{\xi}'|^2+(z-z')^2}\,\doteq\,\Delta(\vec{\xi},\vec{\xi}',z,z')$. Since the potentials $\Phi\,,\Psi\,\ll\,1$, around the lens, the solution of (\ref{geodesic_motion_1}) with the initial condition $\mathbf{u}_i\,=\,(0,0,1)$ can be expressed as follows

\begin{eqnarray}\label{param}
\mathbf{u}\,=\,(\mathcal{O}(\Phi,\Psi),\mathcal{O}(\Phi,\Psi),1+\mathcal{O}(\Phi,\Psi))
\end{eqnarray}
and we can substitute the integration with respect to the affine parameter $\lambda$ with $z$. In fact we note

\begin{eqnarray}\label{param1}
d\lambda\,=\,\frac{dz}{dz/d\lambda}\,=\,\frac{dz}{1+\mathcal{O}(\Phi,\Psi)}\,\sim\,dz
\end{eqnarray}
and the deflection angle (\ref{angle_vect_1}) becomes

\begin{eqnarray}\label{angle_vect_2}
\vec{\alpha}\,=\,\int_{z_i}^{z_f}\biggl[\nabla_{\vec{\xi}}(\Phi+\Psi)+\hat{z}\,\partial_z(\Phi-\Psi)\biggr]dz
\end{eqnarray}
From the expression of potentials for a generic matter distribution (see potential (\ref{superposition_phi}) where $f_X(0)\,=\,1$ for example) we find the relations

\begin{eqnarray}\label{sys}
\begin{array}{ll}
\Phi+\Psi\,=\,-2\,G\int
d^2\vec{\xi}'dz'\frac{\rho(\vec{\xi}',z')}{\Delta(\vec{\xi},\vec{\xi}',z,z')}+2\,G\int
d^2\vec{\xi}'dz'\frac{\rho(\vec{\xi}',z')}{\Delta(\vec{\xi},\vec{\xi}',z,z')}\,e^{-\mu_2\Delta(\vec{\xi},\vec{\xi}',z,z')}
\\\\
\Phi-\Psi\,=\,-\,\frac{2G}{3}\int
d^2\vec{\xi}'dz'\frac{\rho(\vec{\xi}',z')}{\Delta(\vec{\xi},\vec{\xi}',z,z')}
\biggl[e^{-\mu_1\Delta(\vec{\xi},\vec{\xi}',z,z')}-e^{-\mu_2\Delta(\vec{\xi},\vec{\xi}',z,z')}\biggr]
\end{array}
\end{eqnarray}
and the deflection angle (\ref{angle_vect_1}) becomes

\begin{eqnarray}\label{angle_vect_3}
\vec{\alpha}=&&2G\int_{z_i}^{z_f}d^2\vec{\xi}'dz'\,dz\frac{\rho(\vec{\xi}',z')(\vec{\xi}-\vec{\xi}')}{{\Delta(\vec{\xi},\vec{\xi}',z,z')}^3}+
\nonumber\\\nonumber\\\nonumber
&&-2G\int_{z_i}^{z_f}d^2\vec{\xi}'dz'\,dz\frac{\rho(\vec{\xi}',z')[1+\mu_2\Delta(\vec{\xi},\vec{\xi}',z,z')]}
{{\Delta(\vec{\xi},\vec{\xi}',z,z')}^3}e^{-\mu_2\Delta(\vec{\xi},\vec{\xi}',z,z')}(\vec{\xi}-\vec{\xi}')+
\nonumber\\\\\nonumber
&&+\frac{2G}{3}\hat{z}\int_{z_i}^{z_f}d^2\vec{\xi}'dz'\,dz\frac{\rho(\vec{\xi}',z')(z-z')}
{{\Delta(\vec{\xi},\vec{\xi}',z,z')}^3}\biggl[\biggl(1+\mu_1\Delta(\vec{\xi},\vec{\xi}',z,z')\biggr)e^{-\mu_1\Delta(\vec{\xi},\vec{\xi}
',z,z')}+
\nonumber\\\nonumber\\\nonumber
&&\qquad\qquad\qquad\qquad\qquad\qquad\qquad\qquad\qquad-\biggl(1+\mu_2\Delta(\vec{\xi},\vec{\xi}',z,z')\biggr)e^{-\mu_2\Delta(\vec{\xi},\vec{\xi}',z,z')}\biggr]
\end{eqnarray}
In the case of hypothesis of thin lens belonging to plane $(x,y)$ we can consider a weak dependence of modulus $\Delta(\vec{\xi},\vec{\xi}',z,z')$ into variable $z'$ so there is only a trivial error if we set $z'\,=\,0$. With this hypothesis the integral into $z'$ is incorporated by definition of two dimensional mass density $\Sigma(\vec{\xi}')\,=\,\int dz'\rho(\vec{\xi}',z')$. Since we are interested only to the GL performed by one lens we can extend the integration range of $z$ between $(-\infty,\infty)$. Now the deflection angle is the following

\begin{eqnarray}\label{angle_vect_4}
\vec{\alpha}\,=\,4G\int d^2\vec{\xi}'\Sigma(\vec{\xi}')\biggl[\frac{1}{|\vec{\xi}-\vec{\xi}'|}-|\vec{\xi}-\vec{\xi}'|\,
\mathcal{F}_{\mu_2}(\vec{\xi},\vec{\xi}')\biggr]
\frac{\vec{\xi}-\vec{\xi}'}{|\vec{\xi}-\vec{\xi}'|}
\end{eqnarray}
where

\begin{eqnarray}\label{fun2}
\mathcal{F}_{\mu_2}(\vec{\xi},\vec{\xi}')\,=\,\int_0^{\infty}dz\frac{(1+\mu_2\Delta(\vec{\xi},\vec{\xi}',z,0))}{{\Delta(\vec{\xi},\vec{\xi}',z,0)}^3}e^{-\mu_2\Delta(\vec{\xi},\vec{\xi}',z,0)}
\end{eqnarray}
The last integral in (\ref{angle_vect_3}) is vanishing because the integrating function is odd with respect to variable $z$. The expression (\ref{angle_vect_4}) is the generalization of outcome (\ref{angle_deflec_2}) and mainly we found a correction term depending only on the $\mu_2$ parameter.

In the case of point-like source $\Sigma(\vec{\xi}')\,=\,M\,\delta^{(2)}(\vec{\xi}')$ we find

\begin{eqnarray}\label{angle_vect_5}
\vec{\alpha}\,=\,2\,r_g\biggl[\frac{1}{|\vec{\xi}|}-|\vec{\xi}|\,\mathcal{F}_{\mu_2}(\vec{\xi},0)\biggr]
\frac{\vec{\xi}}{|\vec{\xi}|}
\end{eqnarray}
and in the case of $f(X,Y,Z)\,\rightarrow\,f(R)$ (\emph{i.e.} $\mu_2\,\rightarrow\,\infty$ and $\mathcal{F}_{\mu_2}(\vec{\xi},\vec{\xi}')\,\rightarrow\,0$) we recover the outcome of GR $\vec{\alpha}\,=\,2\,r_g\,\vec{\xi}/|\vec{\xi}|^2$. From the theory of GL in GR we know that the deflection angle $2\,r_g/r_0$ is formally equal to $2\,r_g/|\vec{\xi}|$ if we suppose $r_0\,=\,|\vec{\xi}|$. Besides both $r_0$, $|\vec{\xi}|$ are not practically measurable, while it is possible to measure the so-called impact parameter $b$ (see Fig. \ref{deflection}). But only in the first approximation these three quantities are equal.

In fact when the photon is far from the gravitational source we can parameterize the trajectory as follows

\begin{eqnarray}
\begin{aligned}
\begin{cases}
t\,=\,\lambda\\x\,=\,-\,t\\y\,=\,b
\end{cases}
&\rightarrow
\begin{cases}
r\,=\,\sqrt{t^2+b^2}\\\phi\,=\,-\arctan\frac{b}{t}
\end{cases}
\end{aligned}
\end{eqnarray}
and from the definition of angular momentum (\ref{motion_constant}) in the case of $t\,\gg\,b$ we have

\begin{eqnarray}
J\,=\,\dot{\phi}\,r^2\,=\,\frac{b/t^2}{1+b^2/t^2}(t^2+b^2)\,\sim\,b
\end{eqnarray}
By using the condition (\ref{Jcond}) $\dot{r}_{\pm}(r_0)\,=\,0$ we find the relation among $b$ and $r_0$\footnote{The constant $\mathcal{T}$ is dimensionless if we consider that $\lambda$ is the length of trajectory of photon. In this case without losing the generality we can choose $\mathcal{T}\,=\,1$.}

\begin{eqnarray}\label{impact_par}
b\,=\,\frac{r_0\,\mathcal{T}}{\sqrt{1-\frac{r_g}{r_0}\,\Xi(r_0)}}\,\sim\,r_0
\end{eqnarray}
justifying then the position $r_0\,=\,|\vec{\xi}|$ in the limit $r_g/r\,\ll\,1$ (but also $r_g/r_0\,\ll\,1$).

\begin{figure}[htbp]
\centering
\includegraphics[scale=.5]{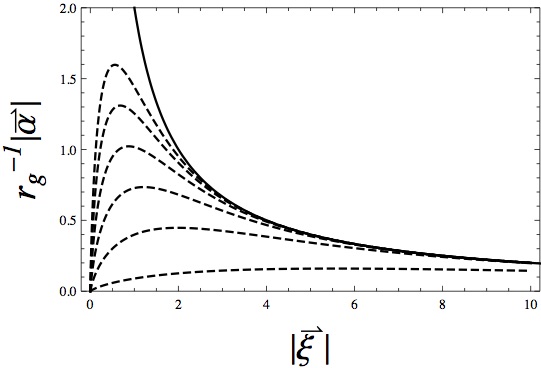}\\
\caption{Comparison between the deflection angle of GR (solid line) and of $f(X,Y,Z)$-gravity (dashed line) (\ref{angle_vect_5}) for $0.2\,<\,\mu_2\,<\,2$.}
\label{plot_2}
\end{figure}

In Fig. \ref{plot_2} we report the plot of deflection angle (\ref{angle_vect_5}). The behaviors shown in figure are parameterized only by $\mu_2$ and we note an equal behavior shown in Fig. \ref{plot_1}.
With the expression (\ref{angle_vect_5}) we have the analytical proof of statement at the end of previous section. In fact in the equation (\ref{angle_vect_5}) we have not any information about the correction induced in the action (\ref{HOGaction}) by a generic function of Ricci scalar ($f_{XX}\,\neq\,0$). This result is very important if we consider only the class of theories $f(R)$-gravity. In this case, since $\mu_2\,\rightarrow\,\infty$, we found the same outcome of GR. From the behavior in Fig. \ref{plot_2} we note that the correction to outcome of GR is deeply different for $r_0\,\rightarrow\,0$, while for $r_0\,\rightarrow\,\infty$ the behavior (\ref{angle_vect_5}) approaches the outcome of GR, but the deviations are smaller. This difference is given by the repulsive correction to the gravitational potential (see metric (\ref{metric_new})) induced by $f(Y,Z)$. Only by leaving the thin lens hypothesis (the lens does not belong to plane $z\,=\,0$) we can have the deflection angle depending by $\mu_1$ (\ref{angle_vect_3}). In fact in this case the third integral in (\ref{angle_vect_3}) is not zero. Then in the case of thin lens we have a complete degeneracy of results: \emph{ $f(R)$-gravity equivalent to the GR in the thin lens case} (see also \cite{jetzer}). In order to find some differences,  we must  include the contributions generated by the squared Ricci tensor. However also in this case we do not get  the right behavior: the deflection angle is smaller than the one in GR. Furthermore $f(X,Y,Z)$-gravity does not  mimic the DM  component by assuming the thin lens hypothesis.

A final remark is order at this point.  The fact that $f(R)$ gravity and GR give the same lensing effects is due to the fact that the corrections induced by $f(R)$ on the $g_{tt}$ and $g_{ij}$ components of the metric  (that is the potentials $\Phi$ and $\Psi$) are opposite and cancel out as demonstrated in details in Ref. \cite{jetzer,stabile_stabile}. This means that the only GR contribution remains in the deflection angle.  


\subsection{The lens equation}\label{GL_lens_equation}

\begin{figure}[htbp]
\centering
\includegraphics[scale=1]{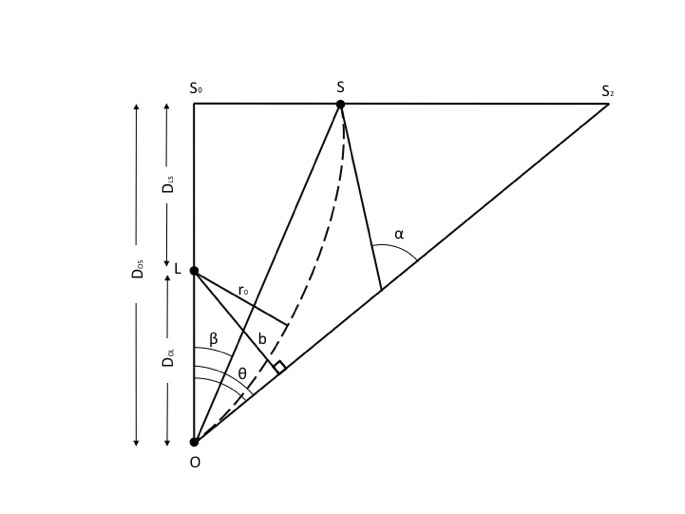}\\
\caption{The gravitational lensing geometric for a point-like source lens $L$ at distance $D_{OL}$ from observer $O$. A source $S$ at distance $D_{OS}$ from $O$ has angular position $\beta$ from the lens. A light ray (dashed line) from $S$ which passes the lens at minimal distance $r_0$ is deflected by $\alpha$; the observer sees an image of the source at angular position $\theta\,=\,b/D_{OL}$ where $b$ is the impact factor. $D_{LS}$ is the distance lens - source.}
\label{deflection}
\end{figure}

In order to demonstrate the effect of a deflecting mass,  let us show the simplest GL configuration in Fig. \ref{deflection}. A point-like mass is located at a distance $D_{OL}$ from the observer $O$. The source is at distance $D_{OS}$ from the observer, and its true angular separation from the lens $L$ is $\beta$, the separation which would be observed in the absence of lensing ($r_g\,=\,0$). The photon which passes the lens at distance $r_0\,\sim\,b$ is deflected with an angle $\alpha$.

Since the deflection angle (\ref{angle_deflec_2}) is equal to (\ref{angle_vect_5}), for sake of simplicity we will use the "vectorial" expression. Then the expression (\ref{angle_vect_5}), by considering the relation (\ref{impact_par}), becomes

\begin{eqnarray}\label{angle_vect_6}
\alpha\,=\,2\,r_g\biggl[\frac{1}{b}-b\,\mathcal{F}_{\mu_2}(b,0)\biggr]
\end{eqnarray}
The condition that this photon reach the observer is obtained from the geometry of Fig. \ref{deflection}. In fact we find
\begin{eqnarray}\label{lens_eq_1}
\beta\,=\,\theta-\frac{D_{LS}}{D_{OS}}\,\alpha
\end{eqnarray}
Here $D_{LS}$ is the distance of the source from the lens. In the simple case with a Euclidean background metric here, $D_{LS}\,=\,D_{OS}-D_{OL}$; however, since the GL occurs in the Universe on large scale, one must use a cosmological model \cite{schneider}. Denoting the angular separation between the deflecting mass and the deflected photon as $\theta\,=\,b/D_{OL}$ the lens equation for $f(X,Y,Z,)$-gravity is the following

\begin{eqnarray}\label{lens_eq_2}
[1+{\theta_E}^2\,\mathcal{F}(\theta)]\,\theta^2-\beta\,\theta-{\theta_E}^2\,=\,0
\end{eqnarray}
where $\theta_E\,=\,\sqrt{\frac{2\,r_g\,D_{LS}}{D_{OL}D_{OS}}}$ is the Einstein angle and

\begin{eqnarray}\label{fun3}
\mathcal{F}(\theta)\,=\,\int_0^{\infty}dz\frac{(1+\mu_2 D_{OL}\sqrt{\theta^2+z^2})}{\sqrt{(\theta^2+z^2)^3}}\,e^{-\mu_2 D_{OL}\sqrt{\theta^2+z^2})}
\end{eqnarray}

\begin{figure}[htbp]
\centering
\includegraphics[scale=.5]{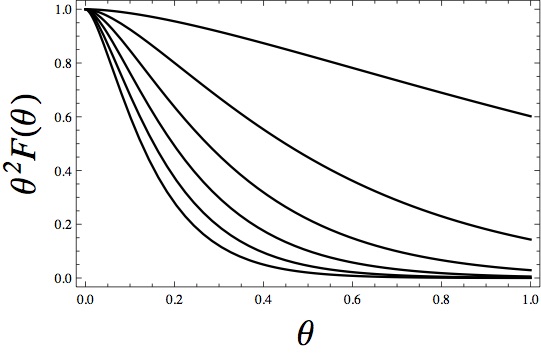}\\
\caption{Plot of function $\theta^2\mathcal{F}(\theta)$ (\ref{fun3}) for $1\,<\,\mu_2D_{OL}\,<\,10$.}
\label{plot_3}
\end{figure}

\begin{figure}[htbp]
\centering
\includegraphics[scale=.5]{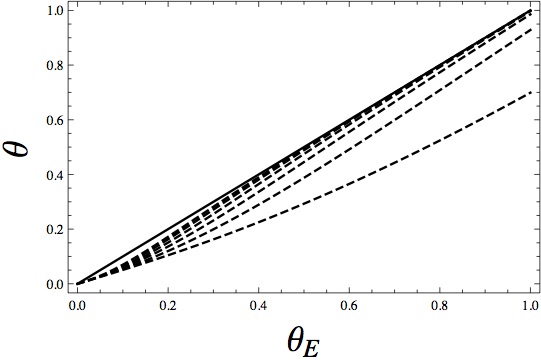}\\
\caption{Plot of the Einstein ring (solid line) and its modification (\ref{lens_sol_ring}) in the $f(X,Y,Z)$-gravity for $1\,<\,\mu_2D_{OL}\,<\,10$ (dashed line).}
\label{plot_4}
\end{figure}

Since we have $0\,<\,\theta^2\mathcal{F}(\theta)\,<\,1$ (Fig. \ref{plot_3}) we can find a perturbative solution of (\ref{lens_eq_2}) by starting from one in GR, $\theta^{GR}_\pm\,=\,\frac{-\beta\pm\sqrt{\beta^2+4{\theta_E}^2}}{2}$. In fact by assuming $\theta\,=\,\theta^{GR}_\pm+\theta^*$ and neglecting ${\theta^*}^2\mathcal{F}(\theta^*)$ in (\ref{lens_eq_2}) we find

\begin{eqnarray}\label{lens_sol}
\theta\,=\,\theta^{GR}_\pm\mp\frac{{\theta_E}^2}{\sqrt{\beta^2+4\,{\theta_E}^2}}\,\mathcal{F}(\theta^{GR}_\pm)\,
{\theta^{GR}_\pm}^2
\end{eqnarray}
and in the case of $\beta\,=\,0$ we find the modification to the Einstein ring

\begin{eqnarray}\label{lens_sol_ring}
\theta\,=\,\pm\theta_E\biggl[1-\frac{{\theta_E}^2}{2}\,\mathcal{F}(\theta_E)\biggr]
\end{eqnarray}
In Fig. \ref{plot_4} we show the angular position of images with respect to the Einstein ring. Both the deflection angle and the position of images assume a smaller value than ones of GR. Then the corrections to the GR quantities are found only for the introduction in the action (\ref{HOGaction}) of curvature invariants $Y$ (or $Z$), while there are no modifications induced by adding a generic function of Ricci scalar $X$. The algebraic signs of terms concerning the parameter $\mu_2$ are ever different with respect to the terms of GR in (\ref{sol_point}) and they can be interpreted as a "repulsive force" giving us a minor curvature of photon. The correction terms concerning the parameter $\mu_1$ have opposite algebraic sign in the metric component $g_{tt}$ and $g_{ij}$ (\ref{sol_point}) and we lose their information in the deflection angle (\ref{angle_vect_2}).


In both approaches we find the same outcomes $\mu_1$-independent because the matter source (in our case it is a point-like mass) is symmetric with respect to $z$-axes and we neglect the second integral in (\ref{angle_vect_3}). Obviously for a generic matter distribution the deflection angle is defined by (\ref{angle_vect_3}) and the choice of second derivative of function of Ricci scalar is not arbitrary anymore.


\section{Conformal transformations in the weak field limit}

As a final issue, let us discuss the conformal transformations  in the weak field limit approximation. It is important to stress that  we developed all the above considerations in the Jordan frame in order to control the behavior of the standard matter that, in this frame, remains minimally coupled. In the Einstein frame, the matter is non-minimally coupled and this fact could give rise to some difficulties in order to control the dynamics. Clearly, this means that all the considered  modifications insist on the geometric part. Controlling the relation between physical quantities passing from a frame to another could give hints to understand if conformal transformations are only a mathematical tool or have some physical meaning (see \cite{book} for a detailed discussion).

\subsection{Scalar tensor gravity in the Jordan frame}\label{ST_JF}


Let us start by considering the action of the scalar-tensor  gravity in 4 dimensions. It is 
\begin{eqnarray}\label{ST_action_1}
\mathcal{A}^{JF}\,=\,\int d^{4}x\sqrt{-g}\biggl[\phi\,R+V(\phi)+\omega(\phi)\,\phi_{;\alpha}\,\phi^{;\alpha}+\mathcal{X}\mathcal{L}_m\biggr]
\end{eqnarray}
Let us  note that the action

\begin{equation}
\mathcal{A}^{JF}\,=\,\int d^{4}x\sqrt{-g}\biggl[F(\phi)R+V(\phi)+\omega(\phi)\,\phi_{;\alpha}\,\phi^{;\alpha}+\mathcal{X}\mathcal{L}_m\biggr]
\end{equation}
is apparently more general than (\ref{ST_action_1}). In fact by substituiting $F(\phi)\,\rightarrow\,\phi$, we obtain only a new definition of functions $\omega(\phi)$ and $V(\phi)$ so the two formulations are essentially equivalent.

The term $\mathcal{L}_m$ is the minimally coupled ordinary matter contribution  considered as a \emph{perfect fluid}; $\omega(\phi)$ is a function of the scalar field and $V(\phi)$ is its potential which specifies  the dynamics. Actually if  $\omega(\phi) =\pm 1,0$ the nature and the dynamics of the scalar field is fixed. It can be a canonical scalar field, a phantom field or a field without dynamics (see e.g. \cite{odi2005,singularity} for details). In the metric approach, the field equations are obtained by varying the action (\ref{ST_action_1}) with respect to $g_{\mu\nu}$ and $\phi$. The field equations are 

\begin{eqnarray}
\label{fieldequation_ST}
&&\phi\,R_{\mu\nu}-\frac{\phi\,R+V(\phi)+\omega(\phi)\,\phi_{;\alpha}\,\phi^{;\alpha}}{2}\,g_{\mu\nu}+\omega(\phi)\,\phi_{;\mu}\,\phi_{;\nu}-\phi_{;\mu\nu}+g_{\mu\nu}\Box\,
\phi\,=\,\mathcal{X}\,T_{\mu\nu}\nonumber\\\\
&&2\,\omega(\phi)\,\Box\,\phi+\omega_{\phi}(\phi)\,\phi_{;\alpha}\phi^{;\alpha}-R-V_{\phi}(\phi)\,=\,0\nonumber
\end{eqnarray}
and the trace equation  is 

\begin{equation}
\phi\,R+2V(\phi)+\omega(\phi)\,\phi_{;\alpha}\phi^{;\alpha}-3\,\Box\,\phi\,=\,-\mathcal{X}\,T
\end{equation}
Here we introduced $V_\phi\,=\,\frac{dV}{d\phi}$ and $\omega_\phi(\phi)\,=\,\frac{d\omega(\phi)}{d\phi}$. 
If we assume that the Lagrangian density $\mathcal{L}_m$ of  matter depends only on the metric components $g_{\mu\nu}$ and not on its derivatives, we obtain $T_{\mu\nu}\,=\,1/2\,\mathcal{L}_m\,g_{\mu\nu}-\delta\mathcal{L}_m/\delta g^{\mu\nu}$. Let us consider a source with mass $M$. The energy-momentum tensor is the same of equation (\ref{emtensor_0}), (\ref{emtensorPPN}). It is useful to get the expression of $\mathcal{L}_m$. In fact from the definition of energy momentum tensor we have

\begin{eqnarray}
\delta\int d^4x\sqrt{-g}\,\mathcal{L}_m\,=\,-\int d^4x\sqrt{-g}\,T_{\mu\nu}\,\delta g^{\mu\nu}\,=\,-\int d^4x\sqrt{-g}\,\rho\,u_\mu u_\nu\,\delta g^{\mu\nu}
\end{eqnarray}
and further from the mathematical properties of metric tensor we have 

\begin{equation}
\delta(\sqrt{-g}\,\rho)\,=\,1/2\,\sqrt{-g}\,\rho\,u^\mu u^\nu\,\delta g_{\mu\nu}\,=\,-1/2\,\sqrt{-g}\,\rho\,u_\mu u_\nu\,\delta g^{\mu\nu}
\end{equation}
 then we find

\begin{eqnarray}
\mathcal{L}_m\,=\,2\,\rho
\end{eqnarray}
The variation of density is given by 

\begin{equation}
\delta\rho\,=\,\frac{\rho}{2}(g_{\mu\nu}-u_\mu u_\nu)\,\delta g^{\mu\nu}
\end{equation}
order to deal with standard self-gravitating systems, any theory of gravity has to be developed in its Newtonian or post-Newtonian limit depending on the order of approximation in terms of squared velocity $v^2$  \cite{Stabile_Capozziello}.

In this context, also the scalar field $\phi$ is approximated as the Ricci scalar. In particular we get $\phi\,=\,\phi^{(0)}\,+\,\phi^{(2)}\,+\dots$ while the functions $V(\phi)$ and $\omega(\phi)$ can be substituted by their corresponding Taylor series.

From the lowest order of field equations (\ref{fieldequation_ST}) we have

\begin{eqnarray}\label{PPN-field-equation-general-theory-ST-O0}
V(\phi^{(0)})\,=\,0\,,\,\,\,\,\,\,\,\,\,\,V_{\phi}(\phi^{(0)})\,=\,0
\end{eqnarray}
and also in the scalar tensor gravity a missing cosmological component in the action (\ref{ST_action_1}) implies that the space-time is asymptotically Minkowskian; moreover the ground value of scalar field $\phi$ must be a stationary point of potential. In the Newtonian limit, we have

\begin{eqnarray}
\label{NL-fieldequation_ST}
&&\triangle\biggl[\Phi-\frac{\phi^{(2)}}{\phi^{(0)}}\biggr]-\frac{R^{(2)}}{2}\,=\,\frac{\mathcal{X}\,\rho}{\phi^{(0)}}\nonumber\\\nonumber\\
&&\biggl\{\triangle\biggl[\Psi+\frac{\phi^{(2)}}{\phi^{(0)}}\biggr]+\frac{R^{(2)}}{2}\biggr\}\delta_{ij}+\biggr\{\Psi-\Phi-\frac{\phi^{(2)}}{\phi^{(0)}}\biggr\}_{,ij}\,=\,0\\\nonumber\\
&&\triangle\phi^{(2)}+\frac{V_{\phi\phi}(\phi^{(0)})}{2\,\omega(\phi^{(0)})}\,\phi^{(2)}+\frac{R^{(2)}}{2\,\omega(\phi^{(0)})}\,=\,0\nonumber\\\nonumber\\
&&R^{(2)}+3\,\frac{\triangle\phi^{(2)}}{\phi^{(0)}}\,=\,-\frac{\mathcal{X}\,\rho}{\phi^{(0)}}\nonumber
\end{eqnarray}
These equations are not simply the merging of field equations of GR and a further massive scalar field, but  come out to the fact that the scalar tensor gravity generates a coupled system of equations with respect to Ricci scalar $R$ and scalar field $\phi$. The gravitational potentials $\Phi$, $\Psi$ and the Ricci scalar $R^{(2)}$ are given by
\begin{eqnarray}\label{new_sol}
&&\Phi(\mathbf{x})\,=\,-\frac{\mathcal{X}}{4\pi\,\phi^{(0)}}\int
d^3\textbf{x}'\frac{\rho(\textbf{x}')}{|\textbf{x}-
\textbf{x}'|}-\frac{1}{8\pi}\int
d^3\textbf{x}'\frac{R^{(2)}(\textbf{x}')}{|\textbf{x}-
\textbf{x}'|}+\frac{\phi^{(2)}(\textbf{x})}{\phi^{(0)}}\nonumber\\\nonumber\\
&&\Psi(\mathbf{x})\,=\,\Phi(\textbf{x})+\frac{\phi^{(2)}(\textbf{x})}{\phi^{(0)}}\\\nonumber\\
&&R^{(2)}(\textbf{x})\,=\,-\frac{\mathcal{X}\,\rho(\textbf{x})}{\phi^{(0)}}-3\,\frac{\triangle\phi^{(2)}(\textbf{x})}{\phi^{(0)}}\nonumber
\end{eqnarray}
and supposing that $2\,\omega(\phi^{(0)})\,\phi^{(0)}-3\,\neq\,0$ we find for the scalar field $\phi^{(2)}$ the Yukawa-like field equation

\begin{eqnarray}
\label{fieldequation_SF}
\biggl[\triangle-{m_\phi}^2\biggr]\phi^{(2)}\,=\,\frac{\mathcal{X}\,\rho}{2\,\omega(\phi^{(0)})\phi^{(0)}-3}
\end{eqnarray}
where we introduced the mass definition

\begin{eqnarray}\label{mass_defin}
{m_\phi}^2\,\doteq\,-\frac{\phi^{(0)}\,V_{\phi\phi}(\phi^{(0)})}{2\,\omega(\phi^{(0)})\,\phi^{(0)}-3}
\end{eqnarray}
It is important to stress that the potential $\Psi$ can be found also as  
\begin{equation}
\Psi(\textbf{x})\,=\,\frac{1}{8\pi}\int
d^3\textbf{x}'\frac{R^{(2)}(\textbf{x}')}{|\textbf{x}-
\textbf{x}'|}-\frac{\phi^{(2)}(\textbf{x})}{\phi^{(0)}}
\end{equation}
see for example  \cite{sta_cap}. By using the Fourier transformation, the solution of Eq. (\ref{fieldequation_SF}) has the following form

\begin{eqnarray}
\label{fieldequation_SF_sol}
\phi^{(2)}(\textbf{x})\,=\,-\frac{\mathcal{X}}{2\,\omega(\phi^{(0)})\phi^{(0)}-3}\int\frac{d^3\textbf{k}}{(2\pi)^{3/2}}\frac{\tilde{\rho}(\textbf{k})\,e^{i\textbf{k}\cdot\textbf{x}}}{\textbf{k}^2+{m_\phi}^2}
\end{eqnarray}
The expressions (\ref{new_sol}) and (\ref{fieldequation_SF_sol}) represent the most general solution of any scalar-tensor gravity in the Newtonian limit. Since the superposition  principle is yet valid (as previously analyzed), it  is sufficient to consider again the solutions generated by the pointlike source. Then the solutions are \cite{Stabile_Capozziello,PRD2,sta_cap} 

\begin{eqnarray}
\label{NL-solution_ST}
&&\phi^{(2)}(\textbf{x})\,=\,-\frac{1}{2\,\omega(\phi^{(0)})\,\phi^{(0)}-3}\frac{r_g}{|\textbf{x}|}\,e^{-m_\phi |\textbf{x}|}\nonumber\\\nonumber\\
&&R^{(2)}(\textbf{x})\,=\,-\frac{4\pi\,r_g}{\phi^{(0)}}\,\delta(\textbf{x})+\frac{3\,{m_\phi}^2}{[2\,\omega(\phi^{(0)})\,\phi^{(0)}-3]\,\phi^{(0)}}\frac{r_g}{|\textbf{x}|}\,e^{-m_\phi |\textbf{x}|}\nonumber\\\\
&&\Phi(\textbf{x})\,=\,-\frac{GM}{\phi^{(0)}|\textbf{x}|}\biggl\{1-\frac{e^{-m_\phi |\textbf{x}|}}{2\,\omega(\phi^{(0)})\,\phi^{(0)}-3}\biggr\}\nonumber\\\nonumber\\
&&\Psi(\textbf{x})\,=\,-\frac{GM}{\phi^{(0)}|\textbf{x}|}\biggl\{1+\frac{e^{-m_\phi |\textbf{x}|}}{2\,\omega(\phi^{(0)})\,\phi^{(0)}-3}\biggr\}\nonumber
\end{eqnarray}
In the case $V(\phi)\,=\,0$, the scalar field is massless and $\omega(\phi)\,=\,-\omega_0/\phi$, we obtain

\begin{eqnarray}\label{sol_point_BD}
\Phi(\textbf{x})\,=\,\Phi_{BD}(\mathbf{x})\,&=&\,-\frac{GM}{\phi^{(0)}|\textbf{x}|}\left[\frac{2(2+\omega_0)}{2\,\omega_0+3}\right]
\,=\,-\frac{G^*M}{|\textbf{x}|}
\nonumber\\\\
\Psi(\textbf{x})\,=\,\Psi_{BD}(\mathbf{x})\,&=&\,-\frac{G^*M}{|\textbf{x}|}\left(\frac{1+\omega_0}{2+\omega_0}\right)
\nonumber
\end{eqnarray}
the well-known  Brans-Dicke solutions \cite{bra-dic} with Eddington's parameter $\gamma\,=\,\frac{1+\omega_0}{2+\omega_0}$ \cite{will} where   the gravitational constat is defined as $G\,\rightarrow\,G^*\,=\,\frac{G}{\phi^{(0)}}\frac{2(2+\omega_0)}{2\,\omega_0+3}$.


\subsection{Scalar tensor gravity in the Einstein frame}\label{ST_EF}


Let us now introduce the conformal transformation 

\begin{eqnarray}\label{transconf}
g_{\mu\nu}\,\longrightarrow\,\tilde{g}_{\mu\nu}\,=\,\Omega^2\,g_{\mu\nu}
\end{eqnarray}
where $\Omega\,=\,\Omega(x)$ is a nowhere vanishing, regular function, called a \emph{Weyl} or \emph{conformal} 
transformation. This transformation has been introduced to show that scalar-tensor theories are, in general, conformally equivalent to the Einstein theory plus minimally coupled scalar fields. However if standard matter is present, the conformal transformation generates the non-minimal coupling between the matter component and the scalar field.

By applying the transformation (\ref{transconf}), the action in (\ref{ST_action_1}) can be
reformulated as follows

\begin{eqnarray}\label{TS-EF-action}
\mathcal{A}^{EF}\,=\,\int d^4x \sqrt{-\tilde{g}}
\biggl[\Xi\,\tilde{R}+W(\tilde{\phi})
+\tilde{\omega}(\tilde{\phi})\tilde{\phi}_{;\alpha}\tilde{\phi}^{;\alpha}+\mathcal{X}
\tilde{\mathcal{L}}_{m}\biggr]
\end{eqnarray}
in which $\tilde{R}$ is the Ricci scalar relative to the metric $\tilde{g}_{\mu\nu}$ and $\Xi$ is a generic constant. The two actions (\ref{ST_action_1}) and (\ref{TS-EF-action}) are mathematically equivalent. In fact the conformal transformation is given by imposing the condition 

\begin{equation}
\sqrt{-g}\biggl[\phi\,R+V(\phi)+\omega(\phi)\,\phi_{;\alpha}\,\phi^{;\alpha}+\mathcal{X}\mathcal{L}_m\biggr]\,=\,\sqrt{-\tilde{g}}
\biggl[\Xi\,\tilde{R}+W(\tilde{\phi})+\tilde{\omega}(\tilde{\phi})\tilde{\phi}_{;\alpha}\tilde{\phi}^{;\alpha}+\mathcal{X}\tilde{\mathcal{L}}_{m}\biggr]
\end{equation}
The relations between the quantities in the two frames are

\begin{eqnarray}\label{transconfTS}
&&\tilde{\omega}(\tilde{\phi})\,{d\tilde{\phi}}^2\,=\,\frac{\Xi}{2}\,[2\,\phi\,\omega(\phi)-3]\biggl(\frac{d\phi}{\phi}\biggr)^2\nonumber\\\nonumber\\
&&W(\tilde{\phi})=\frac{\Xi^2}{\phi(\tilde{\phi})^2}\,V(\phi(\tilde{\phi}))\nonumber\\\\
&&\tilde{\mathcal{L}}_m\,=\,\frac{\Xi^2}{\phi(\tilde{\phi})^2}\,\mathcal{L}_m\biggl(\frac{\Xi\,\tilde{g}_{\rho\sigma}}{\phi(
\tilde{\phi})}\biggr)\nonumber\\\nonumber\\
&&\phi\,\Omega^{-2}\,=\,\Xi\nonumber
\end{eqnarray}
The field equations for the new fields $\tilde{g}_{\mu\nu}$ and $\tilde{\phi}$ are

\begin{eqnarray}\label{fieldequation_ST_EF}
&&\Xi\,\tilde{R}_{\mu\nu}-\frac{\Xi\,\tilde{R}+W(\tilde{\phi})+\tilde{\omega}(\tilde{\phi})\,\tilde{\phi}_{;\alpha}\tilde{\phi}^{;\alpha}}{2}\,\tilde{g}_{\mu\nu}+\tilde{\omega}(\tilde{\phi})\,\tilde{\phi}_{;\mu}\tilde{\phi}_{;\nu}\,=\,\mathcal{X}\,\tilde{T}_{\mu\nu}\nonumber
\\\nonumber\\
&&2\,\tilde{\omega}(\tilde{\phi})\,\tilde{\Box}\tilde{\phi}+\tilde{\omega}_{\tilde{\phi}}(\tilde{\phi})\,\tilde{\phi}_{;\alpha}\tilde{\phi}^{;\alpha}-W_{\tilde{\phi}}(\tilde{\phi})-\mathcal{X}\frac{\delta\,\tilde{\mathcal{L}}_m}{\delta\,\tilde{\phi}}\,=\,0
\\\nonumber\\
&&\Xi\,\tilde{R}+2W(\tilde{\phi})+\tilde{\omega}(\tilde{\phi})\,\tilde{\phi}_
{;\alpha}\tilde{\phi}^{;\alpha}\,=\,-\mathcal{X}\,\tilde{T}\nonumber
\end{eqnarray}
where $\tilde{T}_{\mu\nu}$ and $\tilde{\Box}$ are the re-definition of the quantities (\ref{emtensor_0}) with respect to the metric $\tilde{g}_{\mu\nu}$. The field equations (\ref{fieldequation_ST_EF}) can be obtained from (\ref{fieldequation_ST}) by substituing all geometrical and physical quantities in terms of conformally transformed ones. In particular we have

\begin{eqnarray}\label{transf_conf_II}
&&R_{\mu\nu}\,=\,\tilde{R}_{\mu\nu}+2\,{\ln \Omega}_{\tilde{;\mu\nu}}+2\,{\ln \Omega}_{;\mu}\,{\ln \Omega}_{;\nu}+[\tilde{\Box}\ln\Omega-2\,\tilde{{\ln \Omega}^{;\sigma}{\ln \Omega}_{;\sigma}}]\,\tilde{g}_{\mu\nu}\nonumber\\\nonumber\\
&&R\,=\,\Omega^2\biggl[\tilde{R}+6\,\tilde{\Box}\ln\Omega-3\,\tilde{{\ln \Omega}^{;\sigma}{\ln \Omega}_{;\sigma}}\biggr]\nonumber
\\\\
&&\phi_{;\mu\nu}\,=\,\phi_{\tilde{;\mu\nu}}+2\,\phi_{;\mu}\phi_{;\nu}-\tilde{{\ln \Omega}^{;\sigma}\phi_{;\sigma}}\,\tilde{g}_{\mu\nu}\nonumber\\\nonumber\\
&&\Box(\cdot)\,=\,\Omega^2\,\tilde{\Box}(\cdot)-2\,\tilde{{\ln \Omega}^{;\sigma}\partial_{\sigma}}(\cdot)\nonumber
\end{eqnarray}

The integration of field equations (\ref{fieldequation_ST_EF}) is only formal because we do not know the analytical expression of the coupling function between the matter and the scalar field $\tilde{\phi}$ (see the third line of (\ref{transconfTS})). We can make some assumptions on the  parameter $\Xi$ and the function $\tilde{\omega}(\tilde{\phi})$ in the minimally coupled Lagrangian (\ref{TS-EF-action}) and on the function $\omega(\phi)$ in the  nonminimally coupled Lagrangian (\ref{ST_action_1}). If we choose $\tilde{\omega}(\tilde{\phi})\,=\,-1/2$, $\Xi\,=\,1$ and $\omega(\phi)\,=\,-\omega_0/\phi$, the transformation between  the  scalar fields $\phi$ and  $\tilde{\phi}$ is given by the first line in (\ref{transconfTS}), that is 

\begin{eqnarray}\label{trans_rule}
\tilde{\phi}(\phi)\,=\,\tilde{\phi}_0+\sqrt{2\omega_0+3}\,\ln\phi\,\,\,\,\,\,\,\,\,\,\,\,\,\,\,\phi(\tilde{\phi})\,=\exp{\left(\frac{\tilde{\phi}-\tilde{\phi}_0}{\sqrt{2\omega_0+3}}\right)}
\end{eqnarray} 
where obviously $\omega_0\,>\,-3/2$ and $\tilde{\phi}_0$ is an integration constant\footnote{Without losing  generality, we can set $\tilde{\phi}_0\,=\,0$.}. The potential $W$ and the matter Lagrangian $\tilde{\mathcal{L}}_m$ are

\begin{eqnarray}\label{transconfTS_1}
W(\tilde{\phi})\,=\exp{\left(-\frac{2\tilde{\phi}}{\sqrt{2\omega_0+3}}\right)}\,V\biggl(e^{\frac{\tilde{\phi}}{\sqrt{2\omega_0+3}}}\biggr)\,\,\,\,\,\,\,\,\,\,\,\,\,\,\
\tilde{\mathcal{L}}_m\,=\,2\,\rho\exp{\left(-\frac{2\tilde{\phi}}{\sqrt{2\omega_0+3}}\right)}
\end{eqnarray}

In both frames, the scalar fields are expressed as perturbative contributions on the cosmological background ($\phi^{(0)}$, $\tilde{\phi}^{(0)}$) with respect to  the dimensionless quantity $v^2$. Then also for the scalar field $\tilde{\phi}$, we can consider the develop $\tilde{\phi}\,=\,\tilde{\phi}^{(0)}+\tilde{\phi}^{(2)}+\dots$. Such a develop can be applied to the transformation rule (\ref{trans_rule}) and we obtain  

\begin{eqnarray}\label{trans_rule_pert}
&&\tilde{\phi}(\phi)\,=\,\sqrt{2\omega_0+3}\,\ln\phi\,=\,\sqrt{2\omega_0+3}\,\ln\phi^{(0)}+\frac{\sqrt{2\omega_0+3}}{\phi^{(0)}}\,\phi^{(2)}+\dots\,\doteq\,\tilde{\phi}^{(0)}+\tilde{\phi}^{(2)}\,+\dots\nonumber\\\\\nonumber
&&\phi(\tilde{\phi})\,=\,e^{\frac{\tilde{\phi}}{\sqrt{2\omega_0+3}}}\,=\,e^{\frac{\tilde{\phi}^{(0)}}{\sqrt{2\omega_0+3}}}\,+\,\frac{e^{\frac{\tilde{\phi}^{(0)}}{\sqrt{2\omega_0+3}}}}{\sqrt{2\omega_0+3}}\,\tilde{\phi}^{(2)}\,+\dots\,\doteq\,\phi^{(0)}+\phi^{(2)}\,+\,\dots
\end{eqnarray} 
Since we are interested in the Newtonian limit of field equations (\ref{fieldequation_ST_EF}), we can assume, for the conformally transformed metric $\tilde{g}_{\mu\nu}$, an expression as (\ref{metric_new}) but with some differences. In fact from the conformal transformation ($\ref{transconf}$) and from the last line of (\ref{transconfTS}), we have

\begin{eqnarray}\label{transconf_NL}\tilde{g}_{\mu\nu}\,=\,\phi\,g_{\mu\nu}\,=\,\phi^{(0)}\eta_{\mu\nu}+[\phi^{(0)}g^{(2)}_{\mu\nu}+\phi^{(2)}\eta_{\mu\nu}]+\dots\,=\,\tilde{\eta}_{\mu\nu}+\tilde{g}^{(2)}_{\mu\nu}+\dots\end{eqnarray}
then the conformally transformed metric becomes
 
\begin{eqnarray}\label{me_EF}
{ds}^2\,=\,(\phi^{(0)}+2\tilde{\Phi})\,dt^2-(\phi^{(0)}-2\tilde{\Psi})\,\delta_{ij}dx^idx^j
\end{eqnarray}
and the relation between the gravitational potentials in the two frames is

\begin{eqnarray}\label{diff_pot}
\tilde{\Phi}-\phi^{(0)}\,\Phi\,=\,\frac{\phi^{(2)}}{2}\,,\,\,\,\,\,\,\,\,\,\,\,\,\,\,\,\tilde{\Psi}-\phi^{(0)}\,\Psi\,=\,-\frac{\phi^{(2)}}{2}
\end{eqnarray}
Then the field equations (\ref{fieldequation_ST_EF}) become\footnote{With the assumptions of the metric (\ref{me_EF}) the Ricci tensor $\tilde{R}_{\mu\nu}$ in the Newtonian limit has the form $\frac{\triangle\tilde{\Phi}}{\phi^{(0)}}$ (a similar behaviour for $\tilde{R}^{(2)}_{ij}$), where the Ricci scalar is scaled by the factor ${\phi^{(0)}}^2$. The same scaling occurs for the Laplacian: $\triangle\,\rightarrow\,\frac{\triangle}{\phi^{(0)}}$.}

\begin{eqnarray}\label{NL_fieldequation_ST_EF}
&&\frac{\triangle\tilde{\Phi}}{\phi^{(0)}}-\frac{\tilde{R}^{(2)}}{2}\,\phi^{(0)}\,=\,\mathcal{X}\,\tilde{T}^{(2)}_{00}\nonumber
\\\nonumber\\
&&\biggl\{\frac{\triangle\tilde{\Psi}}{\phi^{(0)}}+\frac{\tilde{R}^{(2)}}{2}\phi^{(0)}\biggr\}\,\delta_{ij}+\frac{(\tilde{\Psi}-\tilde{\Phi})_{,ij}}{\phi^{(0)}}\,=\,0\nonumber
\\\\
&&\frac{\triangle\tilde{\phi}^{(2)}}{\phi^{(0)}}-W_{\tilde{\phi}\tilde{\phi}}(\tilde{\phi}^{(0)})\,\tilde{\phi}^{(2)}-\mathcal{X}\biggl[\frac{\delta\,\tilde{\mathcal{L}}_m}{\delta\,\tilde{\phi}}\biggr]^{(2)}\,=\,0
\nonumber\\\nonumber\\
&&\tilde{R}^{(2)}\,=\,-\mathcal{X}\,\tilde{T}^{(2)}\nonumber
\end{eqnarray}
where also in this case we have $W(\tilde{\phi}^{(0)})\,=\,0$ and $W_{\tilde{\phi}}(\tilde{\phi}^{(0)})\,=\,0$.  However these conditions are an obvious consequence of the conformal transformation of conditions $V(\phi^{(0)})\,=\,0$ and $V_\phi(\phi^{(0)})\,=\,0$. In fact we can figure out  that $V(\phi)\,\propto\,(\phi-\phi^{(0)})^2$ and then $W(\tilde{\phi})\,\propto\,\biggl(e^{\frac{\tilde{\phi}}{\sqrt{2\omega_0+3}}}-\phi^{(0)}\biggr)^2$ which, by using  relations (\ref{trans_rule_pert}), satisfies the above conditions. Finally, we note that $W_{\tilde{\phi}\tilde{\phi}}(\tilde{\phi}^{(0)})\,=\,\frac{V_{\phi\phi}\biggl(e^{\frac{\tilde{\phi}^{(0)}}{\sqrt{2\omega_0+3}}}\biggr)}{2\omega_0+3}\,=\,\frac{V_{\phi\phi}(\phi^{(0)})}{2\omega_0+3}$ and by the definition of mass ${m_\phi}^2$, given in Eq. (\ref{mass_defin}),  we obtain $W_{\tilde{\phi}\tilde{\phi}}(\tilde{\phi}^{(0)})\,=\,{m_\phi}^2/\phi^{(0)}$. Finally,  the energy-momentum tensor $\tilde{T}_{\mu\nu}$ is given by the following expression

\begin{eqnarray}
\tilde{T}_{\mu\nu}\,=\,\rho\exp\left(-\frac{2\,\tilde{\phi}}{\sqrt{2\,\omega_0+3}}\right)\tilde{u}_\mu\tilde{u}_\nu
\end{eqnarray}
where $\tilde{g}_{\mu\nu}\tilde{u}^\mu\tilde{u}^\nu\,=\,1$ then $\tilde{u}_0\,=\,\sqrt{\phi^{(0)}+2\,\tilde{\Phi}}$. In the Newtonian limit, we find $\tilde{T}^{(2)}_{00}\,=\,\rho/\phi^{(0)}$ and $\tilde{T}^{(2)}\,=\,\rho/{\phi^{(0)}}^2$. It remains only to calculate the source term $\delta\tilde{\mathcal{L}}_m/\delta\tilde{\phi}$  of the scalar field $\tilde{\phi}^{(2)}$. From the third line of (\ref{transconfTS}) and, by using the transformation rules (\ref{trans_rule}), we find the  coupling between the scalar field and the ordinary matter 

\begin{eqnarray}\label{var_matter}
\frac{\delta\tilde{\mathcal{L}}_m}{\delta\tilde{\phi}}\,&=&\,\frac{\delta}{\delta\tilde{\phi}}\biggl\{e^{-\frac{2\tilde{\phi}}{\sqrt{2\omega_0+3}}}\mathcal{L}_m\biggl(e^{-\frac{\tilde{\phi}}{\sqrt{2\omega_0+3}}}\,\tilde{g}_{\rho\sigma}\biggr)\biggr\}\,=\,\frac{-2\,e^{-\frac{2\,\tilde{\phi}}{\sqrt{2\omega_0+3}}}}{\sqrt{2\omega_0+3}}\,\mathcal{L}_m(\cdot)+e^{-\frac{2\,\tilde{\phi}}{\sqrt{2\omega_0+3}}}\,\frac{\delta\mathcal{L}_m(\cdot)}{\delta g^{\mu\nu}}\,\frac{\delta g^{\mu\nu}}{\delta \tilde{\phi}}
\nonumber\\\nonumber\\
&=&\frac{-2\,e^{-\frac{2\,\tilde{\phi}}{\sqrt{2\omega_0+3}}}}{\sqrt{2\omega_0+3}}\,\mathcal{L}_m(\cdot)+e^{-\frac{2\,\tilde{\phi}}{\sqrt{2\omega_0+3}}}\,\frac{\mathcal{L}_m(\cdot)}{2}(g_{\mu\nu}-u_\mu u_\nu)\,\frac{\tilde{g}^{\mu\nu}\delta\,\phi(\tilde{\phi})}{\delta\,\tilde{\phi}}
\\\nonumber\\
&=&\nonumber\frac{-2\,e^{-\frac{2\,\tilde{\phi}}{\sqrt{2\omega_0+3}}}}{\sqrt{2\omega_0+3}}\,\mathcal{L}_m(\cdot)+e^{-\frac{2\,\tilde{\phi}}{\sqrt{2\omega_0+3}}}\,\frac{3\,\mathcal{L}_m(\cdot)}{2}\frac{1}{\sqrt{2\omega_0+3}}\,=\,-\frac{1}{2}\frac{e^{-\frac{2\,\tilde{\phi}}{\sqrt{2\omega_0+3}}}}{\sqrt{2\omega_0+3}}\,\mathcal{L}_m(\cdot)\,=\,-\frac{e^{-\frac{2\,\tilde{\phi}}{\sqrt{2\omega_0+3}}}}{\sqrt{2\omega_0+3}}\,\rho
\end{eqnarray}
Then the system of Eqs. (\ref{NL_fieldequation_ST_EF}) becomes

\begin{eqnarray}\label{NL_fieldequation_ST_EF_2}
&&\triangle\tilde{\Phi}\,=\,\frac{\mathcal{X}\,\rho}{2}\,\,\,\,\,\,\,\,\,\,\,\,\,\,\,\,\tilde{\Psi}\,=\,\tilde{\Phi}\nonumber
\\\\
&&\biggl[\triangle-{m_\phi}^2\biggr]\tilde{\phi}^{(2)}\,=\,-\frac{\mathcal{X}\,\rho}{\phi^{(0)}\sqrt{2\omega_0+3}}
\nonumber
\end{eqnarray}
and their solutions in the case of pointlike source are

\begin{eqnarray}\label{NL_fieldequation_ST_EF_2_sol}
\tilde{\Phi}\,=\,-\frac{ G\,M}{|\textbf{x}|}\,\,\,\,\,\,\,\,\,\,\,\,\,\,\,\,\tilde{\Psi}\,=\,\tilde{\Phi}\,\,\,\,\,\,\,\,\,\,\,\,\,\,\,\,\tilde{\phi}^{(2)}\,=\,\frac{1}{\phi^{(0)}\sqrt{2\omega_0+3}}\frac{r_g}{|\textbf{x}|}\,e^{-m_\phi|\textbf{x}|}
\end{eqnarray}
The difference in Eqs.(\ref{diff_pot}) between the gravitational potentials is satisfied by using the expression of scalar field in the Jordan frame (first line of (\ref{NL-solution_ST})) where, obviously, we set $\omega(\phi)\,=\,-\omega_0/\phi$. In fact we find

\begin{eqnarray}
\tilde{\Phi}-\phi^{(0)}\Phi\,=\,\frac{GM}{2\,\omega_0+3}\frac{e^{-m_\phi |\textbf{x}|}}{|\textbf{x}|}\,=\,\frac{\phi^{(2)}}{2}
\end{eqnarray}
and an analogous relations is found also for the couples $\Psi,\,\tilde{\Psi}$. Furthermore we can check  also the transformation rules (\ref{trans_rule}) and (\ref{trans_rule_pert}) for the solutions (\ref{NL-solution_ST}) and (\ref{NL_fieldequation_ST_EF_2_sol}) of the scalar fields $\phi,\,\tilde{\phi}$.

The redefinition of the gravitational constant $G$ (as performed in the Jordan frame $G\,\rightarrow\,G^*$ in the case of Brans-Dicke theory \cite{bra-dic}) is not available when we are interested to compare the outcomes in both frame. In fact the couple of potentials $\Phi, \tilde{\Phi}$ differs not only from the dynamical contribution of the scalar field ($\phi^{(2)}$) but also from the definition of the gravitational constant. Furthermore,  in the Einstein frame, the scalar field $\tilde{\phi}$ does not contribute (the coupling constant between $R$ and $\tilde{\phi}$ is vanishing), then we find the same outcomes of GR with ordinary matter. However by supposing the Jordan frame as starting point and  coming back via conformal transformation, we find that the gravitational constant is not invariant and depends on the background value of the scalar field in the Einstein frame, that is  $G\,\rightarrow\,G_{eff}\,\propto\,e^{-\tilde{\phi}^{(0)}}G$.


\subsection{The analogy between the scalar tensor gravity and $f(R)$-gravity}\label{ST_f(R)}


Recently,  several authors claimed that higher-order theories of gravity and among them, $f(R)$ gravity, are characterized by an ill defined behavior in the Newtonian regime. In particular, it is discussed  that Newtonian corrections of the gravitational potential violate experimental constraints since these quantities can be recovered by a direct analogy with Brans-Dicke gravity  simply supposing the Brans-Dicke characteristic parameter $\omega_0$ vanishing (see \cite{olmo1,olmo2,olmo3} for a discussion). Actually, the calculations of the Newtonian limit of $f(R)$-gravity, directly performed in a rigorous manner, have showed that this is not the case \cite{Stabile_Capozziello, PRD1, PRD2, dick} and it is possible  to discuss also the analogy with Brans-Dicke gravity. The issue is easily overcome once the correct analogy between $f(R)$-gravity and the corresponding scalar-tensor framework is taken into account. It is important  mentioning  that several important results already achieved in the Newtonian regime, see e.g.\cite{hans,stelle}, are confirmed by these rigorous approach and only the wrong analogies are ruled out.

In literature, it is shown that $f(R)$ gravity models can be rewritten in term of a scalar-field Lagrangian non-minimally coupled with gravity but without  kinetic term implying that the Brans-Dicke parameter is $\omega(\phi)\,=\,0$. This fact is considered the reason for the ill-definition of the weak field limit that should be $\omega\rightarrow \infty$ inside the Solar System.

Let us deal with the $f(R)$ gravity formalism in order to set correctly the problem. The field equations (\ref{fieldequationFOG}) can be recast in the framework of scalar-tensor gravity as son as  we select a particular expression for the  free parameters of  the theory. The result is the so-called O'Hanlon theory \cite{ohanlon} which can be written as
\begin{eqnarray}\label{ohanlon}
\mathcal{A}^{JF}_{OH}=\int d^4x\sqrt{-g}\biggl[\phi
R+V(\phi)+\mathcal{X}\mathcal{L}_m\biggr]
\end{eqnarray}
The field equations are obtained by starting from ones of (\ref{fieldequation_ST})

\begin{eqnarray}
\label{fieldequation_OH}
&&\phi\,R_{\mu\nu}-\frac{\phi\,R+V(\phi)}{2}\,g_{\mu\nu}-\phi_{;\mu\nu}+g_{\mu\nu}\Box\,
\phi\,=\,\mathcal{X}\,T_{\mu\nu}\nonumber\\\nonumber\\
&&R+V_{\phi}(\phi)\,=\,0\\\nonumber\\\nonumber
&&\phi\,R+2V(\phi)-3\,\Box\,\phi\,=\,-\mathcal{X}\,T
\end{eqnarray}
By supposing that the Jacobian of the transformation $\phi\,=\,f_R$ is non-vanishing, the two representations  can be mapped one into the other considering the following equivalence

\begin{eqnarray}
\label{equiv}
&&\omega(\phi)\,=\,0\nonumber\\\nonumber\\
&&V(\phi)\,=\,f-f_RR\\\nonumber\\\nonumber
&&\phi\,V_\phi(\phi)-2V(\phi)\,=\,f_RR-2f
\end{eqnarray}
From the definition of the mass (\ref{mass_defin}) we have $\phi\,V_\phi(\phi)-2V(\phi)\,=\,3\,{m_\phi}^2\phi^{(2)}$, then we have also $f_RR-2f\,=\,3\,{m_\phi}^2\phi^{(2)}$ and by performing the Newtonian limit on the function $f$ \cite{Stabile_Capozziello}, we get $f_R(0)R^{(2)}\,=\,-3\,{m_\phi}^2\phi^{(2)}$. The spatial evolution of Ricci scalar is obtained by solving the field equations (\ref{fieldequationFOG})

\begin{eqnarray}\label{Ricci_sol}
R^{(2)}\,=\,-\frac{3\,{m_\phi}^2\phi^{(2)}}{f_R(0)}\,=\,-\frac{{m_\phi}^2\,r_g}{f_R(0)}\frac{e^{-m_\phi|\textbf{x}|}}{|\textbf{x}|}
\end{eqnarray}
without using the conformal transformation \cite{Stabile_Capozziello, PRD2}. The solution for the potentials $\Phi, \Psi$ are obtained simply by setting $\omega(\phi)\,=\,0$ in Eqs. (\ref{NL-solution_ST}) and $\phi^{(0)}\,=\,f_R(0)$. In the case $f(R) \rightarrow\,R$, from the second line of (\ref{equiv}), $V(\phi)\,=\,0\,\rightarrow\,m_\phi\,=\,0$ and the solutions (\ref{NL-solution_ST}) become the standard Schwarzschild solution in the Newtonian limit.

Finally,  we can consider a Taylor expansion\footnote{The terms resulting from $R^n$ with $n \geq 3$ do {\it not} contribute to the Newtonian order.} of the form $f\,=\,f_R(0)\,R^{(2)} +\frac{f_{RR}(0)}{2}\, {R^{(2)}}^2$ so that the associated scalar field reads $\phi\,=\,f_R(0)+f_{RR}(0)\,R^{(2)}$. The relation between $\phi$ and $R^{(2)}$ is $R^{(2)}\,=\,\frac{\phi-f_R(0)}{f_{RR}(0)}$ while the self-interaction potential (second line of (\ref{equiv})) turns out the be $V(\phi)\,=\,-\frac{(\phi-f_R(0))^2}{2\,f_{RR}(0)}$ satisfying the conditions $V(f_R(0))\,=\,0$ and $V_\phi(f_R(0))\,=\,0$. In relation to the definition of the scalar field, we can opportunely identify $f_R(0)$ with a constant value $\phi^{(0)}\,=\,f_R(0)$ which  justifies the previous ansatz for  matching  solutions in the limit of GR. Furthermore, the mass of the scalar field can be expressed in term of the Lagrangian parameters as ${m_\phi}^2\,=\,\frac{1}{3}\phi^{(0)}\,V_{\phi\phi}(\phi^{(0)})\,=\,-\frac{f_R(0)}{3f_{RR}(0)}$. Also in this case the value of mass is the same obtained by solving the problem without invoking the scalar tensor analogy \cite{Stabile_Capozziello,PRD2}. However with this  remark, it is  clear the analogy between $f(R)$-gravity and a particular class of scalar tensor theories \cite{ohanlon}. 

Finally, the above results have to be interpreted also with respect to  results reported in  \cite{olmo1,olmo2,olmo3}. In those works, the relation between $f(R)$ gravity  and scalar-tensor theories was used to compute the weak-field limit. Though $f(R)$ gravity,  in metric formalism, is related to the Brans-Dicke theory with $\omega_0\,=\,0$, the equivalence is not complete because $f(R)$ has a non-trivial potential, which is not present in the original Brans-Dicke theory. The role of the potential is to introduce a length scale (as we found here) and then the post-Newtonian parameter $\gamma$ develops a spatial dependence. In Refs.\cite{olmo1,olmo2,olmo3}  this spatial dependence is discussed despite the fact that 
$\omega_0\,=\,0$. Note that formulas presented in \cite{olmo1,olmo2,olmo3} for the weak-field limit are in complete agreement with those presented here also if achieved with a different approach. 


\section{Discussion and Conclusions}\label{concl}


The weak field limit of Extended Theories of Gravity has been  discussed in view of some relevant astrophysical issues. In particular, we have considered  the hydrostatic equilibrium of  stars, the galactic rotation curves and the gravitational lensing. Finally we have analyzed the relations between the Jordan and Einstein frames in the same limit and  focused our attention on the relations linking the  gravitational potentials in both frames.

Specifically,   we have considered theories containing  generic functions of the Ricci scalar $R$, the squared Ricci tensor  $R_{\alpha\beta}R^{\alpha\beta}$ and the squared Riemann tensor $R_{\alpha\beta\gamma\delta}R^{\alpha\beta\gamma\delta}$.  We obtain the analogy, in the Newtonian limit, with the so-called {\it Quadratic Lagrangian} containing the squared Ricci scalar and the squared Ricci tensor in addition to the linear term $R$. All contributions to the field equations related to the squared Riemann  curvature invariant can be expressed by the other two curvature invariants (squared Ricci tensor  and squared Ricci scalar) via the Gauss-Bonnet invariant. It is straightforward to show that  the spherically symmetric solutions show a Yukawa-like dependence  with  two characteristic lengths. An important result is that, for  generic Fourth Order Gravity models,   two non-equivalent  metric potentials come out. In the limit $f\,\rightarrow\,R$,   only the Newtonian potential is present as it has to be in GR. The presence of two gravitational potentials, together with  the non-validity of  the Birkhoff and Gauss theorems, are the main differences  between  Fourth Order Gravity and GR.

Coming to the astrophysical applications, we adopted a polytropic equation of state relating the mass density to the  pressure and derived a \emph{modified Lan\'{e}-Emden equation} whose solutions  can be compared to the analogous solutions coming from  GR. As soon as one considers the limit $f(R)\,\rightarrow\,R$, the standard hydrostatic equilibrium theory  is fully recovered. Since the Gauss theorem is not valid in this context and the \emph{modified Lan\'{e}-Emden equation} is an integro-differential equation, the mass distribution plays a crucial role. The correlation between two points in the star is given by a Yukawa-like term of the corresponding Green function. These feature opens the possibility to address the structure of anomalous stars that, in standard stellar theory could not be consistently achieved \cite{farinelli}.

As further astrophysical application,  we have considered  the rotation curves of spiral galaxies. Starting from  the pointlike solutions and having formulated the expression of the rotation curves, we  considered   the principal galactic components (i.e. the bulge, the disk and the halo). The theoretical curves have been compared with the experimental data.
The rotation curves have been evaluated by considering the bulge and the DM component spherically symmetric and a circular disk symmetric with respect to a plane where the radius is larger than the thickness. Also in the case of Fourth Order Gravity with standard  matter, we find that the rotation curve has the Keplerian behavior and only if one adds some DM component, a reliable matching with experimental data is achieved. However the DM hypothesis  gives rise to two serious problems: since the DM distribution is diverging when we consider the whole amount of mass, it  is crucial the choice of the cut-off inside the integral. Another critical point is the choice of the mass model and the values of free parameters. These shortcomings can be consistently addressed by introducing a further scalar field as in \cite{sta_cap}. Such scalar field, however, can be reinterpreted in terms of curvature invariants \cite{sta_cap}.

The  gravitational lensing approach  in Fourth Order Gravity has been pursued on two steps: firstly we considered a point-like source and the motion of photons.  In the second step, we took into account  the geodesic motion and  reformulated the light deflection  for a generic matter distribution. In the case of an axially symmetric matter distribution, we obtained the standard relation between the deflection angle and the orthogonal gradient of metric potentials. Otherwise, the angle is depending also on the travel direction of the photon. In particular, if there is a $z$-symmetry, the deflection angle does not depend explicitly on the parameters of Fourth Order Gravity. Starting from the definition of the masses,  one have to  note that the contribution of a generic function of the Ricci scalar  is only in the missing parameter. Then  $f(R)$-gravity admits the same geodesic trajectory of GR. If one wants to take into account   corrections to GR,  one needs to add a generic function of the squared Ricci tensor  into the Hilbert-Einstein action. In this case,  we find the deflection angle smaller than the one of GR. The mathematical motivation is a consequence of the algebraic sign of  the parameter in front of the squared Ricci tensor. In fact it is  different with respect to the GR term and we can interpret it as a "repulsive force" giving  a lower curvature for the photon trajectory.

A similar result  is found for the galactic rotation curve, where the contribution of the squared Ricci tensor   gives  a lower rotation velocity profile than the one derived from the weak field limit of  GR.  However, if we  estimate the weight of the corrections induced by $f(R)$-gravity,   we have a perfect agreement with the GR from the point of view of gravitational lensing. Only by adding $f(R_{\alpha\beta}R^{\alpha\beta})$ in the action, we induce  modifications in both gravitational lensing and galactic rotation profiles.

Finally,  we have tackle   the debate of selecting the physical frame by conformal transformations. Specifically,  we have shown  how the Newtonian limit behaves in the Jordan and in the Einstein frame. The general result is that Newtonian potentials, masses and other physical quantities can be compared in both frames once the perturbative analysis is correctly performed.  The main result is that if such an analysis is carefully developed in a frame, the perturbative process can be controlled step by step leading to coherent results in both frames. In other words, also if the gauge invariance is broken, there is the possibility to control conformal quantities and fix the observables.


\acknowledgements{Acknowledgements}

SC acknowledges  financial support of INFN (initiative specifiche QGSKY  and TEONGRAV).
The authors acknowledge M. De Laurentis, S.D. Odintsov, G. Scelza and An. Stabile for  useful discussions and comments.


%


\bibliographystyle{mdpi}

\begin{thebibliography}{1}


\bibitem{riess}
         Riess A.G. {\it et al.}, Observational Evidence from Supernovae for an Accelerating Universe. \emph{The Astronomical Journal} 1998, {\bf 116}, 1009-1038.
         
\bibitem{ast}
          Perlmutter S. {\it et al.}, Masurements of $\Omega$ and $\Lambda$ form 42 High-redshift Supernovae. \emph{The Astrophysical Journal} 1999, {\bf 517}, 565-586.
          
\bibitem{clo}
          Cole S. {\it et al.}, The 2dF Galaxy Redshift Survey: Power-spectrum analysis of the final dataset and cosmological implications. {\it Monthly Notices of the Royal Astronomical Society} 2005, {\bf 362}, 505-534.
          
\bibitem{spe}
         Spergel D.N. {\it et al.}, Three-Year Wilkinson Microwave Anisotropy Probe (WMAP) Observations: Implications for Cosmology. {\it Astrophysical Journal Supplement Series} 2007, {\bf 170}, 377-408.

\bibitem{carrol}
         Carroll S.M.; Press W.H.; Turner E.L., The cosmological Constant. {\it Annual Review of Astronomy and Astrophysics} 1992, {\bf 30}, 499-542.
         
\bibitem{sahini}
         Sahni V.; Starobinski A., The Case for a Positive Cosmological $\Lambda$-term. {\it International Journal of Modern Physics D} 2000, {\bf 9}, 373-443.

\bibitem{NFW}
         Navarro J.F.; Frenk C.S.; White S.D.M., The Structure of Cold Dark Matter Halos. {\it Astrophysical Journal} 1996, {\bf 462}, 563-575.
      
\bibitem{report}   
        Capozziello S.; De Laurentis M., Extended theories of gravity. {\it Physics Reports} 2011, {\bf 509}, 167-321.
        
\bibitem{repsergei}
        Nojiri S.; Odintsov S.D., Properties of singularities in the (phantom) dark energy universe. {\it Physics Reports} 2011, {\bf 505}, 59-104.
        
\bibitem{olmo_palatini}
        Olmo G., Palatini Approach to Modified Gravity: $f(R)$ Theories and Beyond. {\it International Journal of Modern Physics D} 2011, {\bf 20}, 413-462.     
        

\bibitem{annalen}
         Capozziello S.; De Laurentis M., The Dark Matter Problem from $f(R)$ Gravity Viewpoint. {\it Annalen der Physik} 2012, {\bf 524}, 545-578. 
         
\bibitem{weyl_1}
        Weyl H., Reine Infinitesimalgeometrie. {\it Mathematische Zeitschrift} 1918, \textbf{2}, 384-411.

\bibitem{weyl_2}
       Weyl H., {\it Raum zeit Materie: Vorlesungen uuber allgemeine Relativitatstheorie}, Springer, Berlin, 1921.

\bibitem{pauli}
       Pauli W., Zur Theorie der Gravitation und der Elektrizitat von Hermann Weyl. {\it Phys. Zeit.} 1919, \textbf{20}, 457-467.

\bibitem{bach}
      Bach R., Zur Weylschen Relativitatstheorie und der Weylschen Erweiterung des Krummungstensorbegriffs. {\it Mathematische Zeitschrift} 1921, \textbf{9}, 110-135.

\bibitem{edd}
       Eddington A.S., \emph{The Mathematical Theory of Relativity}, Cambridge University Press London, 1924.

\bibitem{lan}
     Lanczos C., Elektromagnetismus als natürliche Eigenschaft der Riemannschen Geometrie. {\it Zeitschrift fur Physik A Hadrons and Nuclei} 1932, \textbf{73}, 147-168.

\bibitem{buc}
     Buchdahl H.A., On the Gravitational Field Equations Arising from the Square of the Gaussian Curvature. {\it Il Nuovo Cimento} 1962, \textbf{23}, 141-157.
         
\bibitem{bic}
     Bicknell G.V., Non-viability of gravitational theory based on a quadratic lagrangian. {\it Journal of Physics A: Mathematical, Nuclear and General} 1974, {\bf 7}, 1061.

\bibitem{guth}
        Guth A., Inflationary universe: A possible solution to the horizon and flatness problems. {\it Physical Review D} 1981, {\bf 23}, 347-356.
         

\bibitem{book}
        Capozziello S.; Faraoni V., {\it Beyond Einstein Gravity}, Fundamental Theories of Physics Vol. 170, Springer, Dordrecht, 2011.

\bibitem{psaltis}
         Psaltis D., Probes and Tests of Strong-Field Gravity with Observations in the Electromagnetic Spectrum. {\it Living Review Relativity} 2008, \textbf{11}.
         
\bibitem{briscese}
         Briscese F.; Elizalde E.; Nojiri S.; Odintsov S.D., Phantom scalar dark energy as modified gravity: Understanding the origin of the Big Rip singularity. {\it Physics Letters B} 2007, \textbf{646}, 105-111.
         
\bibitem{babi_1}
         Babichev E.; Langlois D., Relativistic stars in $f(R)$ gravity. {\it Physical Review D} 2009, {\bf 80}, 121501(5).
         
\bibitem{babi_2}
         Babichev E.; Langlois D., Relativistic stars in $f(R)$ and scalar-tensor theories. {\it Physical Review D} 2010, {\bf 81}, 124051(15).
         
\bibitem{mag} 
       Muno, M.P. {\it et al}., A Neutron Star with a Massive Progenitor in Westerlund 1. {\it The Astrophysical Journal} 2006, {\bf 636}, L41-L44.
       
\bibitem{neutron1} 
 Astashenok A.V.; Capozziello S.; Odintsov S.D., Further Stable Neutron Star Models from $f(R)$ Gravity. {\it Journal of Cosmology and Astroparticle Physics} 2013, \textbf{12}, 040.

\bibitem{neutron2} 
Astashenok A.V.; Capozziello S.; Odintsov S.D., Maximal Neutron Star Mass and the Resolution of the Hyperon Puzzle in Modified Gravity. {\it Physical Review D} 2014, {\bf 89}, 103509(8).

\bibitem{jaime}  
Jaime L.G.; Patino L.; Salgado M., Robust approach to $f(R)$ gravity. {\it Physical Review D} 2011, {\bf 83}, 024039(5).

\bibitem{santos1}
Santos E., Neutron Stars in Generalized $f(R)$ Gravity. {\it Astrophysics and Space Science} 2012, {\bf 341}, 411-416.


\bibitem{yazadjiev}
Yazadjev S.S.; Doneva D.D.; Kokkotas, K.D.; Staykov K.V., Non-perturbative and Self-consistent Models of Neutron Stars in R-squared Gravity. {\it Journal of Cosmology and Astroparticle Physics} 2014, \textbf{1406}, 003.

\bibitem{staykov}
Staykov K.V.; Doneva D.D.; Yazadjiev S.S.; Kokkotas K.D., Slowly Rotating Neutron and Strange Stars in $R^2$ Gravity. arXiv 2014, 1407.2180 [gr-qc].

\bibitem{orellana}
Orellana M.; Garcia F.; Pannia F.; Romero G., Structure of Neutron Stars in $R$-squared Gravity. {\it General Relativity and Gravitation} 2013, {\bf 45}, 771-783.

\bibitem{cooney}
Cooney A.; De Deo S.; Psaltis D., Neutron Stars in $f(R)$ Gravity with Perturbative Constraints. {\it Physical Review D} 2010, {\bf 82}, 064033(7).

\bibitem{arapoglu}
Arapoglu A.S.; Deliduman C.; Eksi K.Y., Constraints on Perturbative $f(R)$ Gravity via Neutron Stars. {\it Journal of Cosmology and Astroparticle Physics} 2011, {\bf 07}, 020.

\bibitem{alavirad}
Alavirad H.; Weller J.M., Modified Gravity with Logarithmic Curvature Corrections and the Structure of Relativistic Stars. {\it Physical Review D} 2013, {\bf 88}, 124034(17).

\bibitem{ganguly}
Ganguly A.; Gannouji R.; Goswami R.; Ray S.; Neutron Stars in the Starobinsky Model. {\it Physical Review D} 2013, {\bf 89}, 064019(11).

\bibitem{farinelli}
Farinelli R.; De Laurentis M.; Capozziello S.; Odintsov S.D., Numerical Solutions of the Modified Lan\'{e} - Emden Equation in $f(R)$-Gravity. {\it Monthly Notices of the Royal Astronomical Society} 2014, {\bf 440}, 2909-2915.

\bibitem{formisano}
Capozziello S.; De Laurentis M.; De Martino I.; Formisano M.; Odintsov S.D., Jeans Analysis of Self-Gravitating Systems in $f(R)$ Gravity. {\it Physical Review D} 2012, {\bf 85}, 044022(8).
         
\bibitem{bondi}
         Bondi H., \emph{Cosmology}, Cambridge University Press London, 1952.
      

\bibitem{bra-dic}
         Brans C.; Dicke R. H., Mach's Principle and a Relativistic Theory of Gravitation. {\it Physical Review} 1961, {\bf 124}, 925-935.

\bibitem{cap-der-rub-scu}
         Capozziello S.; de Ritis R.; Rubano C.; Scudellaro P., Nother Symmetries in Cosmology. {\it La Rivista del Nuovo Cimento} 1996, \textbf{19}, 1-114.
 
\bibitem{sciama}
         Sciama D. W., On the Origin of Inertia. {\it Monthly Notices of the Royal Astronomical Society} 1953, \textbf{113}, 34-42.
         
\bibitem{mag-fer-fra}
         Magnano G.; Ferraris M.; Francaviglia M., Nonlinear Gravitational Lagrangians. {\it General Relativity and Gravitation} 1987, \textbf{19}, 465-479.            
 
\bibitem{ama-elg-mot-mul}
         Amarzguioui M.; Elgaroy O.; Mota D. F.; Multamaki T., Cosmological Constraints on $f(R)$ Gravity Theories within the Palatini Approach. {\it Astronomy and Astrophysics} 2006, \textbf{454}, 707-714.   
         
\bibitem{all-bor-fra}
         Allemandi G.; Borowiec A.; Francaviglia M., Accelerated Cosmological Models in Ricci Squared Gravity. {\it Physical Review D} 2004, \textbf{70}, 103503(13).

\bibitem{sot1}
         Sotiriou T.P., Constraining $f(R)$ Gravity in the Palatini Formalism. {\it Classical and Quantum Gravity} 2006, \textbf{23}, 1253.

\bibitem{sot-lib}
         Sotiriou T.P.; Liberati S., Metric-affine $f(R)$ Theories of Gravity. {\it Annals of Physics} 2007, \textbf{322}, 935-966.
              
\bibitem{dam-esp}
         Damour T.; Esposito - Far\`{e}se G., Tensor-multi-scalar Theories of Gravitation. {\it Classical and Quantum Gravity} 1992, \textbf{9}, 2093.

\bibitem{olmo1}
         Olmo G.J., Post-Newtonian constraints on $f(R)$ cosmologies in metric and Palatini formalism. {\it Physical Review D} 2005, \textbf{72}, 08350589(17).
         

\bibitem{olmo2}
         Olmo G.J., The Gravity Lagrangian According to Solar System Experiments. {\it Physics Review Letters} 2005, \textbf{95}, 261102(4).

\bibitem{olmo3}
         Olmo G.J., Limit to General Relativity in $f(R)$ Theories of Gravity. {\it Physical Review D} 2007, \textbf{75}, 023511(8).

\bibitem{clifton}
         Clifton T., Parametrized post-Newtonian Limit of Fourth-order Theories of Gravity. {\it Physical Review D} 2008, \textbf{77}, 024041(11).

\bibitem{odintsov}
         Nojiri S.; Odintsov S.D., Modified Gauss - Bonnet Theory as Gravitational Alternative for Dark Energy. {\it Physics Letters B} 2005, \textbf{631}, 1.

\bibitem{PRD}
         Capozziello S.; Stabile A.; Troisi A., Newtonian Limit of $f(R)$ Gravity. {\it Physical Review D} 2007, {\bf 76}, 104019(12).      
      
\bibitem{PRD1}
         Stabile A., Post-Newtonian Limit of $f(R)$ Gravity in the Harmonic Gauge. {\it Physical Review D} 2010, {\bf 82}, 064021(12).

\bibitem{PRD2}
         Stabile A., Most General Fourth-Order Theory of Gravity at Low Energy. {\it Physical Review D} 2010, {\bf 82}, 124026(7).

\bibitem{Stabile_Capozziello}
         Capozziello S.; Stabile A., The Newtonian Limit of Metric Gravity Theories with Quadratic Lagrangians. {\it Classical and Quantum Gravity} 2009, \textbf{26}, 085019.

\bibitem{CCCT}
         Capozziello S.; Cardone V.F.; Carloni S.; Troisi A.; Can Higher Order Curvature Theories Explain Rotation Curves of Galaxies?. {\it Physicis Letters A} 2004, \textbf{326}, 292-296.

\bibitem{minko}
         Capozziello S.; Stabile A.; Troisi A.; The Post-Minkowskian Limit of $f(R)$ Gravity. {\it International Journal of Theoretical Physics} 2010, \textbf{49}, 1251-1261.
   
         
\bibitem{quadrupolo}   
 De Laurentis M.; Capozziello S., Quadrupolar Gravitational Radiation as a Test-Bed for $f(R)$ Gravity. {\it Astroparticle Physics}, {\bf 35}, 257-265.
         
\bibitem{CCT}
         Capozziello S.; Cardone V.F.; Troisi A., Low Surface Brightness Galaxy Rotation Curves in the Low Energy Limit of $R^n$ Gravity: No Need for Dark Matter? {\it Monthly Notices of the Royal Astronomical Society} 2007, \textbf{375}, 1423-1440. 
 
\bibitem{BHL}
         Boehmer C.G.; Harko T; Lobo F.S.N., Dark Matter as a Geometric Effect in $f(R)$ Gravity. {\it Astroparticle Physics} 2008, \textbf{29}, 386-392.

\bibitem{BHL1}
         Boehmer C.G.; Harko T; Lobo F.S.N., The Generalized Virial Theorem in $f(R)$ Gravity. {\it Journal of Cosmology and Astroparticle Physics} 2008, \textbf{0803}, 024. 
      
\bibitem{stabile_scelza}
          Stabile A.; Scelza G., Rotation Curves of Galaxies by Fourth Order Gravity. {\it Physical Review D} 2012, {\bf 84}, 124023(11).  
         
\bibitem{hydrostatic}
          Capozziello S.; De Laurentis M.; Odintsov S.D.; Stabile A., Hydrostatic Equilibrium and Stellar Structure in $f(R)$ Gravity. {\it Physical Review D} 2011, {\bf 83}, 064004(6).
     
\bibitem{stabile_stabile}
          Stabile A.; Stabile An., Weak Gravitational Lensing in Fourth Order Gravity. {\it Physical Review D} 2012, {\bf 85}, 044014(10).
          
\bibitem{jetzer}
         Lubini M.; Tortora C.; Naef J.; Jetzer Ph.; Capozziello S., Probing the Dark Matter Issue in $f(R)$ Gravity via Gravitational Lensing. {\it The European Physical Journal C} 2011, {\bf 71}, 1834.        

\bibitem{stabile_stabile_cap}
          Stabile A.; Stabile An.; Capozziello S., Conformal Transformations and Weak Field Limit of Scalar-Tensor Gravity. {\it Physical Review D} 2013, {\bf 88}, 124011(9).

\bibitem{will}
         Will, C.M., \emph{Theory and Experiment in Gravitational Physics}, Cambridge University Press, London, 1993.


\bibitem{landau}
         Landau L.D., Lif\v{s}its E.M., \emph{Theoretical Physics} vol. II, Butterworth Heinemann, University of Minnesota, 1987.
         
\bibitem{santos}
         Santos E., Quantum Vacuum Effects as Generalized $f(R)$ Gravity: Application to Stars. {\it Physical Review D} 2010, \textbf{81}, 064030(14).   
      
\bibitem{spher_symm_fR}
         Capozziello S.; Stabile A.; Troisi A., Spherical Symmetry in $f(R)$-Gravity. {\it Classical and Quantum Gravity} 2008, {\bf 25}, 085004.

\bibitem{newtonian_limit_fR}
         Capozziello S.; Stabile A.; Troisi A., A General Solution in the Newtonian Limit of $f(R)$ Gravity. {\it Modern physics letters A} 2009, {\bf 24}, 659.

\bibitem{weinberg}
         Weinberg S., \emph{Gravitation and Cosmology}, Wiley, New York, 1972.
         
\bibitem{mio1}
         Capozziello S.; Stabile A., The Weak Field Limit of Fourth Order Gravity. \emph{Classical and Quantum Gravity: Theory, Analysis and Applications}, chapter 2, 1 Nova Science Publishers, Inc 2010, New York, 2010.

\bibitem{dewitt_book}
         de Witt B.S., \emph{Dynamical Theory of Groups and Fields}, Gordon and Breach, New York, 1965.
         
\bibitem{kippe}
         Kippenhahn R.; Weigert A., {\it Stellar Structures and Evolution}, Springer-Verlag, Berlin, 1990.
         
\bibitem{CST}
         Capozziello S.; Stabile A.; Troisi A.,  Comparing Scalar–Tensor Gravity and $f(R)$-Gravity in the Newtonian Limit. {\it Physics Letters  B} 2010,  {\bf 686}, 79-83.
         
\bibitem{deh}
         Dehnen W.; Binney J., Local Stellar Kinematics from HIPPARCOS Data. {\it Monthly Notice of the Royal Astronomical Society} 1998, \textbf{298}, 387-394.
         

\bibitem{kui}
         Kuijken K.; Gilmore, G., The Mass Distribution in the Galactic Disc - II - Determination of the Surface Mass Density of the Galactic Disc Near the SunThe Mass Distribution in the Galactic Disc - II - Determination of the Surface Mass Density of the Galactic Disc Near the Sun. {\it Monthly Notices of the Royal Astronomical Society} 1989, \textbf{239}, 605-649.
         
\bibitem{navarro}
        Navarro I.; Van Acoleyen K., On the Newtonian Limit of Generalized Modified Gravity Models. {\it Phyiscs Letters B} 2005, {\bf 622}, 1-5.
        
\bibitem{wyse}
         Wyse R.F.G., Gilmore G., Franx M., Galactic Bulges. {\it Annual Review of Astronomy and Astrophysics} 1997, \textbf{35}, 637-675.

\bibitem{noo}
         Noordermeer E., The Rotation Curves of Flattened S\'{e}rsic Bulges. {\it Monthly Notices of the Royal Astronomical Society} 2008, \textbf{385}, 1359-1364.

\bibitem{bin}
         Binney J.; Tremaine S., \emph{Galactic Dynamics} - Princeton University Press, Princeton, 1987.

\bibitem{vau}
         de Vaucouleurs G., Photoelectric Photometry of the Andromeda Nebula in the UBV System. {\it The Astrophysical Journal} 1958, \textbf{128}, 465.

\bibitem{bur}
         Burkert A., The Structure of Dark Matter Halos in Dwarf Galaxies. {\it The Astrophysical Journal Letters} 1995, \textbf{447}, L25-L28.

\bibitem{sofue}
         Sofue Y.; Honma M.; Omodaka T., Unified Rotation Curve of the Galaxy. Decomposition into de Vaucouleurs Bulge, Disk, Dark Halo, and the $9$ kpc Rotation Dip. {\it The Astronomical Society of Japan} 2009, \textbf{61}, 227-236.

\bibitem{BG}
         Burton W.B.; Gordon M.A., Carbon Monoxide in the Galaxy. III - The Overall Nature of its Distribution in the Equatorial Plane. {\it Astronomy and Astrophysics} 1978,  \textbf{63}, 7-27.
         
\bibitem{CL}
         Clemens D. P., Massachusetts Stony Brook Galactic Plane CO Survey: The Galactic Disk Rotation Curve. The Galactic Disk Rotation Curve.  {\it The Astrophysical Journal} 1985, \textbf{295}, 422-436.


\bibitem{FBS}
         Fich M.; Blitz L.; Stark A.A., The Rotation Curve of the Milky Way to $2R_0$. {\it The Astrophysical Journal} 1989, \textbf{342}, 272-284.

\bibitem{BFS}
         Blitz L.; Fich M.; Stark A.A., Catalog of CO Radial Velocities Toward Galactic H II Regions. {\it Astrophysical Journal Supplement Series} 1982, \textbf{49}, 183-206.
         
\bibitem{DB}
         Demers S.; Battinelli P., C Stars as Kinematic Probes of the Milky Way Disk from 9 to 15 kpc. {\it Astronomy and Astrophysics} 2007, \textbf{473}, 143-148.

\bibitem{sofue_2}
         Honma M.; Sofue Y., Rotation Curve of the Galaxy. {\it The Astronomical Society of Japan} 1997, \textbf{49}, 453-460.
        
\bibitem{sofue_3}
         Honma M.; Sofue Y., On the Keplerian Rotation Curves of Galaxies. {\it The Astronomical Society of Japan} 1997, \textbf{49}, 539-545.        

\bibitem{HBCHI}
         Sato M. {\it et al.}, Absolute Proper Motions of $H_2O$ Masers Away from the Galactic Plane Measured with VERA in the Superbubble Region NGC 281. {\it The Astronomical Society of Japan} 2007, \textbf{59}, 743-751.

\bibitem{data_MW}
         Sofue Y.; Honma M.; Omodaka T., Errata: Unified Rotation Curve of the Galaxy Decomposition into de Vaucouleurs Bulge, Disk, Dark Halo, and the 9-kpc Rotation Dip. {\it The Astronomical Society of Japan} 2010, \textbf{62}, 1367-1367.

\bibitem{data_3198}
         van Albada T.S.; Bahcall J.N.; Begeman K.; Sanscisi R., Distribution of Dark Matter in the Spiral Galaxy NGC 3198. {\it The Astrophysical Journal} 1985, \textbf{295}, 305-313.

\bibitem{sugg1}
         Pogosian L.; Silvestri A., Pattern of Growth in Viable $f(R)$ Cosmologies. {\it Physical Review D} 2008, \textbf{77}, 023503(15). [Erratum-ibid. D 81 (2010) 049901]
         
\bibitem{sugg2}
        Bean R.; Bernat D.; Pogosian L.; Silvestri A.; Trodden M., Dynamics of Linear Perturbations in $f(R)$ Gravity. {\it Physical Review D} 2007, \textbf{75} 064020(12).


\bibitem{sugg3}
         Bertschinger E.; Zukin P.; Distinguishing mMdified Gravity from Dark Energy. {\it Physical Review D} 2008, \textbf{78}, 024015(13).

\bibitem{sugg4}
         Carloni S.; Dunsby P.K.S.; Troisi A., Evolution of Density Perturbations in $f(R)$ Gravity. {\it Physical Review D} 2008, \textbf{77}, 024024(17).

\bibitem{sugg5}
         Nzioki A.M.; Dunsby P.K.S.; Goswami R.; Carloni S., Geometrical Approach to Strong Gravitational Lensing in $f(R)$ Gravity. {\it Physical Review D} 2011, \textbf{83}, 024030(10).
         
\bibitem{schneider}
        Schneider P.; Ehlers J.; Falco E.E., \emph{Gravitational Lenses}, Springer, New York, 1999. 
         
\bibitem{odi2005}
         Nojiri S.; Odintsov S.D.; Tsujikawa S.; Properties of Singularities in the (phantom) Dark Energy Universe. {\it Physical Review D} 2005, {\bf 71}, 063004(16).
         
\bibitem{singularity}
         Capozziello S.; De Laurentis M.F.; Nojiri S.; Odintsov S.D., Classifying and Avoiding Singularities in the Alternative Gravity Dark Energy Models. {\it Physical Review D} 2009, {\bf 79}, 124007(16).
         
\bibitem{sta_cap}
         Stabile A.; Capozziello S., Galaxy Rotation Curves in $ f(R,\phi)$ Gravity. {\it Physical Review D} 2013, {\bf 87}, 064002(13).
                   
\bibitem{dick} 
         Dick R., Letter: On the Newtonian Limit in Gravity Models with Inverse Powers of $R$. {\it General Relativity and Gravitation} 2004, {\bf 36}, 217-224.
         
\bibitem{hans}
         Schmidt H.J., The Newtonian Limit of Fourth-order Gravity . {\it Astronomische Nachrichten} 1986, {\bf 307}, 339-340.

\bibitem{stelle}
         Stelle K., Classical Gravity with Higher Derivatives. {\it General Relativity and Gravitation} 1978, {\bf 9}, 353-371.


\bibitem{ohanlon}
         O'Hanlon J., Intermediate-Range Gravity: A Generally Covariant Model. {\it Physics Review Letters} 1972, {\bf 29}, 137-138.


\end{thebibliography}
\makeatletter
\renewcommand\@biblabel[1]{#1. }
\makeatother

\end{document}